\documentclass[10pt,DIV=18,twoside=no,BCOR=0pt,abstract]{scrartcl}
\usepackage{etex}
\pdfoutput=1

\usepackage[utf8]{inputenc}
\usepackage[english]{babel}

\usepackage[T1]{fontenc}

\usepackage{enumitem}

\usepackage[svgnames]{xcolor}

\usepackage[sumlimits]{amsmath}
\usepackage{amsthm,amsfonts,amssymb}
\usepackage[scaled=1.0]{rsfso}

\usepackage{latexsym,mathrsfs,ifpdf,wasysym}
\usepackage{esint}

\usepackage{bbold}
\usepackage{dsfont}

\ifpdf
\usepackage[pdftex]{graphicx}
\usepackage{ae,aeguill}
\else
\usepackage{graphicx}
\fi

\usepackage{pifont,mathtools}

\usepackage{tcolorbox}
\tcbuselibrary{skins,breakable,theorems}

\usepackage{rotating}
\usepackage{colortbl}
\usepackage[math]{cellspace}  
\usepackage{multirow}
\addparagraphcolumntypes{X}
\newcolumntype{W}{>{$}Sc<{$}}
\makeatletter
\def \bcolumn #1#2\@nil {%
  \cellspace@lrtrue\global\cellspace@false
  \@expandtwoargs \in@{#1}{\cellspace@parcoltypes}%
  \ifin@
    \cellspace@lrfalse
  \fi
  \ifcellspace@lr
    \begingroup \lrbox {\@tempboxa}%
  \else
    \global\cellspace@true
  \fi
}
\makeatother

\usepackage{ulem}

\newcommand{\dd}{\text{d}}

\newcommand{\norm}[1]{\lVert#1\rVert}

\newcommand{\vecsq}[1]{\vec{#1}^{\,2}}
\allowdisplaybreaks

\DeclarePairedDelimiter\abs{\lvert}{\rvert}   
\DeclarePairedDelimiterX\brakket[3]{\langle}{\rangle}{#1\,\delimsize\vert\,#2\,\delimsize\vert\,#3}%
\DeclarePairedDelimiterX\braket[2]{\langle}{\rangle}{#1\,\delimsize\vert\,#2}%
\DeclarePairedDelimiterX\bra[1]{\langle}{\rvert}{#1}%
\DeclarePairedDelimiterX\ket[1]{\lvert}{\rangle}{#1}%
\DeclarePairedDelimiterX\innerprod[2]{\bigl(}{\bigr)}{{#1},\hspace*{0.2ex}{#2}}%
\DeclarePairedDelimiterX\CG[3]{\langle}{\rangle}{#1\,;\,#2\,\delimsize\vert\,#3}%
\DeclarePairedDelimiterX\CGY[3]{\langle}{\rangle_Y}{#1\,;\,#2\,\delimsize\vert\,#3}%
\DeclarePairedDelimiterX\scalprod[2]{\langle}{\rangle}{\,\vec#1\,,\,#2\,}%

\newlist{order}{enumerate}{3}
\setlist[order]{label=\arabic*${}^\text{o}$)}

\newlist{maliste}{enumerate}{3}
\setlist[maliste]{label=\arabic*${}^\text{o}$)}

\newlist{malistealph}{enumerate}{3}
\setlist[malistealph]{label=\alph*)}

\newlist{orderalph}{enumerate}{3}
\setlist[orderalph]{label=\alph*$)$}

\makeatletter
\renewcommand{\thesection}{\@Roman\c@section}
\renewcommand{\thesubsubsection}{\@Roman\c@section.\@arabic\c@subsection.\@alph\c@subsubsection}
\makeatother

\theoremstyle{break}

\theoremstyle{definition}

\theoremstyle{remark}

\newfont{\titreb}{ecsx1440 scaled \magstep1}
\newfont{\titrec}{ecsx1728 scaled \magstep3}
\newfont{\titrea}{ecsx1095 scaled \magstep0}
\newfont{\titred}{ecsx0600 scaled \magstep1}

\newcommand{\reel}{\ensuremath\mathbb{R}}

\DeclareMathAlphabet{\mathscrbf}{OMS}{mdugm}{b}{n}

\newcommand{\field}[1]{\mathbb{#1}}

\newcommand{\llvec}[1]{\overrightarrow{{#1}}}

\def\dr{\text{\Pisymbol{psy}{182}}}

\newcommand{\ra}[1]
{\mathchoice%
{\overset{\mbox{\xymatrix{*{\hphantom{\displaystyle #1}}%
\ar[]+L;[]+R}}}{\displaystyle #1}}%
{\overset{\mbox{\xymatrix{*{\hphantom{\textstyle #1}}%
\ar[]+L;[]+R}}}{\textstyle #1}}%
{\overset{\mbox{\xymatrix{*{\hphantom{\scriptstyle #1}}%
\ar[]+L;[]+R}}}{\scriptstyle #1}}%
{\overset{\mbox{\xymatrix{*{\hphantom{\scriptscriptstyle #1}}%
\ar[]+L;[]+R}}}{\scriptscriptstyle #1}}}

\renewcommand{\ra}[1]%
{\overset{\raisebox{-0pt}{\mbox{\xymatrix{*{\hphantom{#1}}%
\ar[]+L;[]+R}}}}{#1}}

\newcommand{\overbar}[1]{\mkern 1.5mu\overline{\mkern-1.5mu#1\mkern-1.5mu}\mkern 1.5mu}%

\makeatletter
\newsavebox\myboxA
\newsavebox\myboxB
\newlength\mylenA

\newcommand*\xoverline[2][0.75]{%
    \sbox{\myboxA}{$\m@th#2$}%
    \setbox\myboxB\null
    \ht\myboxB=\ht\myboxA%
    \dp\myboxB=\dp\myboxA%
    \wd\myboxB=#1\wd\myboxA
    \sbox\myboxB{$\m@th\overline{\copy\myboxB}$}
    \setlength\mylenA{\the\wd\myboxA}
    \addtolength\mylenA{-\the\wd\myboxB}%
    \ifdim\wd\myboxB<\wd\myboxA%
       \rlap{\hskip 0.5\mylenA\usebox\myboxB}{\usebox\myboxA}%
    \else
        \hskip -0.5\mylenA\rlap{\usebox\myboxA}{\hskip 0.5\mylenA\usebox\myboxB}%
    \fi}
\makeatother

\def\dr{\text{\Pisymbol{psy}{182}}}
\def\fmslash#1{%
\setbox1=\hbox{$#1$}%
\setbox2=\hbox{$\slash$}%
\setbox3=\hbox{$#1\hskip-.5\wd1\hskip-.5\wd2\slash$}%
\box3\hskip.5\wd1\hskip-.5\wd2}

\newcommand{\boostu}[1]{\dfrac{1 + \fmslash{#1}\,\gamma^0}{\sqrt{2\,(1 + #1^o)}}}%
\newcommand{\recboostu}[1]{\dfrac{1 + \gamma^0\,\fmslash{#1}}{\sqrt{2\,(1 + #1^o)}}}%

\definecolor{pastelgray}{rgb}{0.81, 0.81, 0.77}
\definecolor{gray}{rgb}{0.75, 0.75, 0.75}
\newcommand{\inv}[1]{\colorbox{pastelgray}{\color{black}$#1$}}%

\def\noi{\noindent}

\roman{enumii}
\begin{document}\noindent
\ifpdf
\DeclareGraphicsExtensions{.pdf, .jpg}
\else
\DeclareGraphicsExtensions{.eps, .jpg}
\fi
\tcbset{breakable,top=-2mm,pad after break=3mm,colback=black!5!white,colframe=black!75!black,boxrule=1.0pt}

\begin{center} 

    {\LARGE \bfseries Isgur-Wise functions for $\boldsymbol{\Lambda_b \to \Lambda_c\left({1 \over 2}^\pm \right)}$ transitions in the Bakamjian-Thomas Relativistic Quark Model} \\

    \par \vskip 8 truemm

    V. Mor\'enas
    
    {\itshape Laboratoire de Physique de Clermont Auvergne (UMR 6533)\\
    Campus Universitaire des Cézeaux\\
    4 Avenue Blaise Pascal, 63178 Aubière Cedex, France\\
    E-mail: morenas@in2p3.fr}
    
    \vskip 8 truemm
    
     {A. Le Yaouanc and L. Oliver}
    
    {\itshape IJCLab, P\^ole Th\'eorie, CNRS/IN2P3 et Universit\'e Paris-Saclay\par
    b\^at. 210, 91405 Orsay, France}
    
\end{center}

    \vskip 7 truemm
    
    \begin{abstract}
    
    \noi We study the transitions ${\Lambda_b \to \Lambda_c\left({1 \over 2}^\pm \right)}$ in the Bakamjian-Thomas (BT) relativistic quark model formalism, which describes hadrons with a fixed number of constituents. In the heavy quark mass limit, the BT model yields covariant form factors and Isgur-Wise (IW) scaling, regardless of the spectroscopic model used to describe the bound states. It has been extensively applied to heavy mesons where subtle properties of the IW limit of QCD, including the Bjorken-Uraltsev sum rules, have been shown to be satisfied. The present paper, where the BT construction is applied to baryons, is unavoidably technical because one is dealing with a three-body problem. The complications originate from the natural choice of the Jacobi coordinates ${\vec \rho}$ (relative space coordinate between the two light spectator quarks) and ${\vec \lambda}$ (relative space coordinate between the center-of-mass of the two light quarks and the heavy quark). The corresponding orbital angular momenta are denoted by ${\vec \ell}_\rho$ and ${\vec \ell}_\lambda$, with ${\vec \ell}_\rho + {\vec \ell}_\lambda = {\vec L}$. For the transitions $\Lambda_b \to \Lambda_c\left({1 \over 2}^\pm\right)$, i.e. $L = 0 \to L = 0$ or $L = 0 \to L = 1$, one can see that the moduli $\ell_\rho$ and $\ell_\lambda$ can take an infinite number of values. For $L = 0 \to L = 0$ one has the constraint $\ell_\lambda = \ell_\rho$ and for $L = 0 \to L = 1$ the constraint is $\ell_\lambda = \ell_\rho \pm 1$. We compute explicitly the IW function $\xi_\Lambda (w)$ in the elastic case $\Lambda_b \left({1 \over 2}^+ \right) \to \Lambda_c \left({1 \over 2}^+ \right)$ and the much more involved IW function $\sigma_\Lambda (w)$ in the inelastic case $\Lambda_b \left({1 \over 2}^+ \right) \to \Lambda_c \left({1 \over 2}^- \right)$. These functions exhibit the expected properties of covariance and IW scaling.
    
\end{abstract}
    
    
    \section{Introduction} \hspace*{\parindent}
    The heavy quark limit of QCD implies poweful constraints on form factors. In the elastic case for transitions ${1 \over 2}^- \to {1 \over 2}^-$ for the light cloud $\overline{B} \to D^{(*)} \ell \nu$, this limit implies that all form factors are given in terms of a single function, the famous Isgur-Wise (IW) function $\xi(w)$ \cite{IW-1}.\par
    A sum rule (SR) formulated by Bjorken \cite{BJORKEN} in the heavy quark limit of QCD implies the lower limit $\rho^2 = -\xi'(1) \geq {1 \over 4}$. This SR was formulated in a transparent way by Isgur and Wise, in terms of the IW functions $\tau_{1/2}(w)$ and $\tau_{3/2}(w)$ for the inelastic transitions ${1 \over 2}^- \to {1 \over 2}^+, {3 \over 2}^+$ at zero recoil  \cite{IW-2}. Ten years later, a new SR was discovered by Uraltsev \cite{URALTSEV-1}, making use of the non-forward amplitude $\overline{B}(v_i) \to D^{(n)}(v') \to \overline{B}(v_f)$ ($v_i \not = v_f$). Uraltsev SR combined with Bjorken's yields the much more powerful lower bound for the elastic slope, $\rho^2 \geq {3 \over 4}$. In a number of papers we generalized Bjorken and Uraltsev SR, and we obtained a whole tower of SR that allow to constrain the higher derivatives of the elastic IW function $\xi(w)$. In particular, we found lower bounds on the successive derivatives \cite{LOR-1,LOR-2}, and also an improved lower bound of the curvature in terms of the slope \cite{LOR-3}. Similar results were also formulated for the baryon case $j^P = 0^+$ \cite{LOR-4}.\par
    With the aim of providing predictions for observables in $\Lambda_b \to \Lambda_c^{(*)} \ell \overline{\nu}$, we are considering the framework of quark models, which can tackle simultaneously the ground state and the excitations. Explicitely, we use the Bakamjian-Thomas (BT) relativistic quark model construction which has the fundamental advantage to generate fully covariant transition amplitudes written in terms of wave functions in the rest frame of the hadron. One remarkable fact is that the mass operator, i.e. the Hamiltonian operator in the rest frame, which drives the wave functions at rest, does not need to be specified for the covariance to be valid: it only needs to be rotationally invariant, to solely depend upon the internal variables and to conserve parity. Moreover, the BT construction yields Lorentz invariant IW function in terms of these internal hadron wave functions at rest, and, being a quark model, it describes hadrons as systems with a fixed number of constituents. In order to produce numerical predictions, we need to use an explicit mass operator, that is a spectroscopic quark model Hamiltonian, which fits the observed spectrum. The resulting IW function provides the leading order of the form factors in a heavy quark mass expansion.\par 
    We have applied this BT scheme to the semileptonic $B$ meson decay case $\overline{B} \to D^{(*,**)} \ell \overline{\nu}$ into ground states ($L=0$)~\cite{COVARIANT-QM} and orbitally excited $(L=1)$ mesons~\cite{MORENAS-1}. In the meson case we used, as spectroscopic Hamitonian, the one of Godfrey and Isgur (GI), which describes a wealth of meson data for the $q \overline{q}$ and $Q \overbar{q}$ systems \cite{GODFREY-ISGUR}. Then, a reasonable and theoretically founded description of IW functions, both elastic and inelastic \cite{MORENAS-2}, was obtained. We also studied within the same framework the decay constants of heavy mesons~\cite{MORENAS-3}. Finally, the BT formalism was used to consider certain aspects of the heavy quark limit of current matrix elements, its relationship to HQET and some properties of the IW scaling functions~\cite{LEYAOUANC}.\par
    It is worth emphasizing that this method allows for the study of excited states which are not easily approached using more modern and non perturbative fundamental techniques such as Lattice QCD, while the predictions for the ground states are consistent with Lattice data (see the discussion in~\cite{MORENAS-3}).\par
      In the baryon case, a rather complicated but complete mass operator could be the spectroscopic quark model of Capstick and Isgur \cite{CAPSTICK-ISGUR}. This model amounts to solve a Schrödinger-like equation with a Hamiltonian decomposed into a relativistic kinetical term and a complicated QCD-inspired potential energy term:
\[
    {\cal  H}^\text{kin} = \sqrt{{\vec p}_1^{\,2} + m_1^2} + \sqrt{{\vec p}_2^{\,2} + m^2} + \sqrt{{\vec p}_3^{\,2} + m^2} \ , \qquad {\cal  H}^\text{pot} = \sum_{i<j} H_{ij}(\vec{r}_i - \vec{r}_j)
\]
    \noi and makes use of the Brody-Moshinsky coefficient technique~\cite{SILVESTRE-BRAC}.\par
    Applying the BT model to baryons is unavoidably technical because one is dealing with a three-body problem. Our aim in this paper is to compute the IW functions for the transitions $L = 0 \to L = 0$ elastic case $\Lambda_b \left({1 \over 2}^+ \right) \to \Lambda_c \left({1 \over 2}^+ \right)$ and also for the $L = 0 \to L = 1$ inelastic case $\Lambda_b \left({1 \over 2}^+ \right) \to \Lambda_c \left({1 \over 2}^- \right)$. Of course, the lowest transition $\Lambda_b \left({1 \over 2}^+ \right) \to \Lambda_c \left({3 \over 2}^- \right)$ is related to the former case because both negative parity excited states ${1 \over 2}^-$ and ${3 \over 2}^-$ are related by heavy quark symmetry. Anyway, we will restrict ourselves to the ${1 \over 2}^-$ case.\par
    One can see the complications of the problem by considering the natural choice of the Jacobi coordinates ${\vec \rho}$ (relative space coordinate between the two light spectator quarks) and ${\vec \lambda}$ (relative space coordinate between the center-of-mass of the two light quarks and the heavy quark). The corresponding momenta are denoted by ${\vec p}_\rho$ and ${\vec p}_\lambda$ and the orbital angular momenta by ${\vec \ell}_\rho$ and ${\vec \ell}_\lambda$, with ${\vec \ell}_\rho + {\vec \ell}_\lambda = {\vec L}$. For the transitions $\Lambda_b \to \Lambda_c\left({1 \over 2}^\pm\right)$, i.e. $L = 0 \to L = 0$ or $L = 0 \to L = 1$, we will see that even in these relatively simple cases the moduli $\ell_\rho$ and $\ell_\lambda$ can take an infinite number of values. For $L = 0 \to L = 0$ one has the constraint $\ell_\lambda = \ell_\rho$ and for $L = 0 \to L = 1$ the constraint is $\ell_\lambda = \ell_\rho \pm 1$. We compute the IW function $\xi_\Lambda (w)$ in the elastic case $\Lambda_b \left({1 \over 2}^+ \right) \to \Lambda_c \left({1 \over 2}^+ \right)$ and the much more involved IW function $\sigma_\Lambda (w)$ in the inelastic case $\Lambda_b \left({1 \over 2}^+ \right) \to \Lambda_c \left({1 \over 2}^- \right)$. As we will see below, these functions explicitely exhibit the expected properties of covariance and IW scaling. Since the formalism is rather involved, we have decided to write a detailed paper to allow for an easier reading.\par
    The paper is organized as follows. In Section 2 we start with some generalities on the wave functions of the baryons $\Lambda_Q$ and some useful considerations on spherical harmonics. In Sections 3 and 4 we present the results of the calculation of the spin and orbital angular momentum part of the wave functions for the $L = 0$ and $L = 1$ cases respectively. In Sections 5, we finally write down the baryon wave functions for the $L = 0$ and the $L = 1$ case. In Section 6 we compute the probability amplitudes for the $L = 0 \to L = 0$ elastic transition $\Lambda_b \left({1 \over 2}^+ \right) \to \Lambda_c \left({1 \over 2}^+ \right)$ at infinite mass, giving the elastic IW function $\xi_\Lambda(w)$. Section 7 is devoted to the $L = 0 \to L = 1$ inelastic transition $\Lambda_b \left({1 \over 2}^+ \right) \to \Lambda_c \left({1 \over 2}^- \right)$ at infinite mass. Finally, in Section 8 we conclude by mentioning future developments within the BT scheme, namely the proof of the Bjorken sum rule and also the consideration of some phenomenological implications.
\section{Generalities on the wave function of the heavy baryons $\boldsymbol{\Lambda_Q}$}\label{sec:harmo}
We want to construct the wave functions of the baryons $\Lambda_Q$ with $J=1/2$ and $L=0\text{ or }1$. The baryons contain a heavy quark $Q$ (mass $m_1$, index 1) and two light quarks $q$ (same mass $m$, indices 2 and 3). We will use the Jacobi coordinates given in the appendix~\ref{ann:jacobi}, which are $\vec\rho$ (momentum $\vec k_\rho$) and $\vec\lambda$ (momentum $\vec k_\lambda$). If one exchanges the two light quarks, then $\vec\rho$ and $\vec k_\rho$ transform into their opposite while $\vec\lambda$ and $\vec k_\lambda$ do not change.
\par
It is also convenient to introduce the quantities:
\begin{gather*}
\vec S_{qq} = \vec s_2 + \vec s_3\qquad;\qquad
\vec S = \vec S_{qq} + \vec s_1\qquad;\qquad
\vec L = \vec\ell_\rho + \vec\ell_\lambda\\
\vec\jmath = \vec L + \vec S_{qq}\qquad;\qquad
\vec J = \vec\jmath + \vec s_1 = \vec L + \vec S
\end{gather*}
\par
The wave function will have the following general structure:
\[
\Psi^{(L)}_M\Bigl(\vec k_\rho,\vec k_\lambda\Bigr) = \sum\limits_{\alpha}\Phi^{(L)}_{\alpha,M}\Bigl(\vec k_\rho,\vec k_\lambda\Bigr)
\qquad\text{with}\qquad
\alpha = \Bigl\{S_{qq},S,\ell_\rho,\ell_\lambda\Bigr\}
\]
where the wave functions $\Phi^{(L)}_{\alpha,M}$ factorize according to:
\[
\Phi^{(L)}_{\alpha,M}\Bigl(\vec k_\rho,\vec k_\lambda\Bigr) = \psi_\text{color}\otimes\biggl[\Xi_{S,S_{qq}}\otimes\phi^{(L)}_{\ell_\rho,\ell_\lambda}\Bigl(\vec k_\rho,\vec k_\lambda\Bigr)\biggr]_M\otimes{\mathcal I}_{I,I_{qq}}
\]
In the previous relation, $\psi_\text{color}$ denotes the color part and ${\mathcal I}_{I,I_{qq}}$ the isospin (flavor) part where $I$ is the isospin of the baryon $\Lambda_Q$ and $I_{qq}$ the isospin of the $qq$ system. The remaining factor in brackets is the spin-space part of the wave function.
\par
The baryon $\Lambda_Q$ is a fermion and its wave function is antisymmetric under the exchange of the two light quarks~$q$.
\begin{description}
\item{$\vartriangleright$ \underline{Color part} :}
by construction, the wave function $\psi_\text{color}$ is antisymmetric.
\item{$\vartriangleright$ \underline{Isospin (flavor) part} :} 
let us recall that for the heavy quark $Q$, $I_Q=0$ and for the quark $q$, $I_q=0$ if $q=s$ and $I_q=1/2$ if $q=u,d\,$ ($I^u_{q,z}=1/2$ and $I^d_{q,z}=-1/2$).\\
For the baryon $\Lambda_Q$, its isospin is 0 and the corresponding wave function factorizes in terms of the isospin of the constituent quarks according to $$\Bigl[\bigl[qq\bigr]_{I_{qq}=0}Q\Bigr]_{I=0}$$
Therefore, because of the combination $\bigl[qq\bigr]_{I_{qq}=0}$ with $q=u,d$, this isospin part is antisymmetric in the exchange of the two light quarks.
\item{$\vartriangleright$ \underline{Implications for the spin-space part} :} 
following the previous remarks, the spin-space part must be antisymmetric under the exchange of the two light quarks.\\
If one constructs a state of orbital angular momentum $L$ from the modes $\rho$\footnote{The modes $\rho$ correspond to the light component of the baryon and $\vec\ell_\rho$ describes the orbital angular momentum of the two light quarks system.} and $\lambda$\footnote{The modes $\lambda$ correspond to the interaction between the light component of the baryon and the heavy quark.}, then this state reads:
\[
\sum\limits_{m_\rho, m_\lambda}
\CG{\ell_\rho,\, m_\rho}{\ell_\lambda,\, m_\lambda}{L,\, m}\,
Y_{\ell_\rho}^{(m_\rho)}(\hat k_\rho)\,
Y_{\ell_\lambda}^{(m_\lambda)}(\hat k_\lambda)
\qquad\qquad\text{where}\qquad\hat k = \dfrac{\vec k}{\norm{\vec k}}
\]
Let us recall that, by exchange of the two light quarks $q_2$ and $q_3$, the variable $\hat k_\lambda$ does not change (it is symmetric), while the variable $\hat k_\rho$ becomes its opposite (it is antisymmetric). In fact, the previous orbital term picks up a factor $(-)^{\ell_\rho}$. Since one wants the space-spin part to be antisymmetric, the light component $\vec\jmath$ will be constructed with a total spin for the light quarks $S_{qq}$ satisfying
\[
\left\{
\ 
\begin{aligned}
&\ell_\rho\ \text{ even }:\ S_{qq}\ =\ 0\quad\text{(antisymmetric)}\\
&\ell_\rho\ \text{ odd }:\ S_{qq}\ =\ 1\quad\text{(symmetric)}
\end{aligned}
\right.
\]
\end{description}
We will construct the states $\ket{J^P,L,j,S_{qq}}$ according to the following specific combination of orbital angular momenta and spins:
\[
    \ket{J^P,L,j,S_{qq}} = \ket{{\Bigl\{\bigl[(\ell_\rho\ell_\lambda)_L\,S_{qq}\bigr]_js_1\Bigr\}_J}}
    \qquad\text{based on the decomposition }\qquad\vec J = \vec\jmath + \vec s_1
\]
because, according to the heavy quark symmetry in the infinite mass limit, the light component $\vec\jmath$ is conserved and, thus, is a ``good'' quantum number as well. Another possible construction could also have been considered:
\[
    \ket{J^P,L,S,S_{qq}} = \ket{\Bigl\{\bigl(\ell_\rho\ell_\lambda\bigr)_L\,\bigl(s_1S_{qq}\bigr)_S\Bigr\}_J}
    \qquad\text{based on the decomposition}\qquad\vec J = \vec L + \vec S
\]
but it does not single out the light component $\vec\jmath$.\\
These two types of states obtained by different couplings are related according to:
\begin{equation*}
    \ket{J^P,L,j,S_{qq}} = (-1)^{1/2+S_{qq}+L+J}\sqrt{2j+1}\,
    \sum\limits_S\sqrt{2S+1}\,\begin{Bmatrix}1/2&S_{qq}&S\\ L&J&j\end{Bmatrix}\,\ket{J^P,L,S,S_{qq}}
\end{equation*}
where $\{\,\}$ denotes the $6J$ Wigner coefficients.
\par\medskip
Finally, the spherical harmonics $Y_\ell^{(m)}(\hat p)$, which enter the $\Lambda_Q$ wave functions, will be considered as spherical tensors of rank $\ell$ constructed from $\ell$ vectors $\hat p$ having coordinates written in the standard basis $\hat p^{(0,\pm 1)}$ (see the appendix~\ref{ann:standardbasis}).
\begin{description}
\item{$\vartriangleright$ \underline{${\ell = 1}$ case} :} one has already $\hat p^{(m)}$, a spherical tensor of order 1. With the right coefficient, the spherical harmonic is:
\[
Y_1^{(m)}(\hat p)\ =\ \sqrt{\dfrac{3}{4\pi}}\,\hat p^{(m)}
\]
\item{$\vartriangleright$ \underline{${\ell = 2}$ case} :} one constructs a spherical tensor of rank 2 by combining two tensors of rank 1. In the present case, one combines two $\hat p^{(m)}$ according to:
\[
T_2^{(m)}\ =\ \sum_\mu\,\CG{1,\mu}{1,\,(m-\mu)}{2,\,m}\,\hat p^{(\mu)}\,\hat p^{(m-\mu)}
\]
With the right normalization coefficient, the spherical harmonic $Y_2^{(m)}(\hat p)$ is given as follows:
\[
Y_2^{(m)}(\hat p)\ =\ \sqrt{\dfrac{15}{8\pi}}\,\sum_\mu\,\CG{1,\mu}{1,\,(m-\mu)}{2,\,m}\,\hat p^{(\mu)}\,\hat p^{(m-\mu)}
\]
\item{$\vartriangleright$ \underline{${\ell = 3}$ case} :} one can construct a spherical tensor of order 3 by combining a spherical tensor of order 2 with a spherical tensor of order 1. In the present situation, one uses the two previous cases to obtain:
\[
T_3^{(m)}\ =\ \sum_\mu\,\CG{2,\mu}{1,\,(m-\mu)}{3,\,m}\,T_2^{(\mu)}\,\hat p^{(m-\mu)}
\]
With the right normalization coefficient, this gives:
\[
    Y_3^{(m)}(\hat p)\ =\ \sqrt{\dfrac{35}{8\pi}}\,\sum_{\mu_1,\mu_2}\,
\CG{1,\,\mu_1}{1,\,(\mu_2-\mu_1)}{2,\,\mu_2}\,\CG{2,\,\mu_2}{1,\,(m-\mu_2)}{3,\,m}\,\hat p^{(\mu_1)}\,\hat p^{(\mu_2-\mu_1)}\,\hat p^{(m-\mu_2)}
\]
\item{$\vartriangleright$ \underline{${\ell \geqslant 4}$ case} :} one proceeds in the same way by combining a $\hat p^{(m)}$ with a $T_{\ell-1}^{(m)}$ to construct $T_{\ell}^{(m)}$\ldots~and then multiplying by the right normalization coefficient to get $Y_\ell^{(m)}(\hat p)$.
\end{description}
The normalization coefficient $N^{(\ell)}$ of the previous decompositions is given in all generality by:
\[
N^{(\ell)}\ =\ (2\ell-1)!!\,\sqrt{\dfrac{2\ell+1}{4\pi}\,\dfrac{2^\ell}{(2\ell)!}}
\ =\ \dfrac{1}{\ell!}\,\sqrt{\dfrac{2\ell+1}{4\pi}\,\dfrac{(2\ell)!}{2^\ell}}
\]
The proof of this relation goes as follows:
\begin{order}
\item We start with the expression of the spherical harmonics $Y_\ell^{(m)}(\hat p)$ in terms of associated Legendre polynomials:
\[
Y_\ell^{(m)}(\hat p)\ =\ \sqrt{\dfrac{2\ell+1}{4\pi}\,\dfrac{(\ell-m)!}{(\ell+m)!}}\,
P_\ell^m(\cos\theta)\,e^{im\varphi}
\]
which will be applied to the case $m=\ell$ since the normalization factor $N^{(\ell)}$ does not depend on $m$.
\item We then use the previous construction of $Y_\ell^{(m)}(\hat p)$ in terms of Clebsch-Gordan coefficients. Since $m=\ell$, all the coefficients $\mu_i$ take a single value (one gets in fact $\mu_i = i$ with $i$ running between 1 and $\ell -1$) and thus the Clebsch-Gordan coefficients are all equal to 1 and the corresponding spherical harmonic reads:
\[
Y_\ell^{(\ell)}(\hat p)\ =\ N^{(\ell)}\,\bigl(\hat p^{(+1)}\bigr)^\ell
\]
\item The associated Legendre polynomial $P_\ell^\ell(x)$ is known in general and its expression is:
\[
P_\ell^\ell(x) = (-)^\ell\,(2\ell-1)!!\,(1-x^2)^{\ell/2}
\]
\item We should not forget also that:
\[
\hat p^{(\pm 1)}\ =\ \mp\,\dfrac{1}{\sqrt{2}}\,\sin\theta\,\exp(\pm i\varphi)
\]
which will imply the factor $\sqrt{2^\ell}$ in the expression of the coefficient $N^{(\ell)}$
\item Finally, we use the properties of the double factorials recalled in the appendix~\ref{ann:facto}.
\end{order}
After identification, the expression given above is obtained.
\section{Spin and orbital angular momentum contribution to the $\boldsymbol{L=0}$ heavy baryon wave function}\label{sec:Lzero}
\subsection{Clebsch-Gordan coefficients and construction}
We need to combine the orbital angular momenta $\ell_\lambda$ and $\ell_\rho$ to obtain an orbital angular momentum $L=0$
\begin{align*}
\sum\limits_{m_\rho, m_\lambda}
\CG{\ell_\rho,\, m_\rho}{\ell_\lambda,\, m_\lambda}{0,\, 0}\,
Y_{\ell_\rho}^{(m_\rho)}(\hat k_\rho)\,
Y_{\ell_\lambda}^{(m_\lambda)}(\hat k_\lambda)& =\ \delta_{\ell_\lambda\ell_\rho}\,
\sum\limits_{m_\rho}
\CG{\ell_\rho,\, m_\rho}{\ell_\rho,\, -m_\rho}{0,\, 0}\,
Y_{\ell_\rho}^{(m_\rho)}(\hat k_\rho)\,
Y_{\ell_\rho}^{(-m_\rho)}(\hat k_\lambda)\\
& =\ \delta_{\ell_\lambda\ell_\rho}\,\dfrac{(-)^{\ell_\rho}}{\sqrt{2\ell_\rho +1}}
\sum\limits_{m_\rho=-\ell_\rho}^{\ell_\rho}
\,(-)^{-m_\rho}\,
Y_{\ell_\rho}^{(m_\rho)}(\hat k_\rho)\,
Y_{\ell_\rho}^{(-m_\rho)}(\hat k_\lambda)
\end{align*}
Consequently:
\begin{description}
\item{$\vartriangleright$ \underline{For even $\ell_\rho$} :} 
the light component is obtained by combining the spin $S_{qq} = 0$ with the orbital angular momentum $L=0$, and this light component then combines with the heavy quark spin to build the baryon $J^P=1/2^+$:
\[
\begin{split}
\sum\limits_{\substack{j,m_2,m_3\\ m_\rho,m_\lambda\\ m,\mu}}
\CG{1/2,\,m_2}{1/2,\,m_3}{0,\,0}\,\chi^{(m_2)}_{s_2}\,\chi^{(m_3)}_{s_3}\,
\CG{\ell_\rho,\, m_\rho}{\ell_\lambda,\, m_\lambda}{0,\, 0}\,
Y_{\ell_\rho}^{(m_\rho)}(\hat k_\rho)\,
Y_{\ell_\lambda}^{(m_\lambda)}(\hat k_\lambda)\qquad\qquad\qquad\\
\times
\CG{0,\,0}{0,\,0}{j,\,m}\,
\CG{j,\,m}{1/2,\,\mu}{1/2,\,M}\,\chi^{(\mu)}_{s_1}
\end{split}
\]
Expanding this expression, we get:
\begin{equation*}
\CG{1/2,\,s_2}{1/2,\,s_3}{0,\,0}\,
\delta_{\ell_\lambda\ell_\rho}\,\dfrac{(-)^{\ell_\rho}}{\sqrt{2\ell_\rho +1}}\,
\sum\limits_{m=-\ell_\rho}^{\ell_\rho}
(-)^{-m}
\,
Y_{\ell_\rho}^{(m)}(\hat k_\rho)\,
Y_{\ell_\rho}^{(-m)}(\hat k_\lambda)\,\chi^{(M)}_{s_1}
\end{equation*}
which constrains $j$ to be equal to $0$.
\item{$\vartriangleright$ \underline{For odd $\ell_\rho$} :} 
one proceeds in the same way but with a light spin $S_{qq} = 1$.
\[
\begin{split}
\sum\limits_{\substack{j,m_2,m_3\\ m,m_\rho,m_\lambda\\ m',\mu}}
\CG{1/2,\,m_2}{1/2,\,m_3}{1,\,m}\,\chi^{(m_2)}_{s_2}\,\chi^{(m_3)}_{s_3}\,
\CG{\ell_\rho,\, m_\rho}{\ell_\lambda,\, m_\lambda}{0,\, 0}\,
Y_{\ell_\rho}^{(m_\rho)}(\hat k_\rho)\,
Y_{\ell_\lambda}^{(m_\lambda)}(\hat k_\lambda)\qquad\qquad\qquad\\
\times
\CG{1,\,m}{0,\,0}{j,\,m'}\,
\CG{j,\,m'}{1/2,\,\mu}{1/2,\,M}\,\chi^{(\mu)}_{s_1}
\end{split}
\]
After some reordering, we get:
\begin{equation*}
\begin{split}
\delta_{\ell_\lambda\ell_\rho}\,\dfrac{(-)^{\ell_\rho}}{\sqrt{2\ell_\rho +1}}\,
\sum\limits_{m}
(-)^{-m}\,
\CG{1/2,\,s_2}{1/2,\,s_3}{1,\,m}\qquad\qquad\qquad\qquad\qquad\qquad\qquad\qquad\qquad
\\
\times
\CG{1,\,m}{1/2,\,(M-m)}{1/2,\,M}\,
Y_{\ell_\rho}^{(m)}(\hat k_\rho)\,
Y_{\ell_\rho}^{(-m)}(\hat k_\lambda)\,
\chi^{(M-m)}_{s_1}
\end{split}
\end{equation*}
which requires $j=1$.
\item{$\vartriangleright$ \underline{Allowed values for $\ell_\lambda$ and $\ell_\rho$}:} 
to obtain $L=0$, the orbital angular momenta $\ell_\lambda$ and $\ell_\rho$ are not arbitrary and must satisfy:
\[
\abs{\ell_\lambda - \ell_\rho}\ \leqslant 0\ \leqslant \ell_\lambda + \ell_\rho
\qquad\Longrightarrow\qquad
\ell_\lambda \ =\ \ell_\rho
\qquad\text{where}\quad
    \ell_\lambda,\,\ell_\rho\ \in\ \field{N}
\]
\item{$\vartriangleright$ \underline{Parity of the states}:} 
the parity of the states built in this way is given by the properties of the spherical harmonics, and thus: 
\[
    P = (-)^{\ell_\rho + \ell_\lambda}
\]
As we are interested in the states $J^P = 1/2^+$, this implies that ${\ell_\rho + \ell_\lambda}$ is even, which is automatically verified by the allowed values of $\ell_\lambda$ and $\ell_\rho$.
\end{description}
\subsection{Some explicit calculations}
We have calculated explicitly the contributions of certain modes $(\ell_\rho,\ell_\lambda)=(\ell_\rho,\ell_\rho)$ to be considered. The results are gathered in the following tables~\ref{fig:pair-bis} and~\ref{fig:impair-bis}. The vector $\vec\Sigma$ denotes $\bigl[\vec\sigma\,\sigma_2\bigr]_{s_2,s_3}$. The two columns $P_X$ et $d_{\ell_\rho}$ give the coefficients by which one has to multiply the vector structure to get the spin and orbital momentum part of the wave function, as explained in the following paragraph. One should not forget also that, in the following structures, the matrices $\vec\sigma$ or $\vec\Sigma$ act on a Pauli spinor $\chi^{(m)}$.
\bigskip
\begin{table}[htb]
\begin{center}
\begin{tabular}{|W|W|W|W|}
\hline
\multicolumn{4}{|>{}c|}{\bfseries$\boldsymbol{\ell_\rho}$ even $\qquad$and$\qquad\boldsymbol{\ell_\lambda = \ell_\rho}$}\\
\hline
\multicolumn{4}{|>{}c|}{$\boldsymbol{j=0}$}\\
\hline
(\ell_\rho,\ell_\lambda)&\text{vector structure}&P_X&d_{\ell_\rho}\\
\hline
(0,0)&1&\dfrac{i}{\sqrt{2}}\,(\sigma_2)_{s_2,s_3}&\dfrac{1}{4\pi}\sqrt{1}\\
\hline
(2,2)&\dfrac12\,\biggl[3\,\bigl(\hat k_\rho\cdot\hat k_\lambda\bigr)^2 -1\biggr]&\dfrac{i}{\sqrt{2}}\,(\sigma_2)_{s_2,s_3}&\dfrac{1}{4\pi}\sqrt{5}\\
\hline
(4,4)&\dfrac18\,\biggl[35\,\bigl(\hat k_\rho\cdot\hat k_\lambda\bigr)^4 -30\,\bigl(\hat k_\rho\cdot\hat k_\lambda\bigr)^2 + 3\biggr]&\dfrac{i}{\sqrt{2}}\,(\sigma_2)_{s_2,s_3}&\dfrac{1}{4\pi}\sqrt{9}\\
\hline
(6,6)&\dfrac{1}{16}\,\biggl[231\,\bigl(\hat k_\rho\cdot\hat k_\lambda\bigr)^6 -315\,\bigl(\hat k_\rho\cdot\hat k_\lambda\bigr)^4 + 105\,\bigl(\hat k_\rho\cdot\hat k_\lambda\bigr)^2 - 5\biggr]&\dfrac{i}{\sqrt{2}}\,(\sigma_2)_{s_2,s_3}&\dfrac{1}{4\pi}\sqrt{13}\\
\hline
\multicolumn{4}{|>{}c|}{$\boldsymbol{j=1}$}\\
\hline
(\ell_\rho,\ell_\lambda)&\text{vector structure}&P_X&d_{\ell_\rho}\\
\hline
(0,0)&-&-&-\\
\hline
(2,2)&-&-&-\\
\hline
(4,4)&-&-&-\\
\hline
(6,6)&-&-&-\\
\hline
\end{tabular}
\end{center}
\caption{Spin and orbital angular momentum part of the wave functions $L=0$ for the even values of $\ell_\rho$.}\label{fig:pair-bis}
\end{table}
\begin{table}[htb]
\begin{center}
\begin{tabular}{|W|W|W|W|}
\hline
\multicolumn{4}{|>{}c|}{\bfseries$\boldsymbol{\ell_\rho}$ odd $\qquad$and$\qquad\boldsymbol{\ell_\lambda = \ell_\rho}$}\\
\hline
\multicolumn{4}{|>{}c|}{$\boldsymbol{j=0}$}\\
\hline
(\ell_\rho,\ell_\lambda)&\text{vector structure}&P_X&d_{\ell_\rho}\\
\hline
(1,1)&-&-&-\\
\hline
(3,3)&-&-&-\\
\hline
(5,5)&-&-&-\\
\hline
\multicolumn{4}{|>{}c|}{$\boldsymbol{j=1}$}\\
\hline
(\ell_\rho,\ell_\lambda)&\text{vector structure}&P_X&d_{\ell_\rho}\\
\hline
(1,1)&\hat k_\rho\cdot\hat k_\lambda&\dfrac{i}{\sqrt{2}}\bigl(\vec\Sigma\cdot\vec\sigma\bigr)&\dfrac{1}{4\pi}\sqrt{\dfrac33}\\
\hline
(3,3)&\dfrac12\,\biggl[5\,\bigl(\hat k_\rho\cdot\hat k_\lambda\bigr)^3 - 3\,\bigl(\hat k_\rho\cdot\hat k_\lambda\bigr) \biggr]&\dfrac{i}{\sqrt{2}}\bigl(\vec\Sigma\cdot\vec\sigma\bigr)&\dfrac{1}{4\pi}\sqrt{\dfrac73}\\
\hline
(5,5)&\dfrac18\,\biggl[63\,\bigl(\hat k_\rho\cdot\hat k_\lambda\bigr)^5 - 70\,\bigl(\hat k_\rho\cdot\hat k_\lambda\bigr)^3 + 15\,\bigl(\hat k_\rho\cdot\hat k_\lambda\bigr)\biggr]&\dfrac{i}{\sqrt{2}}\bigl(\vec\Sigma\cdot\vec\sigma\bigr)&\dfrac{1}{4\pi}\sqrt{\dfrac{11}{3}}\\
\hline
\multicolumn{4}{|>{}c|}{\rule{0pt}{15pt}{{NOTATION : }${\vec\Sigma\equiv\bigl[\vec\sigma\,\sigma_2\bigr]_{s_2,s_3}}$}}\\[-1em]
\multicolumn{4}{|>{}c|}{}\\
\hline
\end{tabular}
\end{center}
\caption{Spin and orbital angular momentum part of the wave functions $L=0$ for the odd values of $\ell_\rho$.}\label{fig:impair-bis}
\end{table}
\subsection{Attempt at generalization}
\subsubsection*{Spin and orbital angular momentum term}
By inspection of the tables~\ref{fig:pair-bis} and~\ref{fig:impair-bis}, one realizes that the spin and orbital angular momentum term has a recurring structure. Indeed, in the column ``vector structure'' of those tables appear the associated Legendre polynomials $P_\ell^0(x)$ where the variable $x$ denotes the scalar product $\hat k_\rho\cdot\hat k_\lambda$. For illustration, the first $P_\ell^0(x)$ are:
\begin{align*}
P_0^0(x)&=1&&P_1^0(x) = x\\[2mm]
P_2^0(x) &= \dfrac12\Bigl(3\,x^2-1\Bigr)&&P_3^0(x) = \dfrac12\Bigl(5\,x^3-3\,x\Bigr)\\[2mm]
P_4^0(x) &= \dfrac18\Bigl(35\,x^4-30\,x^2+3\Bigr)
&&P_5^0(x) = \dfrac18\Bigl(63\,x^5-70\,x^3+15\,x\Bigr)\\[2mm]
P_6^0(x) &= \dfrac{1}{16}\Bigl(231\,x^6-315\,x^4+105\,x^2-5\Bigr)
\end{align*}
\subsubsection*{The coefficients}
The term which multiplies the vector structure has the form:
\[
\text{coeff}\ =\ d\times P_X
\]
where $d$ is a coefficient that characterizes the structure in associated Legendre polynomials and $P_X$ is the term linked with the Pauli matrices ($X=\sigma,\Sigma\cdot\sigma$) such that
\[
P_\sigma\ =\ \dfrac{i}{\sqrt{2}}\,(\sigma_2)_{s_2,s_3}
\qquad\text{and}\qquad
P_{\Sigma\cdot\sigma}\ =\ \dfrac{i}{\sqrt{2}}\bigl({\vec\Sigma\cdot\vec\sigma}\bigr)
\]
\subsubsection*{Summary}
To summarize, the spin and orbital angular momentum parts of the wave functions have the structures gathered in the following table:
\begin{center}
\begin{tabular}{|Sc|Sc|Sc|Sc|}
\hline
\multicolumn{4}{|>{}c|}{\bfseries Spin and orbital angular momentum part of the wave functions $\boldsymbol{L=0}$}\\
\hline
\bfseries $\boldsymbol{(\ell_\rho,\ell_\lambda)}$&\bfseries$\boldsymbol{j}$&\bfseries$\boldsymbol{\ell_\rho}$ even&\bfseries$\boldsymbol{\ell_\rho}$ odd\\
\hline
\multirow{3}{*}{$(\ell_\rho,\ell_\rho)$}&0&${{\dfrac{i}{\sqrt{2}}}\,(\sigma_2)_{s_2,s_3}}\,{\dfrac{1}{4\pi}\sqrt{2\,\ell_\rho+1}}\,P_{\ell_\rho}^0\bigl(\hat k_\rho\cdot\hat k_\lambda\bigr)$&0\\
\cline{2-4}
\multirow{-3}{*}{$(\ell_\rho,\ell_\rho)$}&1&0&${{\dfrac{i}{\sqrt{2}}}\,\bigl({\vec\Sigma\cdot\vec\sigma}\bigr)}\,{\dfrac{1}{4\pi\sqrt{3}}\sqrt{2\,\ell_\rho+1}}\,P_{\ell_\rho}^0\bigl(\hat k_\rho\cdot\hat k_\lambda\bigr)$\\
\hline
\end{tabular}
\end{center}
\section{Spin and orbital angular momentum contribution to the $\boldsymbol{L=1}$ heavy baryon wave function}\label{sec:Lun}
\subsection{Clebsch-Gordan coefficients and construction}
In this case, we now combine the orbital angular momenta $\ell_\lambda$ and $\ell_\rho$ to build an orbital angular momentum $L=1$:
\[
\sum\limits_{m_\rho, m_\lambda}
\CG{\ell_\rho,\, m_\rho}{\ell_\lambda,\, m_\lambda}{1,\, m}\,
Y_{\ell_\rho}^{(m_\rho)}(\hat k_\rho)\,
Y_{\ell_\lambda}^{(m_\lambda)}(\hat k_\lambda)
\]
From the previous remarks (leaving aside the radial part of the wave functions for the moment):
\begin{description}
\item{$\vartriangleright$ \underline{For even $\ell_\rho$} :} 
we first construct the light component by combining the spin $S_{qq} = 0$ with the orbital angular momentum $L=1$ and then this light component is coupled to the heavy quark spin to build the baryon $1/2^-$.
\[
    \begin{split}
        \sum\limits_{\substack{j,m_2,m_3\\ m,m_\rho,m_\lambda\\ m^\prime,m'',\mu}}
        \CG{1/2,\,m_2}{1/2,\,m_3}{0,\,m}\,\chi^{(m_2)}_{s_2}\,\chi^{(m_3)}_{s_3}\,
        \CG{\ell_\rho,\, m_\rho}{\ell_\lambda,\, m_\lambda}{1,\, m'}\,
        Y_{\ell_\rho}^{(m_\rho)}(\hat k_\rho)\,
        Y_{\ell_\lambda}^{(m_\lambda)}(\hat k_\lambda)\qquad\qquad\qquad\\
        \times
        \CG{0,\,m}{1,\,m'}{j,\,m''}\,
        \CG{j,\,m''}{1/2,\,\mu}{1/2,\,M}\,\chi^{(\mu)}_{s_1}
    \end{split}
\]
Expanding this expression, we get:
\begin{equation}\label{eq:pair}
    \begin{split}
    \sum\limits_{j,m,m^\prime}
    \CG{1/2,\,s_2}{1/2,\,s_3}{0,\,0}\,
    \CG{0,\,0}{1,\,m'}{j,\,m'}\,
    \CG{\ell_\rho,\, m}{\ell_\lambda,\, (m^\prime - m)}{1,\, m^\prime}\qquad\qquad\qquad
    \\
    \qquad\qquad\qquad\times
    \CG{j,\,m'}{1/2,\,(M - m')}{1/2,\,M}\,
    Y_{\ell_\rho}^{(m)}(\hat k_\rho)\,
    Y_{\ell_\lambda}^{(m'-m)}(\hat k_\lambda)\,\chi^{(M-m')}_{s_1}
    \end{split}
\end{equation}
Notice that this implies in fact $j=1$.
\item{$\vartriangleright$ \underline{For odd $\ell_\rho$ }:} 
we proceed in the same way but with a light spin $S_{qq} = 1$.
\[
\begin{split}
\sum\limits_{\substack{j,m_2,m_3\\ m,m_\rho,m_\lambda\\ m^\prime,m'',\mu}}
\CG{1/2,\,m_2}{1/2,\,m_3}{1,\,m}\,\chi^{(m_2)}_{s_2}\,\chi^{(m_3)}_{s_3}\,
\CG{\ell_\rho,\, m_\rho}{\ell_\lambda,\, m_\lambda}{1,\, m'}\,
Y_{\ell_\rho}^{(m_\rho)}(\hat k_\rho)\,
Y_{\ell_\lambda}^{(m_\lambda)}(\hat k_\lambda)\qquad\qquad\qquad\\
\times
\CG{1,\,m}{1,\,m'}{j,\,m''}\,
\CG{j,\,m''}{1/2,\,\mu}{1/2,\,M}\,\chi^{(\mu)}_{s_1}
\end{split}
\]
After some reordering, we get:
\begin{multline}\label{eq:impair}
\sum\limits_{\substack{j,m\\ m',m''}}
\CG{1/2,\,s_2}{1/2,\,s_3}{1,\,m}\,
\CG{\ell_\rho,\, m'}{\ell_\lambda,\, (m''-m-m')}{1,\,(m''-m)}\,
\CG{1,\,m}{1,\,(m''-m)}{j,\,m''}\quad\\
\times
\CG{j,\,m''}{1/2,\,(M-m'')}{1/2,\,M}\,
Y_{\ell_\rho}^{(m')}(\hat k_\rho)\,
Y_{\ell_\lambda}^{(m''-m-m')}(\hat k_\lambda)\,
\chi^{(M-m'')}_{s_1}
\end{multline}
\item{$\vartriangleright$ \underline{Allowed values for $\ell_\lambda$ and $\ell_\rho$} :} 
to get $L=1$, the orbital angular momenta $\ell_\lambda$ and $\ell_\rho$ are not arbitrary and must satisfy
\[
\abs{\ell_\lambda - \ell_\rho}\ \leqslant 1\ \leqslant \ell_\lambda + \ell_\rho
\qquad\Longrightarrow\qquad
\left\{
\begin{aligned}
	\text{i.e. }\ \ell_\lambda &\ =\ \ell_\rho\ \neq\ 0\\[2mm]
	\text{i.e. }\ \ell_\lambda &\ =\ \ell_\rho\ \pm 1
\end{aligned}
\right.
\qquad\text{where}\qquad
    \ell_\lambda,\,\ell_\rho\ \in\ \field{N}
\]
\item{$\vartriangleright$ \underline{Parity of the states} :} 
the parity of the states built in this way is still given by the properties of the spherical harmonics so that: 
\[
    P = (-)^{\ell_\rho + \ell_\lambda}
\]
Since we are interested in the states $J^P = 1/2^-$, this implies that ${\ell_\rho + \ell_\lambda}$ is odd. Therefore, this additional constraint forbids the possibility $\ell_\lambda = \ell_\rho$ and the allowed values are finally:
\[
    \ell_\lambda \ =\ \ell_\rho\ \pm 1
\]
\end{description}

\subsection{Some explicit calculations}\label{para:lesCl}
Using the previous relations, we have computed explicitly the contributions of certain modes $(\ell_\rho,\ell_\lambda)=(\ell_\rho,\ell_\rho\pm 1)$ to be considered. The results are gathered in the following tables~\ref{fig:pair}~and~\ref{fig:impair}, with the same conventions as in the tables of the $L=0$ case. A distinction has been done between the even and odd values of $\ell_\rho$. The appendix~\ref{ann:lesCl} describes the calculation of the ${c_{\ell_\rho;\pm 1}}$ coefficients and an explicit calculation for the mode $(\ell_\rho,\ell_\lambda)=(0,1)$ of table~\ref{fig:pair} can be found in the appendix~\ref{ann:example}.
\bigskip
\begin{table}[htb]
\begin{center}
\begin{tabular}{|W|W|W|W|}
\hline
\multicolumn{4}{|>{}c|}{\bfseries$\boldsymbol{\ell_\rho}$ even $\qquad$and$\qquad\boldsymbol{\ell_\lambda = \ell_\rho\pm 1}$}\\
\hline
\multicolumn{4}{|>{}c|}{$\boldsymbol{j=0}$}\\
\hline
(\ell_\rho,\ell_\lambda)&\text{vector structure}&P_X&c_{\ell_\rho:\pm 1}\\
\hline
(0,1)&0&-&-\\
\hline
(2,1)&0&-&-\\
\hline
(2,3)&0&-&-\\
\hline
(4,3)&0&-&-\\
\hline
(4,5)&0&-&-\\
\hline
\multicolumn{4}{|>{}c|}{$\boldsymbol{j=1}$}\\
\hline
(\ell_\rho,\ell_\lambda)&\text{vector structure}&P_X&c_{\ell_\rho:\pm 1}\\
\hline
(0,1)&\hat k_\lambda\cdot\vec\sigma&\dfrac{i}{\sqrt{2}}\,(\sigma_2)_{s_2,s_3}&-\dfrac{1}{4\pi}\\
\hline
(2,1)&\Bigl[- 3\bigl(\hat k_\rho\cdot\hat k_\lambda\bigr)\hat k_\rho + \hat k_\lambda\Bigr]\cdot\vec\sigma&\dfrac{i}{\sqrt{2}}\,(\sigma_2)_{s_2,s_3}&-\dfrac{1}{4\pi}\dfrac{1}{\sqrt{2}}\\
\hline
(2,3)&\dfrac32\,\biggl[-2\,\bigl(\hat k_\rho\cdot\hat k_\lambda\bigr)\hat k_\rho + \Bigl[5\bigl(\hat k_\rho\cdot\hat k_\lambda\bigr)^2 - 1\Bigr]\,\hat k_\lambda\biggr]\cdot\vec\sigma&\dfrac{i}{\sqrt{2}}\,(\sigma_2)_{s_2,s_3}&-\dfrac{1}{4\pi}\dfrac{1}{\sqrt{3}}\\
\hline
(4,3)&\dfrac12\,\biggl[-5\,\bigl(\hat k_\rho\cdot\hat k_\lambda\bigr)\Bigl(7\,\bigl(\hat k_\rho\cdot\hat k_\lambda\bigr)^2-3\Bigr)\,\hat k_\rho + 3\,\Bigl(5\,\bigl(\hat k_\rho\cdot\hat k_\lambda\bigr)^2-1\Bigr)\,\hat k_\lambda \biggr]\cdot\vec\sigma&\dfrac{i}{\sqrt{2}}\,(\sigma_2)_{s_2,s_3}&-\dfrac{1}{8\pi}\\
\hline
(4,5)&\dfrac58\,\biggl[- 4\,\bigl(\hat k_\rho\cdot\hat k_\lambda\bigr)\Bigl(7\,\bigl(\hat k_\rho\cdot\hat k_\lambda\bigr)^2-3\Bigr)\,\hat k_\rho + 3\,\Bigl(21\,\bigl(\hat k_\rho\cdot\hat k_\lambda\bigr)^4 - 14\,\bigl(\hat k_\rho\cdot\hat k_\lambda\bigr)^2+1\Bigr)\,\hat k_\lambda \biggr]\cdot\vec\sigma&\dfrac{i}{\sqrt{2}}\,(\sigma_2)_{s_2,s_3}&-\dfrac{1}{4\pi}\dfrac{1}{\sqrt{5}}\\
\hline
\end{tabular}
\end{center}
\caption{Spin and orbital angular momentum part of the wave functions $L=1$ for the even values of $\ell_\rho$.}\label{fig:pair}
\end{table}
\begin{table}[htb]
\begin{center}
\begin{tabular}{|W|W|W|W|}
\hline
\multicolumn{4}{|>{}c|}{\bfseries$\boldsymbol{\ell_\rho}$ odd $\qquad$and$\qquad\boldsymbol{\ell_\lambda = \ell_\rho\pm 1}$}\\
\hline
\multicolumn{4}{|>{}c|}{$\boldsymbol{j=0}$}\\
\hline
(\ell_\rho,\ell_\lambda)&\text{vector structure}&P_X&c_{\ell_\rho:\pm 1}\\
\hline
(1,0)&-\,\hat k_\rho\,\cdot\,\vec\Sigma&\dfrac{i}{\sqrt{2}}&\dfrac{1}{4\pi}\\
\hline
(1,2)&\Bigl[ - \hat k_\rho + 3\bigl(\hat k_\rho\cdot\hat k_\lambda\bigr)\hat k_\lambda\Bigr]\cdot\vec\Sigma&\dfrac{i}{\sqrt{2}}&\dfrac{1}{4\pi}\dfrac{1}{\sqrt{2}}\\
\hline
(3,2)&\dfrac32\,\biggl[ - \Bigl[5\bigl(\hat k_\rho\cdot\hat k_\lambda\bigr)^2 - 1\Bigr]\,\hat k_\rho + 2\,\bigl(\hat k_\rho\cdot\hat k_\lambda\bigr)\hat k_\lambda\biggr]\cdot\vec\Sigma&\dfrac{i}{\sqrt{2}}&\dfrac{1}{4\pi}\dfrac{1}{\sqrt{3}}\\
\hline
(3,4)&\dfrac12\,\biggl[- 3\,\Bigl(5\,\bigl(\hat k_\rho\cdot\hat k_\lambda\bigr)^2-1\Bigr)\,\hat k_\rho + 5\,\bigl(\hat k_\rho\cdot\hat k_\lambda\bigr)\Bigl(7\,\bigl(\hat k_\rho\cdot\hat k_\lambda\bigr)^2-3\Bigr)\,\hat k_\lambda  \biggr]\cdot\vec\Sigma&\dfrac{i}{\sqrt{2}}&\dfrac{1}{8\pi}\\
\hline
(5,4)&\dfrac58\,\biggl[-3\,\Bigl(21\,\bigl(\hat k_\rho\cdot\hat k_\lambda\bigr)^4 - 14\,\bigl(\hat k_\rho\cdot\hat k_\lambda\bigr)^2+1\Bigr)\,\hat k_\rho + 4\,\bigl(\hat k_\rho\cdot\hat k_\lambda\bigr)\Bigl(7\,\bigl(\hat k_\rho\cdot\hat k_\lambda\bigr)^2-3\Bigr)\,\hat k_\lambda\biggr]
\cdot\vec\Sigma&\dfrac{i}{\sqrt{2}}&\dfrac{1}{4\pi}\dfrac{1}{\sqrt{5}}\\
\hline
\multicolumn{4}{|>{}c|}{$\boldsymbol{j=1}$}\\
\hline
(\ell_\rho,\ell_\lambda)&\text{vector structure}&P_X&c_{\ell_\rho:\pm 1}\\
\hline
(1,0)&-\hat k_\rho\cdot\bigl(\vec\Sigma\wedge\vec\sigma\bigr)&-\dfrac12&\dfrac{1}{4\pi}\\
\hline
(1,2)&\Bigl[ - \hat k_\rho + 3\bigl(\hat k_\rho\cdot\hat k_\lambda\bigr)\hat k_\lambda\Bigr]\cdot\bigl(\vec\Sigma\wedge\vec\sigma\bigr)&-\dfrac12&\dfrac{1}{4\pi}\dfrac{1}{\sqrt{2}}\\
\hline
(3,2)&\dfrac32\,\biggl[- \Bigl[5\bigl(\hat k_\rho\cdot\hat k_\lambda\bigr)^2 - 1\Bigr]\,\hat k_\rho + 2\,\bigl(\hat k_\rho\cdot\hat k_\lambda\bigr)\hat k_\lambda \biggr]\cdot\bigl(\vec\Sigma\wedge\vec\sigma\bigr)&-\dfrac12&\dfrac{1}{4\pi}\dfrac{1}{\sqrt{3}}\\
\hline
(3,4)&\dfrac12\,\biggl[- 3\,\Bigl(5\,\bigl(\hat k_\rho\cdot\hat k_\lambda\bigr)^2-1\Bigr)\,\hat k_\rho + 5\,\bigl(\hat k_\rho\cdot\hat k_\lambda\bigr)\Bigl(7\,\bigl(\hat k_\rho\cdot\hat k_\lambda\bigr)^2-3\Bigr)\,\hat k_\lambda \biggr]\cdot\bigl(\vec\Sigma\wedge\vec\sigma\bigr)&-\dfrac12&\dfrac{1}{8\pi}\\
\hline
(5,4)&\dfrac58\,\biggl[-3\,\Bigl(21\,\bigl(\hat k_\rho\cdot\hat k_\lambda\bigr)^4 - 14\,\bigl(\hat k_\rho\cdot\hat k_\lambda\bigr)^2+1\Bigr)\,\hat k_\rho + 4\,\bigl(\hat k_\rho\cdot\hat k_\lambda\bigr)\Bigl(7\,\bigl(\hat k_\rho\cdot\hat k_\lambda\bigr)^2-3\Bigr)\,\hat k_\lambda\biggr]\cdot\bigl(\vec\Sigma\wedge\vec\sigma\bigr)&-\dfrac12&\dfrac{1}{4\pi}\dfrac{1}{\sqrt{5}}\\
\hline
\multicolumn{4}{|>{}c|}{\rule{0pt}{15pt}{{NOTATION : }${\vec\Sigma\equiv\bigl[\vec\sigma\,\sigma_2\bigr]_{s_2,s_3}}$}}\\[-1em]
\multicolumn{4}{|>{}c|}{}\\
\hline
\end{tabular}
\end{center}
\caption{Spin and orbital angular momentum part of the wave functions $L=1$ for the odd values of $\ell_\rho$.}\label{fig:impair}
\end{table}
\subsection{Attempt at generalization}
\subsubsection*{Spin and orbital angular momentum term}
Looking at the previous tables, one realizes, as in the $L=0$ case, that the spin and orbital angular momentum terms have a recurring structure. Roughly speaking, we are dealing with the scalar product of a vector function (which depends on the unitary vectors $\hat k_\rho$ and $\hat k_\lambda$) with Pauli matrices ($\vec\sigma$, $\vec\Sigma$ and $\vec\Sigma\wedge\vec\sigma$). Moreover, the coefficients of the vectors $\hat k_\rho$ and $\hat k_\lambda$ in the vector function correspond to the associated Legendre polynomials $P_\ell^1(x)$ where the variable $x$ denotes the scalar product $\hat k_\rho\cdot\hat k_\lambda$. For example, the first $P_\ell^1(x)$ are:
\begin{align*}
P_0^1(x)&=0&&P_1^1(x) = -\,\sqrt{1-x^2}\\
P_2^1(x) &= -\,3\,x\,\sqrt{1-x^2}&&P_3^1(x) = -\,\dfrac32\,\bigl(5\,x^2 - 1\bigr)\sqrt{1-x^2}\\
P_4^1(x) &= -\,\dfrac52\,x\,\bigl(7\,x^2 - 3\bigr)\sqrt{1-x^2}
&&P_5^1(x) = -\,\dfrac{15}{8}\,\bigl(21\,x^4 - 14\,x^2 + 1\bigr)\sqrt{1-x^2}\\
P_6^1(x) &= -\,\dfrac{21}{8}\,x\,\bigl(33\,x^4 - 30\,x^2 + 5\bigr)\sqrt{1-x^2}
\end{align*}
knowing that here one also has:
\[
x\ =\ \hat k_\rho\cdot\hat k_\lambda
\qquad\Longrightarrow\qquad
\sqrt{1-x^2}\ =\ \norm{\hat k_\rho\wedge\hat k_\lambda}
\]
\subsubsection*{The coefficients}
The coefficients also exhibit a structure. According to the initial form of the spin and orbital angular momentum parts, a Pauli matrix $\vec\sigma$ always appears multiplied by a factor $i/\sqrt{2}$ and the same applies to the matrices $\vec\Sigma$ (the term $\vec\Sigma\wedge\vec\sigma$ is therefore associated to a factor $-1/2$). In summary, this means that they can be written as: 
\[
\text{coeff}\ =\ c\times P_X
\]
where $c$ is a coefficient characterizing the structure in associated Legendre polynomials and $P_X$ is the factor of the Pauli matrices ($X=\sigma,\Sigma,\Sigma\wedge\sigma$) such that:
\[
P_\sigma\ =\ \dfrac{i}{\sqrt{2}}\,(\sigma_2)_{s_2,s_3}
\qquad\text{and}\qquad
P_\Sigma\ =\ \dfrac{i}{\sqrt{2}}
\qquad\text{and also}\qquad
P_{\Sigma\wedge\sigma}\ =\ -\dfrac12
\]
\subsubsection*{Summary}
To summarize, the spin and orbital angular momentum parts of the wave functions exhibit the following different structures:
\begin{center}
\begin{tabular}{|Sc|Sc|Sc|Sc|}
\hline
\multicolumn{4}{|>{}c|}{\bfseries Spin and orbital angular momentum parts of the $\boldsymbol{L=1}$ wave functions}\\
\hline
\bfseries$\boldsymbol{(\ell_\rho,\ell_\lambda)}$&\bfseries$\boldsymbol{j}$&\bfseries$\boldsymbol{\ell_\rho}$ even&\bfseries$\boldsymbol{\ell_\rho}$ odd\\
\hline
\multirow{3}{*}{$(\ell_\rho,\ell_\rho\pm 1)$}&0&0&${\dfrac{i}{\sqrt{2}}}\,\vec F_{\ell_\rho;\pm 1}\bigl(\hat k_\rho\,;\,\hat k_\lambda\bigr){\cdot\vec\Sigma}$\\
\cline{2-4}
\multirow{-3}{*}{$(\ell_\rho,\ell_\rho\pm 1)$}&1&${{\dfrac{i}{\sqrt{2}}}\,(\sigma_2)_{s_2,s_3}}\,\vec F_{\ell_\rho;\pm 1}\bigl(\hat k_\rho\,;\,\hat k_\lambda\bigr){\cdot\vec\sigma}$&${-\dfrac12}\,\vec F_{\ell_\rho;\pm 1}\bigl(\hat k_\rho\,;\,\hat k_\lambda\bigr){\cdot\bigl(\vec\Sigma\wedge\vec\sigma\bigr)}
$\\
\hline
\end{tabular}
\end{center}
where the vector functions of the type $\vec F_{n;\pm 1}$ are expressed in terms of the associated Legendre polynomials $P_\ell^m(x)$
\[
\vec F_{n;\pm 1}\bigl(\hat k_\rho\,;\,\hat k_\lambda\bigr)\ =\ c_{n;\pm 1}\,\dfrac{1}{\norm{\hat k_\rho\wedge\hat k_\lambda}}\,\Bigl[P_n^1\bigl(\hat k_\rho\cdot\hat k_\lambda\bigr)\,\hat k_\rho \ -\ P_{n\pm 1}^1\bigl(\hat k_\rho\cdot\hat k_\lambda\bigr)\,\hat k_\lambda\Bigr]
\]
The coefficients $c_{n;\pm 1}$ can be calculated explicitly (see appendix~\ref{ann:lesCl}):
\[
c_{n;+1}\ =\ \dfrac{(-)^{n+1}}{4\pi\,\sqrt{n+1}}
\qquad\text{and}\qquad
c_{n;-1}\ =\ \dfrac{(-)^{n+1}}{4\pi\,\sqrt{n}}
\qquad\text{satisfying}\qquad
c_{n;+1}\ =\ -\,c_{n+1;-1}
\]
Another expression of the vector functions $\vec F_{n;\pm 1}$ can be given. The associated Legendre polynomials $P_\ell^1(x)$ and the Legendre polynomials $P_\ell(x)$, which also turn out to be the associated Legendre polynomials $P_\ell^0(x)$, are related in the following way:
\[
\dfrac{1}{\sqrt{1-x^2}}\,P_\ell^1(x)\ =\ -\,\dfrac{\dd\,P_\ell^0(x)}{\dd x}
\]
Therefore, one also has:
\[
\vec F_{n;\pm 1}\bigl(\hat k_\rho\,;\,\hat k_\lambda\bigr)\ =\ c_{n;\pm 1}\,\Biggl[\dfrac{\dd\,P_{n\pm 1}^0(x)}{\dd x}\,\hat k_\lambda\ -\ \dfrac{\dd\,P_n^0(x)}{\dd x}\,\hat k_\rho\Biggr]
\qquad\text{where}\qquad
x\ =\ \hat k_\rho\cdot\hat k_\lambda
\]

\section{Baryon wave functions}
In this section, the heavy baryon wave functions are listed. They will be given in the rest-frame of the baryon and in terms of the internal momenta $k_i$ of the three quarks defined by:
\[
k_i = {\boldsymbol B}^{-1}_{_{\sum p_j}} p_i
\qquad\qquad\text{with}\qquad\qquad\sum\limits_{i=1}^3\vec k_i = \vec 0
\]
where ${\boldsymbol B}_{p}$ is the Lorentz boost $(\sqrt{p^2},\vec 0)\ \leadsto\ p$.\\
As a consequence, the choice of the independant internal momentum variables will be those of the light quarks $\vec k_2$ and $\vec k_3$, which are also used in the definition of the Jacobi variables $\vec k_\rho$ and $\vec k_\lambda$ (see appendix~\ref{ann:jacobi}).
\subsection{The $\boldsymbol{L=0}$ case}\label{sec:WFzero}
For the modes $(\ell_\rho,\ell_\lambda=\ell_\rho)$, the notation of the rest-frame internal $L=0$ wave functions will be:
\[
\prescript{(j)}{(\ell_\rho)}\Phi^{(M)}_{s_1,s_2,s_3}(\vec k_\lambda,\,\vec k_\rho)
\]
where $j$ represents the spin of the light component.\\
The spin and angular momentum part has already been studied in section~\ref{sec:Lzero}. The remaining radial part, being rotationally invariant, is written generically as follows:
\[
    \phi\bigl(\norm{\vec k_2}^2,\,\norm{\vec k_3}^2\bigr)
\]
whose dependency becomes, when the Jacobi variables are used:
\[
    \phi\bigl(\norm{\vec k_\rho}^2,\,\norm{\vec k_\lambda}^2,\,\vec k_\rho\cdot\vec k_\lambda\bigr)
\]
Gathering all these results, the non-vanishing wave functions are:
\begin{center}
\begin{tabular}{|Sc|Sc|}
\hline
$\boldsymbol{\ell_\rho\ \text{\bfseries even}}$&$\prescript{(0)}{(\ell_\rho)}\Phi^{(M)}_{s_1,s_2,s_3}(\vec k_\lambda,\,\vec k_\rho)\ =\ 
{\dfrac{i}{\sqrt{2}}}\,(\sigma_2)_{s_2,s_3}\,\chi^{(M)}_{s_1}\,\dfrac{1}{4\pi}\sqrt{2\,\ell_\rho+1}\,P_{\ell_\rho}^0\bigl(\hat k_\rho\cdot\hat k_\lambda\bigr)
\,\phi^{(\ell_\rho)}\bigl(\norm{\vec k_\lambda}^2,\,\norm{\vec k_\rho}^2,\,\vec k_\rho\cdot\vec k_\lambda\bigr)$\\
\hline
$\boldsymbol{\ell_\rho\ \text{\bfseries odd}}$&$\prescript{(1)}{(\ell_\rho)}\Phi^{(M)}_{s_1,s_2,s_3}(\vec k_\lambda,\,\vec k_\rho)\ =\ 
{\dfrac{i}{\sqrt{2}}}\,\bigl(\vec\sigma\,\sigma_2\bigr)_{s_2,s_3}\cdot\Bigl[\vec\sigma\,\chi^{(M)}\Bigr]_{s_1}\,
\dfrac{1}{4\pi}\sqrt{\dfrac{2\,\ell_\rho+1}{3}}\,P_{\ell_\rho}^0\bigl(\hat k_\rho\cdot\hat k_\lambda\bigr)
\,\phi^{(\ell_\rho)}\bigl(\norm{\vec k_\lambda}^2,\,\norm{\vec k_\rho}^2,\,\vec k_\rho\cdot\vec k_\lambda\bigr)$\\
\hline
\end{tabular}
\end{center}
\par\medskip
Moreover, the $L=0$ wave functions are normalized according to:
\begin{equation*}
    \sum\limits_{s_1,s_2,s_3}\int\dfrac{\dd\vec k_2}{(2\pi)^3}\dfrac{\dd\vec k_3}{(2\pi)^3} 
    \Phi^{(M)}_{s_1,s_2,s_3}(\vec k_\lambda,\,\vec k_\rho)\,\Phi^{(M')*}_{s_1,s_2,s_3}(\vec k_\lambda,\,\vec k_\rho) \ =\ \delta^{MM'}
\end{equation*}
knowing that:
\[
\Phi^{(M)}_{s_1,s_2,s_3}(\vec k_\lambda,\,\vec k_\rho) =
\sum\limits_\text{$\ell_\rho$ even}\prescript{(0)}{(\ell_\rho)}\Phi^{(M)}_{s_1,s_2,s_3}(\vec k_\lambda,\,\vec k_\rho)\ + \sum\limits_\text{$\ell_\rho$ odd}\prescript{(1)}{(\ell_\rho)}\Phi^{(M)}_{s_1,s_2,s_3}(\vec k_\lambda,\,\vec k_\rho)
\]
\subsection{The $\boldsymbol{L=1}$ case}\label{sec:WFun}
Following the same line of reasoning as in the $L=0$ case, the rest-frame internal $L=1$ wave functions for the modes $(\ell_\rho,\ell_\rho\pm 1)$ will be denoted as follows:
\[
\prescript{(j)}{(\ell_\rho;\pm 1)}\Psi^{(m)}_{s_1,s_2,s_3}(\vec k_\lambda,\,\vec k_\rho)
\]\par
With the help of the results of section~\ref{sec:Lun}, the non vanishing wave functions are:
\begin{center}
\begin{tabular}{|Sc|Sc|}
\hline
$\boldsymbol{\text{\bfseries even}\ \ell_\rho}$&$\prescript{(1)}{(\ell_\rho;\pm 1)}\Psi^{(m)}_{s_1,s_2,s_3}(\vec k_\lambda,\,\vec k_\rho)\ =\ 
\dfrac{i}{\sqrt{2}}\,(\sigma_2)_{s_2,s_3}\,\biggl[\Bigl(\vec F_{\ell_\rho;\pm 1}\cdot\vec\sigma\Bigr)\,\chi^{(m)}\biggr]_{s_1}
\,\psi^{(\ell_\rho;\pm 1)}\bigl(\norm{\vec k_\lambda}^2,\,\norm{\vec k_\rho}^2,\,\vec k_\rho\cdot\vec k_\lambda\bigr)$\\
\hline
\multirow{3}{*}{}&$\prescript{(0)}{(\ell_\rho;\pm 1)}\Psi^{(m)}_{s_1,s_2,s_3}(\vec k_\lambda,\,\vec k_\rho)\ =\ 
\dfrac{i}{\sqrt{2}}\,\vec F_{\ell_\rho;\pm 1}\cdot\bigl(\vec\sigma\,\sigma_2\bigr)_{s_2,s_3}\,\chi^{(m)}_{s_1}
\,\psi^{(\ell_\rho;\pm 1)}\bigl(\norm{\vec k_\lambda}^2,\,\norm{\vec k_\rho}^2,\,\vec k_\rho\cdot\vec k_\lambda\bigr)$\\
\cline{2-2}
\multirow{-3}{*}{$\boldsymbol{\text{\bfseries odd}\ \ell_\rho}$}&$\prescript{(1)}{(\ell_\rho;\pm 1)}\Psi^{(m)}_{s_1,s_2,s_3}(\vec k_\lambda,\,\vec k_\rho)\ =\ 
\dfrac{1}{2}\,
\bigl(\vec\sigma\,\sigma_2\bigr)_{s_2,s_3}\cdot\biggl[
\Bigl(
\vec F_{\ell_\rho;\pm 1}\wedge\vec\sigma\Bigr)
\,\chi^{(m)}\biggr]_{s_1}
\,\psi^{(\ell_\rho;\pm 1)}\bigl(\norm{\vec k_\lambda}^2,\,\norm{\vec k_\rho}^2,\,\vec k_\rho\cdot\vec k_\lambda\bigr)
$\\
\hline
\end{tabular}
\end{center}
where $\psi^{(\ell_\rho;\pm 1)}$ represents the radial parts of the wave functions.
\par\medskip
And finally, the $L=1$ wave functions are normalized according to:
\begin{equation*}
    \sum\limits_{s_1,s_2,s_3}\int\dfrac{\dd\vec k_2}{(2\pi)^3}\dfrac{\dd\vec k_3}{(2\pi)^3} 
    \Psi^{(M)}_{s_1,s_2,s_3}(\vec k_\lambda,\,\vec k_\rho)\,\Psi^{(M')*}_{s_1,s_2,s_3}(\vec k_\lambda,\,\vec k_\rho) \ =\ \delta^{MM'}
\end{equation*}
    where
\begin{multline*}
    \Psi^{(M)}_{s_1,s_2,s_3}(\vec k_\lambda,\,\vec k_\rho) =
    \sum\limits_\text{$\ell_\rho$ pair}\prescript{(1)}{(\ell_\rho;+1)}\Psi^{(M)}_{s_1,s_2,s_3}(\vec k_\lambda,\,\vec k_\rho)\ + \sum\limits_\text{$\ell_\rho$ pair}\prescript{(1)}{(\ell_\rho;-1)}\Psi^{(M)}_{s_1,s_2,s_3}(\vec k_\lambda,\,\vec k_\rho)\\[2mm]
    + \sum\limits_\text{$\ell_\rho$ impair}\prescript{(0)}{(\ell_\rho;+1)}\Psi^{(M)}_{s_1,s_2,s_3}(\vec k_\lambda,\,\vec k_\rho)\ +\ \sum\limits_\text{$\ell_\rho$ impair}\prescript{(0)}{(\ell_\rho;-1)}\Psi^{(M)}_{s_1,s_2,s_3}(\vec k_\lambda,\,\vec k_\rho)\\[2mm]
    + \sum\limits_\text{$\ell_\rho$ impair}\prescript{(1)}{(\ell_\rho;+1)}\Psi^{(M)}_{s_1,s_2,s_3}(\vec k_\lambda,\,\vec k_\rho)\ +\ \sum\limits_\text{$\ell_\rho$ impair}\prescript{(1)}{(\ell_\rho;-1)}\Psi^{(M)}_{s_1,s_2,s_3}(\vec k_\lambda,\,\vec k_\rho)
\end{multline*}
%
%
\section{The $\boldsymbol{1/2^+ \to 1/2^+}$ transitions}
In this section, the detailed calculation of the ${1/2^+ \to 1/2^+}$ transitions in the Bakamjian-Thomas framework will be presented. 
\subsection{The BT formalism}
In the BT formalism~\cite{COVARIANT-QM}, the transition amplitude from a state $\Psi$ to a state $\Psi^\prime$ under some vector/axial interaction generically denoted by $J$ is given by (unless said otherwise, the sums and products concern the three momenta):
\begin{multline*}
\brakket{\Psi^\prime,\,m^\prime}{J}{\Psi,\,m} = \int\dfrac{\dd\vec p_2}{(2\pi)^3}\dfrac{\dd\vec p_3}{(2\pi)^3}\,\sqrt{\dfrac{\left(\sum\limits_{j=1}^3 {p^{\prime o}_j}\right)\left(\sum\limits_{j=1}^3 {p_j^o}\right)}{M^\prime_oM_o}}\left(\prod\limits_{j=1}^3\sqrt{\dfrac{ {k^{\prime o}_j} {k^{o}_j} }{ {p^{\prime o}_j} {p^{o}_j} }}\right) \\
\times\sum\limits_{s_1^\prime,s_2^\prime,s_3^\prime}\sum\limits_{s_1,s_2,s_3}
\Psi^{\prime\,(m')}_{s_1^\prime,s_2^\prime,s_3^\prime}(\vec k^\prime_2,\,\vec k^\prime_3)^*
\left[D({\boldsymbol R}_1^{\prime -1})\,J(\vec p^{\,\prime}_1,\,\vec p_1)\,D({\boldsymbol R}_1)\right]_{s_1^\prime,s_1}\\
\times
\left[D({\boldsymbol R}_2^{\prime -1}{\boldsymbol R}_2)\right]_{s_2^\prime,s_2}
\left[D({\boldsymbol R}_3^{\prime -1}{\boldsymbol R}_3)\right]_{s_3^\prime,s_3}
\Psi^{(m)}_{s_1,s_2,s_3}(\vec k_2,\,\vec k_3)
\end{multline*}
In this expression, the index 1 corresponds to the heavy quark quantities while the indices 2 and 3 correspond to the two spectator quarks of the baryon.\\
The momenta ${\vec p}^{\,(\prime)}_1$ are related to the total momenta of the hadron ${\vec P}^{(\prime)}$ by
\[
{\vec p}^{\,(\prime)}_1 = {\vec P}^{(\prime)} - ({\vec p}^{\,(\prime)}_2 + {\vec p}^{\,(\prime)}_3)
\]
Since the quarks 2 and 3 are spectators, then:
\[
{\vec p}^{\,\prime}_2\ =\ {\vec p}_2\qquad\text{and}\qquad{\vec p}^{\,\prime}_3\ =\ {\vec p}_3
\]
The momenta $k_i$ are the internal momenta of each spectator quark in the rest frame of the baryon.\\
The notation ${\boldsymbol R}$ stands for the Wigner rotation which describes the behaviour of the spins when going from an arbitrary reference frame to the internal reference frame and $D({\boldsymbol R})$ for the corresponding matrix representation.\\
One will also introduce the following quadrivectors $u^{(\prime)}$ such that:
\[
m_{\Psi^{(\prime)}}\,u^{(\prime)} = \sum\limits_{j=1}^3 {p_j^{(\prime)}}\qquad\text{with}\qquad u^{(\prime)}\cdot u^{(\prime)} = 1
\]
where $m_{\Psi^{(\prime)}}$ is the mass of the baryon $\Psi^{(\prime)}$.\par
An important remark is the following. In the previous expression of the transition amplitude, the baryon states are normalized according to (non-relativistic normalization):
\[
\ket{\Psi(p)}\qquad\text{such that}\qquad
\braket{\Psi(p^\prime)}{\Psi(p)}\ =\ (2\pi)^3\,\delta^{(3)}({\vec p}^{\,\prime} - \vec p)
\]
while the usual relativistic normalization is given by
\[
\ket{\Psi(p)}_{R}\qquad\text{where}\qquad
{}_R\braket{\Psi(p^\prime)}{\Psi(p)}_R\ =\ (2\pi)^3\,2p^0\,\delta^{(3)}({\vec p}^{\,\prime} - \vec p)
\]
The relationship between our normalized BT states and the usual states is therefore 
\[
{\ket{\Psi(p)}\ =\ \dfrac{1}{\sqrt{2p^0}}\,\ket{\Psi(p)}_R\ =\ \dfrac{1}{\sqrt{2\,m_\Psi}}\,\dfrac{1}{\sqrt{v^0}}\,\ket{\Psi(p)}_R}
\]
where the last equality holds in the limit of infinite mass.
\subsection{Determination of the transition amplitude}
The main contribution to the $1/2^+$ ground state is expected to be the one obtained with $\ell_\rho=\ell_\lambda=0$ and the light quark spin $S_{qq}=0$. Hence we will only consider this case in the present paper. From the section~\ref{sec:WFzero}, one then has: 
\[
{
\Psi^{(m)}_{s_1,s_2,s_3}(\vec k_\lambda,\,\vec k_\rho)
= \dfrac{1}{4\pi}\dfrac{i}{\sqrt{2}}\,\chi^{(m)}_{s_1}\,(\sigma_2)_{s_2,s_3}\,\varphi(\norm{\vec k_\lambda}^2,\,\norm{\vec k_\rho}^2,\,{\vec k_\lambda}\cdot{\vec k_\rho})
}
\]
It follows :
\[
{
\Psi^{\prime(m^\prime)*}_{s_1^\prime,s_2^\prime,s_3^\prime}(\vec k_\lambda^\prime,\,\vec k_\rho^\prime)
= \dfrac{1}{4\pi}\dfrac{i}{\sqrt{2}}\,\chi^{(m^\prime)\dagger}_{s_1^\prime}\,(\sigma_2)_{s_2^\prime,s_3^\prime}\,\varphi^*(\norm{\vec k_\lambda^\prime}^2,\,\norm{\vec k_\rho^\prime}^2,\,{\vec k^\prime_\lambda}\cdot{\vec k^\prime_\rho})
}
\]
because
\[
\sigma_2^* = -\,\sigma_2
\qquad\text{and also}\qquad
\chi^{(m^\prime)*}_{s_1^\prime} = {}^t\!\left(\chi^{(m^\prime)\dagger}\right)_{s_1^\prime} = \chi^{(m^\prime)\dagger}_{s_1^\prime}
\]
Following the Bakamjian-Thomas construction of the transition amplitudes, the starting point then reads: 
\begin{align*}
\brakket{\Psi^\prime,\,m^\prime}{J}{\Psi,\,m}	
	& = \begin{multlined}[t][145mm]
	- \dfrac{1}{32\pi^2}\,\int\dfrac{\dd\vec p_2}{(2\pi)^3}\dfrac{\dd\vec p_3}{(2\pi)^3}\,
	\sqrt{u^{\prime o}\,u^o}\,
	\sqrt{\dfrac{ {k^{\prime o}_1} {k^{o}_1} }{ {p^{\prime o}_1} {p^{o}_1} }}\,
	\sqrt{\dfrac{ {k^{\prime o}_2} {k^{o}_2} }{ {p^{\prime o}_2} {p^{o}_2} }}\,
	\sqrt{\dfrac{ {k^{\prime o}_3} {k^{o}_3} }{ {p^{\prime o}_3} {p^{o}_3} }}\,
	\\
	\times	\varphi^{*}(\norm{\vec k^\prime_\lambda}^2,\,\norm{\vec k^\prime_\rho}^2,\,{\vec k^\prime_\lambda}\cdot{\vec k^\prime_\rho})\,
	\varphi(\norm{\vec k_\lambda}^2,\,\norm{\vec k_\rho}^2,\,{\vec k_\lambda}\cdot{\vec k_\rho})
\sum\limits_{s_1^\prime,s_2^\prime,s_3^\prime}\sum\limits_{s_1,s_2,s_3}
	\chi^{(m)}_{s_1}\,(\sigma_2)_{s_2,s_3}\,
	\chi^{(m')\dagger}_{s_1^\prime}(\sigma_2)_{s^\prime_2,s^\prime_3}
	\\
	\times
	\left[D({\boldsymbol R}_1^{\prime -1})\,J(\vec p^{\,\prime}_1,\,\vec p_1)\,D({\boldsymbol R}_1)\right]_{s_1^\prime,s_1}\,
	\left[D({\boldsymbol R}_2^{\prime -1}{\boldsymbol R}_2)\right]_{s_2^\prime,s_2}\,
	\left[D({\boldsymbol R}_3^{\prime -1}{\boldsymbol R}_3)\right]_{s_3^\prime,s_3}
	\end{multlined}\\[2mm]
	& = \begin{multlined}[t][145mm]
	- \dfrac{1}{32\pi^2}\,\int\dfrac{\dd\vec p_2}{(2\pi)^3}\dfrac{\dd\vec p_3}{(2\pi)^3}\,
	\sqrt{u^{\prime o}\,u^o}\,
	\sqrt{\dfrac{ {k^{\prime o}_1} {k^{o}_1} }{ {p^{\prime o}_1} {p^{o}_1} }}\,
	\sqrt{\dfrac{ {k^{\prime o}_2} {k^{o}_2} }{ {p^{\prime o}_2} {p^{o}_2} }}\,
	\sqrt{\dfrac{ {k^{\prime o}_3} {k^{o}_3} }{ {p^{\prime o}_3} {p^{o}_3} }}\,
	\\
	\times\varphi^{*}(\norm{\vec k^\prime_\lambda}^2,\,\norm{\vec k^\prime_\rho}^2,\,{\vec k^\prime_\lambda}\cdot{\vec k^\prime_\rho})\,
	\varphi(\norm{\vec k_\lambda}^2,\,\norm{\vec k_\rho}^2,\,{\vec k_\lambda}\cdot{\vec k_\rho})
	\left[
	\chi^{(m')\dagger}\,
	D({\boldsymbol R}_1^{\prime -1})\,J(\vec p^{\,\prime}_1,\,\vec p_1)\,D({\boldsymbol R}_1)
	\chi^{(m)}
	\right]\\
	\times\sum\limits_{s_2^\prime,s_3^\prime}\sum\limits_{s_2,s_3}
	\left[D({\boldsymbol R}_2^{\prime -1}{\boldsymbol R}_2)\right]_{s_2^\prime,s_2}\,
	\left[D({\boldsymbol R}_3^{\prime -1}{\boldsymbol R}_3)\right]_{s_3^\prime,s_3}
	\,(\sigma_2)_{s_2,s_3}\,(\sigma_2)_{s^\prime_2,s^\prime_3}
	\end{multlined}\\[2mm]
	& = \begin{multlined}[t][145mm]
	\dfrac{1}{32\pi^2}\,\int\dfrac{\dd\vec p_2}{(2\pi)^3}\dfrac{\dd\vec p_3}{(2\pi)^3}\,
	\sqrt{u^{\prime o}\,u^o}\,
	\sqrt{\dfrac{ {k^{\prime o}_1} {k^{o}_1} }{ {p^{\prime o}_1} {p^{o}_1} }}\,
	\sqrt{\dfrac{ {k^{\prime o}_2} {k^{o}_2} }{ {p^{\prime o}_2} {p^{o}_2} }}\,
	\sqrt{\dfrac{ {k^{\prime o}_3} {k^{o}_3} }{ {p^{\prime o}_3} {p^{o}_3} }}\,
	\\
	\times\varphi^{*}(\norm{\vec k^\prime_\lambda}^2,\,\norm{\vec k^\prime_\rho}^2,\,{\vec k^\prime_\lambda}\cdot{\vec k^\prime_\rho})\,
	\varphi(\norm{\vec k_\lambda}^2,\,\norm{\vec k_\rho}^2,\,{\vec k_\lambda}\cdot{\vec k_\rho})
	\\
	\times
	\left[
	\chi^{(m')\dagger}\,
	D({\boldsymbol R}_1^{\prime -1})\,J(\vec p^{\,\prime}_1,\,\vec p_1)\,D({\boldsymbol R}_1)
	\chi^{(m)}
	\right]\,
	\text{Tr}\left[
	D({\boldsymbol R}_2^{\prime -1}{\boldsymbol R}_2)\,\sigma_2\,
	\bigl(\sigma_2\,D({\boldsymbol R}_3^{-1}{\boldsymbol R}^\prime_3)\,\sigma_2\bigr)\,
	\sigma_2
	\right]
	\end{multlined}
\end{align*}
where one has used the known properties:
\[
\left[D({\boldsymbol R}_3^{\prime -1}{\boldsymbol R}_3)\right]_{s_3^\prime,s_3} = 
\left[\sigma_2\,D({\boldsymbol R}_3^{-1}{\boldsymbol R}^\prime_3)\,\sigma_2\right]_{s_3,s_3^\prime}
\qquad\text{as well as}\qquad
{}^{t}\sigma_2 = -\sigma_2
\]
Finally, the expression of the amplitude in two dimensions becomes:
\[
{
\begin{multlined}[\textwidth]
\brakket{\Psi^\prime,\,m^\prime}{J}{\Psi,\,m} = \dfrac{1}{32\pi^2}\,\int\dfrac{\dd\vec p_2}{(2\pi)^3}\dfrac{\dd\vec p_3}{(2\pi)^3}\,
\sqrt{u^{\prime o}\,u^o}\,
\sqrt{\dfrac{ {k^{\prime o}_1} {k^{o}_1} }{ {p^{\prime o}_1} {p^{o}_1} }}\,
\sqrt{\dfrac{ {k^{\prime o}_2} {k^{o}_2} }{ {p^{\prime o}_2} {p^{o}_2} }}\,
\sqrt{\dfrac{ {k^{\prime o}_3} {k^{o}_3} }{ {p^{\prime o}_3} {p^{o}_3} }}\,
\\[2mm]
\times\varphi^{*}(\norm{\vec k^\prime_\lambda}^2,\,\norm{\vec k^\prime_\rho}^2,\,{\vec k^\prime_\lambda}\cdot{\vec k^\prime_\rho})\,
	\varphi(\norm{\vec k_\lambda}^2,\,\norm{\vec k_\rho}^2,\,{\vec k_\lambda}\cdot{\vec k_\rho})
\\[2mm]
\times
\left[
\chi^{(m')\dagger}\,
D({\boldsymbol R}_1^{\prime -1})\,J(\vec p^{\,\prime}_1,\,\vec p_1)\,D({\boldsymbol R}_1)
\chi^{(m)}
\right]\,
\text{Tr}\left[
D({\boldsymbol R}_2^{\prime -1})\,D({\boldsymbol R}_2)\,
D({\boldsymbol R}_3^{-1})\,D({\boldsymbol R}^\prime_3)
\right]
\end{multlined}
}
\]
\subsection{Four-dimensional formulation of the amplitude}
\subsubsection*{Going to the 4D formalism}\label{para:bracketUN}
The next step of the Bakamjian-Thomas method is to transfer the transition amplitude written in terms of two-dimensional objects such as the Pauli matrices and the Wigner rotations into an expression which involves four-dimensional objects such as Dirac matrices and Lorentz boosts (see appendix~\ref{para:DDform}).
\begin{maliste}
\item{\underline{Calculation of the trace} :}\par
Using the expressions of the Wigner rotations in terms of boosts (see appendix~\ref{ann:wigner}), the trace reads: 
\[
{\mathscr T}_1 = \text{Tr}
\left[
{\boldsymbol B}_{k^\prime_2}^{-1}\,{\boldsymbol B}_{u^\prime}^{-1}\,{\boldsymbol B}_{p^\prime_2}\,
{\boldsymbol B}_{p_2}^{-1}\,{\boldsymbol B}_{u}\,{\boldsymbol B}_{k_2}\,
{\boldsymbol B}_{k_3}^{-1}\,{\boldsymbol B}_{u}^{-1}\,{\boldsymbol B}_{p_3}\,
{\boldsymbol B}_{p^\prime_3}^{-1}\,{\boldsymbol B}_{u^\prime}\,{\boldsymbol B}_{k_3^\prime}
\right]
\]
Following the method described in the appendix, the formulation in 4D is obtained by introducing the usual projector:
\[
{\mathscr T}_1 \quad\leadsto\quad{\mathscr T}_1 = \text{Tr}
\left[
\dfrac{1 + \gamma^0}{2}\,
{\boldsymbol B}_{k^\prime_2}^{-1}\,{\boldsymbol B}_{u^\prime}^{-1}\,{\boldsymbol B}_{p^\prime_2}\,
{\boldsymbol B}_{p_2}^{-1}\,{\boldsymbol B}_{u}\,{\boldsymbol B}_{k_2}\,
{\boldsymbol B}_{k_3}^{-1}\,{\boldsymbol B}_{u}^{-1}\,{\boldsymbol B}_{p_3}\,
{\boldsymbol B}_{p^\prime_3}^{-1}\,{\boldsymbol B}_{u^\prime}\,{\boldsymbol B}_{k_3^\prime}
\right]
\]
Then, with some algebra :
\begin{align*}
{\mathscr T}_1 & = \dfrac{1}{16}\text{Tr}\left[
(1 + \gamma^0)^4\,
{\boldsymbol B}_{k^\prime_2}^{-1}\,{\boldsymbol B}_{ u^\prime}^{-1}\,{\boldsymbol B}_{ p^\prime_2}\,
{\boldsymbol B}_{ p_2}^{-1}\,{\boldsymbol B}_{ u}\,{\boldsymbol B}_{ k_2}\,
{\boldsymbol B}_{ k_3}^{-1}\,{\boldsymbol B}_{ u}^{-1}\,{\boldsymbol B}_{ p_3}\,
{\boldsymbol B}_{ p^\prime_3}^{-1}\,{\boldsymbol B}_{ u^\prime}\,{\boldsymbol B}_{ k_3^\prime}
\right] \qquad\text{since}\quad(1 + \gamma^0)^4 = 8(1 + \gamma^0)\\[2mm]
& = \begin{multlined}[t][160mm]\dfrac{1}{16}\text{Tr}\left[
(1 + \gamma^0)\,
{\boldsymbol B}_{ k^\prime_2}^{-1}\,{\boldsymbol B}_{ u^\prime}^{-1}\,{\boldsymbol B}_{ p^\prime_2}\,
{\boldsymbol B}_{ p_2}^{-1}\,{\boldsymbol B}_{ u}\,{\boldsymbol B}_{ k_2}\,(1 + \gamma^0)^2\,
{\boldsymbol B}_{ k_3}^{-1}\,{\boldsymbol B}_{ u}^{-1}\,{\boldsymbol B}_{ p_3}\,
{\boldsymbol B}_{ p^\prime_3}^{-1}\,{\boldsymbol B}_{ u^\prime}\,{\boldsymbol B}_{ k_3^\prime}\,(1 + \gamma^0)
\right]\\ \quad\text{since}\quad(1 + \gamma^0)\,D({\boldsymbol R}_i) = D({\boldsymbol R}_i)\,(1 + \gamma^0)\end{multlined}\\[2mm]
& = \dfrac{1}{16}\text{Tr}\left[
(1 + \gamma^0)\,
{\boldsymbol B}_{ k^\prime_2}^{-1}\,{\boldsymbol B}_{ u^\prime}^{-1}\,
{\boldsymbol B}_{ u}\,{\boldsymbol B}_{ k_2}\,(1 + \gamma^0)^2\,
{\boldsymbol B}_{ k_3}^{-1}\,{\boldsymbol B}_{ u}^{-1}\,
{\boldsymbol B}_{ u^\prime}\,{\boldsymbol B}_{ k_3^\prime}\,(1 + \gamma^0)
\right]\qquad\text{since $\ {\vec p}^{\,\prime}_2\ =\ {\vec p}_2\ $ and $\ {\vec p}^{\,\prime}_3\ =\ {\vec p}_3$}\\[2mm]
& = \begin{multlined}[t][160mm]\dfrac{1}{16}\text{Tr}\left[
{\boldsymbol B}_{ u^\prime}\,(1 + \gamma^0)\,
{\boldsymbol B}_{ k^\prime_2}^{-1}\,{\boldsymbol B}_{ u^\prime}^{-1}\,
{\boldsymbol B}_{ u}\,{\boldsymbol B}_{ k_2}\,(1 + \gamma^0)\,{\boldsymbol B}_{ u}^{-1}{\boldsymbol B}_{ u}\,(1 + \gamma^0)\,
{\boldsymbol B}_{ k_3}^{-1}\,{\boldsymbol B}_{ u}^{-1}\,
{\boldsymbol B}_{ u^\prime}\,{\boldsymbol B}_{ k_3^\prime}\,(1 + \gamma^0)\,{\boldsymbol B}_{ u^\prime}^{-1}
\right]\\ \quad\text{inserting in the right places $\ {\boldsymbol B}_{ u^{(\prime)}}^{-1}{\boldsymbol B}_{ u^{(\prime)}} = {\boldsymbol 1}$}\end{multlined}\\[2mm]
& = \begin{multlined}[t][160mm]\dfrac{1}{64}\dfrac{1}{m^2}\,
\dfrac{1}{\sqrt{m + k^{\prime o}_2}}\,
\dfrac{1}{\sqrt{m + k^{o}_2}}\,
\dfrac{1}{\sqrt{m + k^{\prime o}_3}}\,
\dfrac{1}{\sqrt{m + k^{o}_3}}\\ \times
\text{Tr}\left[
(1+\fmslash{ u}^\prime)(m + \fmslash{ p}_2)
(m + \fmslash{ p}_2)(1+\fmslash{ u})
(1+\fmslash{ u})(m + \fmslash{ p}_3)
(m + \fmslash{ p}_3)(1+\fmslash{ u}^\prime)
\right]
\end{multlined}\\[2mm]
& = \begin{multlined}[t][160mm]\dfrac{1}{4}\,
\dfrac{1}{\sqrt{m + k^{\prime o}_2}}\,
\dfrac{1}{\sqrt{m + k^{o}_2}}\,
\dfrac{1}{\sqrt{m + k^{\prime o}_3}}\,
\dfrac{1}{\sqrt{m + k^{o}_3}}\,
\text{Tr}\left[
(1+\fmslash{ u}^\prime)(m + \fmslash{ p}_2)
(1+\fmslash{ u})(m + \fmslash{ p}_3)
\right]\\
\quad\text{since $\ (1+\fmslash{ u})^2 = 2(1+\fmslash{ u})\ $ and $\ (m+\fmslash{ p})^2 = 2\,m(m+\fmslash{ p})$}\end{multlined}
\end{align*}
The last trace can be easily calculated to obtain:
\[
{
\begin{multlined}[170mm]
{\mathscr T}_1 = \dfrac{1}{\sqrt{m + k^{\prime o}_2}}\,\dfrac{1}{\sqrt{m + k^{o}_2}}\,
\dfrac{1}{\sqrt{m + k^{\prime o}_3}}\,\dfrac{1}{\sqrt{m + k^{o}_3}}\\[2mm]
\times
\left[
{
m^2(1 +  u\cdot u^\prime) + m( p_2 +  p_3)( u +  u^\prime)
+ ( p_2\cdot u)( p_3\cdot u^\prime) + ( p_3\cdot u)( p_2\cdot u^\prime)
+ ( p_2\cdot p_3)(1 -  u\cdot u^\prime)
}
\right]
\end{multlined}
}
\]
\item{\underline{Term between brackets} :}\par
In this term, we will use the expression of the current provided in the appendix~\ref{ann:current} where the momentum used is the one of the heavy quark:
\[
J(\vec p^{\,\prime}_1,\,\vec p_1) = \sqrt{\dfrac{m^{\prime}_1\,m_1}{p^{\prime o}_1\,p^o_1}}\,
\dfrac{1 + \gamma^0}{2}\,{\boldsymbol B}_{p^\prime_1}^{-1}\,J\,{\boldsymbol B}_{p_1}\,
\dfrac{1 + \gamma^0}{2}
\]
One then starts from (see appendix~\ref{para:DDform}):
\begin{multline*}
{\mathscr T}_2 = \left[
\chi^{\dagger(m')}\,
D({\boldsymbol R}_1^{\prime -1})\,J(\vec p^{\,\prime}_1,\,\vec p_1)\,D({\boldsymbol R}_1)
\chi^{(m)}
\right]\\
\leadsto\quad{\mathscr T}_2 = \sqrt{\dfrac{m^{\prime}_1\,m_1}{p^{\prime o}_1\,p^o_1}}\,\left[
\chi^{(m')\dagger}\,\dfrac{1 + \gamma^0}{2}\,
D({\boldsymbol R}_1^{\prime -1})\,
\dfrac{1 + \gamma^0}{2}\,{\boldsymbol B}_{ p^\prime_1}^{-1}\,J\,{\boldsymbol B}_{ p_1}\,
\dfrac{1 + \gamma^0}{2}
\,D({\boldsymbol R}_1)\,
\dfrac{1 + \gamma^0}{2}\,\chi^{(m)}
\right]
\end{multline*}
Replacing the expressions of the Wigner rotations (see appendix~\ref{ann:wigner}):
\begin{align*}
{\mathscr T}_2 &= \sqrt{\dfrac{m^{\prime}_1\,m_1}{p^{\prime o}_1\,p^o_1}}\,\left[
\chi^{(m')\dagger}\,\dfrac{1 + \gamma^0}{2}\,
{\boldsymbol B}_{ k^\prime_1}^{-1}\,{\boldsymbol B}_{ u^\prime}^{-1}\,{\boldsymbol B}_{ p^\prime_1}
\,
\dfrac{1 + \gamma^0}{2}\,{\boldsymbol B}_{ p^\prime_1}^{-1}\,J\,{\boldsymbol B}_{ p_1}\,
\dfrac{1 + \gamma^0}{2}
\,
{\boldsymbol B}_{ p_1}^{-1}\,{\boldsymbol B}_{ u}\,{\boldsymbol B}_{ k_1}
\,
\dfrac{1 + \gamma^0}{2}\,\chi^{(m)}
\right]\\[2mm]
& = \begin{multlined}[t][160mm]\dfrac{1}{16}\,\sqrt{\dfrac{m^{\prime}_1\,m_1}{p^{\prime o}_1\,p^o_1}}\,\left[
\chi^{(m')\dagger}\,(1 + \gamma^0)\,(1 + \gamma^0)\,
{\boldsymbol B}_{ k^\prime_1}^{-1}\,{\boldsymbol B}_{ u^\prime}^{-1}\,{\boldsymbol B}_{ p^\prime_1}
\,
{\boldsymbol B}_{ p^\prime_1}^{-1}\,J\,{\boldsymbol B}_{ p_1}\,
{\boldsymbol B}_{ p_1}^{-1}\,{\boldsymbol B}_{ u}\,{\boldsymbol B}_{ k_1}
\,(1 + \gamma^0)
\,
(1 + \gamma^0)\,\chi^{(m)}
\right]\\
\text{since}\quad(1 + \gamma^0)\,D({\boldsymbol R}_i) = D({\boldsymbol R}_i)\,(1 + \gamma^0)\end{multlined}\\[2mm]
& = \dfrac{1}{4}\,\sqrt{\dfrac{m^{\prime}_1\,m_1}{p^{\prime o}_1\,p^o_1}}\,
\left[
\chi^{(m')\dagger}\,(1 + \gamma^0)\,
{\boldsymbol B}_{ k^\prime_1}^{-1}\,{\boldsymbol B}_{ u^\prime}^{-1}\,J\,
{\boldsymbol B}_{ u}\,{\boldsymbol B}_{ k_1}
\,(1 + \gamma^0)
\,\chi^{(m)}
\right]\qquad\text{since}\quad(1 + \gamma^0)^2 = 2\,(1 + \gamma^0)\\[2mm]
& = \dfrac{1}{4}\,\sqrt{\dfrac{m^{\prime}_1\,m_1}{p^{\prime o}_1\,p^o_1}}\,
\left[
\chi^{(m')\dagger}\,\gamma^0\,(1 + \gamma^0)\,
{\boldsymbol B}_{ k^\prime_1}^{-1}\,{\boldsymbol B}_{ u^\prime}^{-1}\,J\,
{\boldsymbol B}_{ u}\,{\boldsymbol B}_{ k_1}
\,(1 + \gamma^0)
\,\chi^{(m)}
\right]\qquad\text{since}\quad\gamma^0\,(1 + \gamma^0) = 1 + \gamma^0\\[2mm]
& = \begin{multlined}[t][160mm]\dfrac{1}{4}\,\sqrt{\dfrac{m^{\prime}_1\,m_1}{p^{\prime o}_1\,p^o_1}}\,
\left[
\chi^{(m')\dagger}\,\gamma^0\,
{\boldsymbol B}_{ u^{\prime}}^{-1}{\boldsymbol B}_{ u^{\prime}}
\,(1 + \gamma^0)\,
{\boldsymbol B}_{ k^\prime_1}^{-1}\,{\boldsymbol B}_{ u^\prime}^{-1}\,J\,
{\boldsymbol B}_{ u}\,{\boldsymbol B}_{ k_1}
\,(1 + \gamma^0)\,
{\boldsymbol B}_{ u}^{-1}{\boldsymbol B}_{ u}
\,\chi^{(m)}
\right]\\
\text{inserting in the right places $\ {\boldsymbol B}_{ u^{(\prime)}}^{-1}{\boldsymbol B}_{ u^{(\prime)}} = {\boldsymbol 1}$}
\end{multlined}\\[2mm]
& = \dfrac{1}{4}\,\sqrt{\dfrac{m^{\prime}_1\,m_1}{p^{\prime o}_1\,p^o_1}}\,
\left[
\chi^{(m')\dagger}\,\gamma^0\,{\boldsymbol B}_{ u^{\prime}}^{-1}\,
\dfrac{(1+\fmslash{ u^{\,\prime}})(m^\prime_1 + \fmslash{ p^{\,\prime}_1})}{\sqrt{2\,m^\prime_1\,(m^\prime_1 + k^{\prime 0}_1)}}
\,J\,
\dfrac{(m_1 + \fmslash{ p_1})(1+\fmslash{ u})}{\sqrt{2\,m_1\,(m_1 + k^0_1)}}\,
{\boldsymbol B}_{ u}
\,\chi^{(m)}
\right]\qquad
\text{using the relation \eqref{ann:eq3}}
\\[2mm]
& = \dfrac{1}{8}\,\dfrac{1}{\sqrt{p^{\prime o}_1\,p^o_1}}\,
\dfrac{1}{\sqrt{m^\prime_1 + k^{\prime 0}_1}}\,
\dfrac{1}{\sqrt{m_1 + k^{0}_1}}\,\left[
\bar\chi^{(m')}\,{\boldsymbol B}_{ u^{\prime}}^{-1}\,
(1+\fmslash{ u^{\,\prime}})\,(m^\prime_1 + \fmslash{ p^{\,\prime}_1})
\,J\,
(m_1 + \fmslash{ p_1})\,(1+\fmslash{ u})\,
{\boldsymbol B}_{ u}
\,\chi^{(m)}
\right]
\end{align*}
Finally, we introduce the boosted spinors $\chi^{(m)}_{u}$ defined by:
\[
{\chi^{(m)}_{u} \ =\ {\boldsymbol B}_{u}\,\chi^{(m)}}
\]
which satisfies the following properties:
\[
\chi^{(m)\dagger}_{u} = \chi^{(m)\dagger}\,{\boldsymbol B}_{u}^\dagger = \chi^{(m)\dagger}\,\gamma^0\,{\boldsymbol B}_{u}^{-1}\,\gamma^0
\quad\Longrightarrow\quad
\chi^{(m)\dagger}_{u}\,\gamma^0 = \chi^{(m)\dagger}\,\gamma^0\,{\boldsymbol B}_{u}^{-1}
\quad\Longrightarrow\quad
{
\bar\chi^{(m)}_{u}\ =\ \bar\chi^{(m)}\,{\boldsymbol B}_{u}^{-1}
}
\]
and one checks immediately that: 
\[
{(1+\fmslash{u})\,\chi^{(m)}_{u}\ =\ 2\,\chi^{(m)}_{u}}
\qquad\text{as well as}\qquad
{\bar\chi^{(m)}_{u}\,(1+\fmslash{u})\ =\ 2\,\bar\chi^{(m)}_{u}}
\]
Hence we obtain:
\[
{
{\mathscr T}_2  = \dfrac{1}{2}\,\dfrac{1}{\sqrt{p^{\prime o}_1\,p^o_1}}\,
\dfrac{1}{\sqrt{m^\prime_1 + k^{\prime 0}_1}}\,
\dfrac{1}{\sqrt{m_1 + k^{0}_1}}\,
\left[
\bar\chi^{(m')}_{ u^{\prime}}\,
(m^\prime_1 + \fmslash{ p}^{\,\prime}_1)
\,J\,
(m_1 + \fmslash{ p}_1)\,
\chi^{(m)}_{ u}
\right]
}
\]
\item{\underline{Conclusion} :}\par
Gathering everything, the transition amplitude is given as follows:
\[{
\begin{multlined}[\textwidth]
\brakket{\Psi^\prime,\,m^\prime}{J}{\Psi,\,m} = \dfrac{1}{64\pi^2}\,\int\dfrac{\dd\vec p_2}{(2\pi)^3}\dfrac{\dd\vec p_3}{(2\pi)^3}\,
\sqrt{u^{\prime o}\,u^o}\,
\sqrt{\dfrac{ {k^{\prime o}_1} {k^{o}_1} }{ {p^{\prime o}_1} {p^{o}_1} }}\,
\dfrac{\sqrt{ {k^{\prime o}_2} {k^{o}_2} }}{ {p^{o}_2} }\,
\dfrac{\sqrt{ {k^{\prime o}_3} {k^{o}_3} }}{ {p^{o}_3} }\,
\\[2mm]
\times\varphi^{*}(\norm{\vec k^\prime_\lambda}^2,\,\norm{\vec k^\prime_\rho}^2,\,{\vec k^\prime_\lambda}\cdot{\vec k^\prime_\rho})\,
	\varphi(\norm{\vec k_\lambda}^2,\,\norm{\vec k_\rho}^2,\,{\vec k_\lambda}\cdot{\vec k_\rho})
\\[2mm]
\times
\dfrac{1}{\sqrt{p^{\prime o}_1\,p^o_1}}\,\dfrac{1}{\sqrt{m^\prime_1 + k^{\prime 0}_1}}\,\dfrac{1}{\sqrt{m_1 + k^{0}_1}}\,
\dfrac{1}{\sqrt{m + k^{\prime o}_2}}\,\dfrac{1}{\sqrt{m + k^{o}_2}}\,\dfrac{1}{\sqrt{m + k^{\prime o}_3}}\,\dfrac{1}{\sqrt{m + k^{o}_3}}\\[2mm]
\times
\Bigl[
{
m^2(1 +  u\cdot u^\prime) + m( p_2 +  p_3)( u +  u^\prime)
+ ( p_2\cdot u)( p_3\cdot u^\prime) + ( p_3\cdot u)( p_2\cdot u^\prime)
+ ( p_2\cdot p_3)(1 -  u\cdot u^\prime)
}
\Bigr]\\[2mm]
\times
\left[
\bar\chi^{(m')}_{ u^{\prime}}\,
(m^\prime_1 + \fmslash{ p}^{\,\prime}_1)
\,J\,
(m_1 + \fmslash{ p}_1)\,
\chi^{(m)}_{ u}
\right]
\end{multlined}
}
\]
\end{maliste}
\subsubsection*{Limit of infinite mass}
The ``infinite mass'' limit for the heavy baryon is defined by:
\begin{gather*}
    \left(\dfrac{ p_1}{m_1},\,\dfrac{ p_1^\prime}{m_1^\prime}\right)\ \to\ ( v,\, v^\prime)
    \qquad\qquad
    ( u,\, u^{\prime})\ \to\ ( v,\, v^{\prime})
    \qquad\qquad
    \left(\dfrac{k^o_1}{m_1},\,\dfrac{k_1^{\prime o}}{m_1^\prime}\right)\ \to\ (1,\,1)\\[2mm]
    \text{for }i=2,\,3\ :\qquad\qquad\qquad
    ( k_i,\, k_i^\prime)\ \to\ \left({\boldsymbol B}_{ v}^{-1}\, p_i,\,{\boldsymbol B}_{ v^\prime}^{-1}\, p_i\right)
    \qquad\qquad
    (k^o_i,\,k^{\prime o}_i)\ \to\ \left( p_i\cdot v,\, p_i\cdot v^\prime\right)
\end{gather*}
One sets also : $\qquad\qquad\qquad\qquad\qquad\qquad w =  v\cdot v^\prime$\par\bigskip
Applying this limit to the previous transition amplitude, one arrives at the following expression:
\[
\begin{multlined}[\textwidth]
\brakket{\Psi^\prime,\,m^\prime}{J}{\Psi,\,m} = \dfrac{1}{8\pi^2}\,\dfrac{1}{\sqrt{v^o\,v^{\prime o}}}\,\int\dfrac{\dd\vec p_2}{(2\pi)^3}\,\dfrac{1}{2p^o_2}\,\dfrac{\dd\vec p_3}{(2\pi)^3}\,
\dfrac{1}{2p^o_3}\,
\varphi^{*}(\norm{\vec k^\prime_\lambda}^2,\,\norm{\vec k^\prime_\rho}^2,\,{\vec k^\prime_\lambda}\cdot{\vec k^\prime_\rho})\,
\varphi(\norm{\vec k_\lambda}^2,\,\norm{\vec k_\rho}^2,\,{\vec k_\lambda}\cdot{\vec k_\rho})
\\[2mm]
\times
\sqrt{\dfrac{ p_2\cdot v}{m +  p_2\cdot v}}\,
\sqrt{\dfrac{ p_2\cdot v^\prime}{m +  p_2\cdot v^\prime}}\,
\sqrt{\dfrac{ p_3\cdot v}{m +  p_3\cdot v}}\,
\sqrt{\dfrac{ p_3\cdot v^\prime}{m +  p_3\cdot v^\prime}}\\[2mm]
\times
\Bigl[
{
m^2(1 + w) + m( p_2 +  p_3)( v +  v^\prime)
+ ( p_2\cdot v)( p_3\cdot v^\prime) + ( p_3\cdot v)( p_2\cdot v^\prime)
+ ( p_2\cdot p_3)(1 - w)
}
\Bigr]\\[2mm]
\times
\left\{
\bar\chi^{(m')}_{ v^{\prime}}
\,J\,
\chi^{(m)}_{ v}
\right\}
\end{multlined}
\]
\subsubsection*{Manifest covariance of the obtained transition amplitude}
From the definition of the internal variables~\eqref{eq:definition}, we have 
\[
\vec k^{\prime}_\lambda = \llvec{{\boldsymbol{B}}^{-1}_{ v^{\prime}} p_\lambda} = {\dfrac12}\left( \llvec{{\boldsymbol{B}}^{-1}_{ v^{\prime}} p_2} + \llvec{{\boldsymbol{B}}^{-1}_{ v^{\prime}} p_3}\right)
\qquad\text{and also}\qquad
\vec k^{\prime}_\rho = \llvec{{\boldsymbol{B}}^{-1}_{ v^{\prime}} p_\rho} = \dfrac{1}{2}\left( \llvec{{\boldsymbol{B}}^{-1}_{ v^{\prime}} p_2} - \llvec{{\boldsymbol{B}}^{-1}_{ v^{\prime}} p_3}\right)
\]
Therefore, using the relations~\eqref{eq:internes} et~\eqref{eq:definition}, one gets
\begin{multline}\label{eq:covaUN}
{{\vec k}_\rho^\prime}{}^2\ =\ \bigl( p_\rho\cdot v^\prime\bigr)^2 -  p_\rho\cdot p_\rho
\ =\ \dfrac14\,\Bigl[\bigl( p_2 -  p_3\bigr)\cdot v^\prime\Bigr]^2 - \dfrac14\,\bigl( p_2 -  p_3\bigr)^2\\[2mm]
\Longrightarrow\qquad{
{\norm{\vec k_\rho^{\prime}}}\ =\ \dfrac{1}{2}\,\sqrt{\Bigl[\bigl( p_2 -  p_3\bigr)\cdot v^\prime\Bigr]^2 - \bigl( p_2 -  p_3\bigr)^2}
\ =\ \dfrac{1}{2}\,\sqrt{\Bigl[\bigl( p_2 -  p_3\bigr)\cdot v^\prime\Bigr]^2 - 2\,\bigl(m^2 -  p_2\cdot p_3\bigr)}
}
\end{multline}
One also has
\begin{multline}\label{eq:covaDEUX}
{{\vec k}_\lambda^\prime}{}^2\ =\ \bigl( p_\lambda\cdot v^\prime\bigr)^2 -  p_\lambda\cdot p_\lambda
\ =\ \dfrac14\,\Bigl[\bigl( p_2 +  p_3\bigr)\cdot v^\prime\Bigr]^2 - \dfrac14\,\bigl( p_2 +  p_3\bigr)^2\\[2mm]
\Longrightarrow\qquad{
{\norm{\vec k_\lambda^{\prime}}}\ =\ \dfrac{1}{2}\,\sqrt{\Bigl[\bigl( p_2 +  p_3\bigr)\cdot v^\prime\Bigr]^2 - \bigl( p_2 +  p_3\bigr)^2}
\ =\ \dfrac{1}{2}\,\sqrt{\Bigl[\bigl( p_2 +  p_3\bigr)\cdot v^\prime\Bigr]^2 - 2\,\bigl(m^2 +  p_2\cdot p_3\bigr)}
}
\end{multline}
and
\begin{multline}\label{eq:covaTROIS}
\vec k^{\prime}_\lambda\cdot\vec k^{\prime}_\rho =\ \bigl(p_\lambda\cdot v^\prime\bigr)\bigl( p_\rho\cdot v^\prime\bigr)\qquad\text{since $\ p_\lambda\cdot p_\rho = 0\quad$ (by a direct calculation using~\eqref{eq:definition})}\\[2mm]
\Longrightarrow\qquad{\vec k^{\prime}_\lambda\cdot\vec k^{\prime}_\rho\ =\ \dfrac{1}{4}\Bigl[\bigl( p_2 + p_3\bigr)\cdot v^\prime\Bigr]\Bigl[\bigl( p_2 -  p_3\bigr)\cdot v^\prime\Bigr]
\ =\ \dfrac{1}{4}\Bigl[\bigl( p_2 \cdot v^\prime\bigr)^2 - \bigl( p_3 \cdot v^\prime\bigr)^2\Bigr]
}
\end{multline}
The obtained expression of the transition amplitude is then manifestly covariant \footnote{The factor $\sqrt{v^{\prime o}\,v^o}$ comes from the normalization of the states in the Bakamjian-Thomas formalism. See the appendix \eqref{ann:normalisation}.}.
\subsubsection*{Consequence of the covariance}
The last expression of the transition amplitude can obviously be factorized according to:
\[
\brakket{\Psi^\prime,\,m^\prime}{J}{\Psi,\,m} = \dfrac{1}{8\pi^2}\,\dfrac{1}{\sqrt{v^o\,v^{\prime o}}}\,{\mathscr A}\,
\left\{
\bar\chi^{(m')}_{ v^{\prime}}
\,J\,
\chi^{(m)}_{ v}
\right\}
\]
where ${\mathscr A}$ stands for the integral 
\[
{
\begin{multlined}[\textwidth]
{\mathscr A} = \int\dfrac{\dd\vec p_2}{(2\pi)^3}\,\dfrac{1}{2p^o_2}\,\dfrac{\dd\vec p_3}{(2\pi)^3}\,
\dfrac{1}{2p^o_3}\,
\varphi^{*}(\norm{\vec k^\prime_\lambda}^2,\,\norm{\vec k^\prime_\rho}^2,\,{\vec k^\prime_\lambda}\cdot{\vec k^\prime_\rho})\,
\varphi(\norm{\vec k_\lambda}^2,\,\norm{\vec k_\rho}^2,\,{\vec k_\lambda}\cdot{\vec k_\rho})
\\[2mm]
\times
\sqrt{\dfrac{ p_2\cdot v}{m +  p_2\cdot v}}\,
\sqrt{\dfrac{ p_2\cdot v^\prime}{m +  p_2\cdot v^\prime}}\,
\sqrt{\dfrac{ p_3\cdot v}{m +  p_3\cdot v}}\,
\sqrt{\dfrac{ p_3\cdot v^\prime}{m +  p_3\cdot v^\prime}}\,
\\[2mm]
\times
\Bigl[
{
m^2(1 + w) + m( p_2 +  p_3)( v +  v^\prime)
+ ( p_2\cdot v)( p_3\cdot v^\prime) + ( p_3\cdot v)( p_2\cdot v^\prime)
+ ( p_2\cdot p_3)(1 - w)
}
\Bigr]
\end{multlined}
}
\]
Since the initial amplitude is written in a covariant way, this means that the term ${\mathscr A}$ is invariant under any Lorentz transformations. Therefore, one knows its structure 
\[
{\mathscr A}\ =\ A(w)
\]
which follows from the unique way to write a relativistic invariant scalar using the objects at our disposal (the two velocities and the metric tensor).
\subsection{The Isgur-Wise function $\boldsymbol{\xi_\Lambda(w)}$}
The usual definition of the Isgur-Wise function $\xi_\Lambda(w)$ is given in terms of the baryon states with the relativistic normalization:
\[
\dfrac{{}_R\brakket{\Psi^\prime,\,m^\prime}{J}{\Psi,\,m}_R}{\sqrt{m_\Psi\,m_{\Psi^\prime}}}\ =\ \xi_\Lambda(w)\,\left\{
\bar\chi^{(m')}_{ v^{\prime}}
\,J\,
\chi^{(m)}_{ v}
\right\}
\]
Since, in the Bakamjian-Thomas formalism, the states are normalized non-relativistically, our definition will be:
\[
{
\brakket{\Psi^\prime,\,m^\prime}{J}{\Psi,\,m}\ =\ \dfrac{1}{2}\,\dfrac{1}{\sqrt{v^o\,v^{\prime o}}}\,\xi_\Lambda(w)\,\left\{
\bar\chi^{(m')}_{ v^{\prime}}
\,J\,
\chi^{(m)}_{ v}
\right\}
}
\]
Identifying both formulas one gets
\[
\xi_\Lambda(w) \ =\ 
\dfrac{1}{4\pi^2}\,{\mathscr A}
\]
and therefore
\begin{multline*}
\xi_\Lambda(w) = \dfrac{1}{4\pi^2}\,\int\dfrac{\dd\vec p_2}{(2\pi)^3}\,\dfrac{1}{2p^o_2}\,\dfrac{\dd\vec p_3}{(2\pi)^3}\,
\dfrac{1}{2p^o_3}\,
\varphi^{*}(\norm{\vec k^\prime_\lambda}^2,\,\norm{\vec k^\prime_\rho}^2,\,{\vec k^\prime_\lambda}\cdot{\vec k^\prime_\rho})\,
\varphi(\norm{\vec k_\lambda}^2,\,\norm{\vec k_\rho}^2,\,{\vec k_\lambda}\cdot{\vec k_\rho})
\\[2mm]
\times
\sqrt{\dfrac{ p_2\cdot v}{m +  p_2\cdot v}}\,
\sqrt{\dfrac{ p_2\cdot v^\prime}{m +  p_2\cdot v^\prime}}\,
\sqrt{\dfrac{ p_3\cdot v}{m +  p_3\cdot v}}\,
\sqrt{\dfrac{ p_3\cdot v^\prime}{m +  p_3\cdot v^\prime}}\,
\\[2mm]
\times
\Bigl[
{
m^2(1 + w) + m( p_2 +  p_3)( v +  v^\prime)
+ ( p_2\cdot v)( p_3\cdot v^\prime) + ( p_3\cdot v)( p_2\cdot v^\prime)
+ ( p_2\cdot p_3)(1 - w)
}
\Bigr]
\end{multline*}
%
%
\section{The $\boldsymbol{1/2^+\to 1/2^-}$ transition}
We will consider in this paragraph the transition amplitude from a $(L=0,\,\ell_\rho=0)$ baryonic state to the full tower of $L=1$ states. Considering the corresponding wave functions presented in the sections~\ref{sec:WFzero} and~\ref{sec:WFun}, we can see that the transition amplitudes will contain three types of terms:
\begin{align*}
(i)&\quad\bigl(\sigma_2\bigr)_{s_2,s_3}\,\Bigl(\hat k_{\rho/\lambda}\cdot\vec\sigma\Bigr)\,\chi^{(m)}\quad(\text{even}\ \ell_\rho)\\
(ii)&\quad\hat k_{\rho/\lambda}\cdot\bigl(\vec\sigma\,\sigma_2\bigr)_{s_2,s_3}\,\chi^{(m)}\quad(\text{odd}\ \ell_\rho,\, j=0)\\
(iii)&\quad\bigl(\vec\sigma\,\sigma_2\bigr)_{s_2,s_3}\cdot\biggl[
\Bigl(\hat k_{\rho/\lambda}\wedge\vec\sigma\Bigr)\,\chi^{(m)}\biggr]\quad(\text{odd}\ \ell_\rho,\, j=1)
\end{align*}
multiplied by real scalar functions that depend on the scalar product $\hat k_\rho\cdot\hat k_\lambda$.
\subsection{Term of the type $\boldsymbol{(i)}$}
The starting point is 
\[
\left\{
\begin{aligned}
\Psi^{(m)}_{s_1,s_2,s_3}(\vec k_\lambda,\,\vec k_\rho)
&= \dfrac{1}{4\pi}\dfrac{i}{\sqrt{2}}\,\chi^{(m)}_{s_1}\,(\sigma_2)_{s_2,s_3}\,\varphi(\norm{\vec k_\lambda}^2,\,\norm{\vec k_\rho}^2,\,{\vec k_\lambda}\cdot{\vec k_\rho})\\[2mm]
\Psi^{\prime(m')}_{s_1^\prime,s_2^\prime,s_3^\prime}(\vec k^\prime_\lambda,\,\vec k^\prime_\rho) &= \dfrac{i}{\sqrt{2}}\,{\mathscr G}^{(\rho/\lambda)}
\,(\sigma_2)_{s_2^\prime,s_3^\prime}
\,\biggl[\Bigl(\hat k^\prime_{\rho/\lambda}\cdot\vec\sigma\Bigr)\,\chi^{(m')}
\biggr]_{s^\prime_1}
\,\psi(\norm{\vec k^\prime_\lambda}^2,\,\norm{\vec k^\prime_\rho}^2,\,{\vec k^\prime_\lambda}\cdot{\vec k^\prime_\rho})
\end{aligned}
\right.
\]
where
\[
\vec F_{\ell_\rho;\pm 1}\quad\leadsto\quad
{\mathscr G}^{(\rho/\lambda)}\quad\leadsto\quad
\left\{
\begin{aligned}
\ {\mathscr G}^{(\rho)}&= c_{\ell_\rho;\pm 1}\,\dfrac{1}{\norm{\hat k^\prime_\rho\wedge\hat k^\prime_\lambda}}\,P_{\ell_\rho}^1\bigl(\hat k^\prime_\rho\cdot\hat k^\prime_\lambda\bigr)\quad\in\reel\\[2mm]
\ {\mathscr G}^{(\lambda)}&= -\,c_{\ell_\rho;\pm 1}\,\dfrac{1}{\norm{\hat k^\prime_\rho\wedge\hat k^\prime_\lambda}}\,P_{\ell_\rho\pm 1}^1\bigl(\hat k^\prime_\rho\cdot\hat k^\prime_\lambda\bigr)\quad\in\reel
\end{aligned}
\right.
\]
Therefore, we have:
\begin{align*}
\Psi^{\prime*\,(m')}_{s_1^\prime,s_2^\prime,s_3^\prime}(\vec k^\prime_\lambda,\,\vec k^\prime_\rho) &= \dfrac{-i}{\sqrt{2}}\,{\mathscr G}^{(\rho/\lambda)}
\,(\sigma_2)^{*}_{s_2^\prime,s_3^\prime}
\,\biggl[\Bigl(\hat k^\prime_{\rho/\lambda}\cdot\vec\sigma\Bigr)\,\chi^{(m')}
\biggr]^{*}_{s^\prime_1}
\,\psi^{*}(\norm{\vec k^\prime_\lambda}^2,\,\norm{\vec k^\prime_\rho}^2,\,{\vec k^\prime_\lambda}\cdot{\vec k^\prime_\rho})\\[2mm]
&= \dfrac{-i}{\sqrt{2}}\,{\mathscr G}^{(\rho/\lambda)}
\,\Bigl[{}^{t}(\sigma_2)^{\dagger}\Bigr]_{s_2^\prime,s_3^\prime}
\,{}^{t}\biggl[\Bigl(\hat k^\prime_{\rho/\lambda}\cdot\vec\sigma\Bigr)\,\chi^{(m')}
\biggr]^{\dagger}_{s^\prime_1}
\,\psi^{*}(\norm{\vec k^\prime_\lambda}^2,\,\norm{\vec k^\prime_\rho}^2,\,{\vec k^\prime_\lambda}\cdot{\vec k^\prime_\rho})\\[2mm]
&= \dfrac{-i}{\sqrt{2}}\,{\mathscr G}^{(\rho/\lambda)}
\,\Bigl[{}^{t}(\sigma_2)^{\dagger}\Bigr]_{s_2^\prime,s_3^\prime}
\,{}^{t}\biggl[\chi^{\dagger(m')}\,\Bigl(\hat k^\prime_{\rho/\lambda}\cdot\vec\sigma^\dagger\Bigr)
\biggr]_{s^\prime_1}
\,\psi^{*}(\norm{\vec k^\prime_\lambda}^2,\,\norm{\vec k^\prime_\rho}^2,\,{\vec k^\prime_\lambda}\cdot{\vec k^\prime_\rho})\\[2mm]
&= \dfrac{-i}{\sqrt{2}}\,{\mathscr G}^{(\rho/\lambda)}
\,\Bigl[{}^{t}(\sigma_2)\Bigr]_{s_2^\prime,s_3^\prime}
\,\biggl[\chi^{\dagger(m')}\,\Bigl(\hat k^\prime_{\rho/\lambda}\cdot\vec\sigma\Bigr)
\biggr]_{s^\prime_1}
\,\psi^{*}(\norm{\vec k^\prime_\lambda}^2,\,\norm{\vec k^\prime_\rho}^2,\,{\vec k^\prime_\lambda}\cdot{\vec k^\prime_\rho})\qquad\text{since $\sigma_i^\dagger = \sigma_i$}\\[2mm]
&= \dfrac{-i}{\sqrt{2}}\,{\mathscr G}^{(\rho/\lambda)}
\,(\sigma_2)_{s_3^\prime,s_2^\prime}
\,\biggl[\chi^{\dagger(m')}\,\Bigl(\hat k^\prime_{\rho/\lambda}\cdot\vec\sigma\Bigr)
\biggr]_{s^\prime_1}
\,\psi^{*}(\norm{\vec k^\prime_\lambda}^2,\,\norm{\vec k^\prime_\rho}^2,\,{\vec k^\prime_\lambda}\cdot{\vec k^\prime_\rho})
\end{align*}
The BT transition amplitude to be computed then reads:
\begin{multline*}
\brakket{\Psi^\prime,\,m^\prime}{J}{\Psi,\,m}_{(\rho/\lambda)} = \dfrac{1}{8\pi}\,\int\dfrac{\dd\vec p_2}{(2\pi)^3}\dfrac{\dd\vec p_3}{(2\pi)^3}\,
\sqrt{u^{\prime o}\,u^o}\,
\sqrt{\dfrac{ {k^{\prime o}_1} {k^{o}_1} }{ {p^{\prime o}_1} {p^{o}_1} }}\,
\sqrt{\dfrac{ {k^{\prime o}_2} {k^{o}_2} }{ {p^{\prime o}_2} {p^{o}_2} }}\,
\sqrt{\dfrac{ {k^{\prime o}_3} {k^{o}_3} }{ {p^{\prime o}_3} {p^{o}_3} }}\\[2mm]
\times
\psi^{*}(\norm{\vec k^\prime_\lambda}^2,\,\norm{\vec k^\prime_\rho}^2,\,{\vec k^\prime_\lambda}\cdot{\vec k^\prime_\rho})\,
\varphi(\norm{\vec k_\lambda}^2,\,\norm{\vec k_\rho}^2,\,{\vec k_\lambda}\cdot{\vec k_\rho})
\\[2mm]
\times{\mathscr G}^{(\rho/\lambda)}\,\sum\limits_{s_1^\prime,s_2^\prime,s_3^\prime}\sum\limits_{s_1,s_2,s_3}
\chi^{(m)}_{s_1}\,(\sigma_2)_{s_2,s_3}\,
\biggl[\chi^{\dagger(m')}\,\Bigl(\hat k^\prime_{\rho/\lambda}\cdot\vec\sigma\Bigr)
\biggr]_{s_1^\prime}(\sigma_2)_{s^\prime_3,s^\prime_2}
\\[2mm]
\times
\left[D({\boldsymbol R}_1^{\prime -1})\,J(\vec p^{\,\prime}_1,\,\vec p_1)\,D({\boldsymbol R}_1)\right]_{s_1^\prime,s_1}\,
\left[D({\boldsymbol R}_2^{\prime -1}{\boldsymbol R}_2)\right]_{s_2^\prime,s_2}\,
\left[D({\boldsymbol R}_3^{\prime -1}{\boldsymbol R}_3)\right]_{s_3^\prime,s_3}
\end{multline*}
i.e.,
\begin{multline*}
\brakket{\Psi^\prime,\,m^\prime}{J}{\Psi,\,m}_{(\rho/\lambda)} = \dfrac{1}{8\pi}\,\int\dfrac{\dd\vec p_2}{(2\pi)^3}\dfrac{\dd\vec p_3}{(2\pi)^3}\,
\sqrt{u^{\prime o}\,u^o}\,
\sqrt{\dfrac{ {k^{\prime o}_1} {k^{o}_1} }{ {p^{\prime o}_1} {p^{o}_1} }}\,
\sqrt{\dfrac{ {k^{\prime o}_2} {k^{o}_2} }{ {p^{\prime o}_2} {p^{o}_2} }}\,
\sqrt{\dfrac{ {k^{\prime o}_3} {k^{o}_3} }{ {p^{\prime o}_3} {p^{o}_3} }}\\[2mm]
\times
\psi^{*}(\norm{\vec k^\prime_\lambda}^2,\,\norm{\vec k^\prime_\rho}^2,\,{\vec k^\prime_\lambda}\cdot{\vec k^\prime_\rho})\,
\varphi(\norm{\vec k_\lambda}^2,\,\norm{\vec k_\rho}^2,\,{\vec k_\lambda}\cdot{\vec k_\rho})
\,{\mathscr G}^{(\rho/\lambda)}\,
\biggl[\chi^{\dagger(m')}\,\Bigl(\hat k^\prime_{\rho/\lambda}\cdot\vec\sigma\Bigr)\,
D({\boldsymbol R}_1^{\prime -1})\,J(\vec p^{\,\prime}_1,\,\vec p_1)\,D({\boldsymbol R}_1)
\chi^{(m)}\biggr]\\[2mm]
\times\sum\limits_{s_2^\prime,s_3^\prime}\sum\limits_{s_2,s_3}
\left[D({\boldsymbol R}_2^{\prime -1}{\boldsymbol R}_2)\right]_{s_2^\prime,s_2}\,
\left[D({\boldsymbol R}_3^{\prime -1}{\boldsymbol R}_3)\right]_{s_3^\prime,s_3}
\,(\sigma_2)_{s_2,s_3}\,(\sigma_2)_{s^\prime_3,s^\prime_2}
\end{multline*}
The Wigner rotations satisfy the following property 
\[
\sigma_2\,\left[D({\boldsymbol R})\right]\,\sigma_2 = {}^t\left[D({\boldsymbol R}^{-1})\right]\qquad\Longrightarrow\qquad
\left[D({\boldsymbol R}_3^{\prime -1}{\boldsymbol R}_3)\right]_{s_3^\prime,s_3} = 
\left[\sigma_2\,D({\boldsymbol R}_3^{-1}{\boldsymbol R}^\prime_3)\,\sigma_2\right]_{s_3,s_3^\prime}
\]
Therefore one arrives at the starting relation :
\[
{
\begin{multlined}
\brakket{\Psi^\prime,\,m^\prime}{J}{\Psi,\,m}_{(\rho/\lambda)} = \dfrac{1}{8\pi}\,\int\dfrac{\dd\vec p_2}{(2\pi)^3}\dfrac{\dd\vec p_3}{(2\pi)^3}\,
\sqrt{u^{\prime o}\,u^o}\,
\sqrt{\dfrac{ {k^{\prime o}_1} {k^{o}_1} }{ {p^{\prime o}_1} {p^{o}_1} }}\,
\sqrt{\dfrac{ {k^{\prime o}_2} {k^{o}_2} }{ {p^{\prime o}_2} {p^{o}_2} }}\,
\sqrt{\dfrac{ {k^{\prime o}_3} {k^{o}_3} }{ {p^{\prime o}_3} {p^{o}_3} }}\\[2mm]
\times
\psi^{*}(\norm{\vec k^\prime_\lambda}^2,\,\norm{\vec k^\prime_\rho}^2,\,{\vec k^\prime_\lambda}\cdot{\vec k^\prime_\rho})\,
\varphi(\norm{\vec k_\lambda}^2,\,\norm{\vec k_\rho}^2,\,{\vec k_\lambda}\cdot{\vec k_\rho})
\\[2mm]
\qquad\qquad\qquad\qquad\qquad\qquad\times{\mathscr G}^{(\rho/\lambda)}\,
\biggl[\chi^{\dagger(m')}\,\Bigl(\hat k^\prime_{\rho/\lambda}\cdot\vec\sigma\Bigr)\,
D({\boldsymbol R}_1^{\prime -1})\,J(\vec p^{\,\prime}_1,\,\vec p_1)\,D({\boldsymbol R}_1)
\chi^{(m)}\biggr]\,\text{Tr}\Bigl[
D({\boldsymbol R}_2^{\prime -1}{\boldsymbol R}_2)\,D({\boldsymbol R}_3^{-1}{\boldsymbol R}^\prime_3)\Bigr]
\end{multlined}
}
\]
Remember that the complete amplitude reads, from the definition ${\norm{\hat k^\prime_\rho\wedge\hat k^\prime_\lambda}}\,\vec F_{\ell_\rho;\pm 1} = {\mathscr G}^{(\rho)}\hat k^\prime_{\rho} + {\mathscr G}^{(\lambda)}\hat k^\prime_{\lambda}$ :
\[
\brakket{\Psi^\prime,\,m^\prime}{J}{\Psi,\,m} = \brakket{\Psi^\prime,\,m^\prime}{J}{\Psi,\,m}_{(\rho)} + \brakket{\Psi^\prime,\,m^\prime}{J}{\Psi,\,m}_{(\lambda)}
\]
\subsubsection*{Formulation in 4D}
\begin{maliste}
\item{\underline{Term between brackets} :}\par
Following the usual procedure, one starts from: 
\begin{multline*}
{\mathscr T}_1 = \biggl[
\chi^{\dagger(m')}\Bigl(\hat k^\prime_{\rho/\lambda}\cdot\vec\sigma\Bigr)\,
D({\boldsymbol R}_1^{\prime -1})\,J(\vec p^{\,\prime}_1,\,\vec p_1)\,D({\boldsymbol R}_1)
\chi^{(m)}
\biggr]\\
\leadsto\quad{\mathscr T}_1 = \sqrt{\dfrac{m^{\prime}_1\,m_1}{p^{\prime o}_1\,p^o_1}}\,\left[
\chi^{(m')\dagger}\,\dfrac{1 + \gamma^0}{2}\,(\hat k^\prime_{\rho/\lambda}\cdot\vec\sigma)
D({\boldsymbol R}_1^{\prime -1})\,
\dfrac{1 + \gamma^0}{2}\,{\boldsymbol B}_{p^\prime_1}^{-1}\,J\,{\boldsymbol B}_{p_1}\,
\dfrac{1 + \gamma^0}{2}
\,D({\boldsymbol R}_1)\,
\dfrac{1 + \gamma^0}{2}\,\chi^{(m)}
\right]
\end{multline*}
Again, one replaces the expression of the Wigner rotations:
\begin{align*}
{\mathscr T}_1 &= \sqrt{\dfrac{m^{\prime}_1\,m_1}{p^{\prime o}_1\,p^o_1}}\,\left[
\chi^{(m')\dagger}\,\dfrac{1 + \gamma^0}{2}\,(\hat k_{\rho/\lambda}^\prime\cdot\vec\sigma)
{\boldsymbol B}_{k^\prime_1}^{-1}\,{\boldsymbol B}_{u^\prime}^{-1}\,{\boldsymbol B}_{p^\prime_1}
\,
\dfrac{1 + \gamma^0}{2}\,{\boldsymbol B}_{p^\prime_1}^{-1}\,J\,{\boldsymbol B}_{p_1}\,
\dfrac{1 + \gamma^0}{2}
\,
{\boldsymbol B}_{p_1}^{-1}\,{\boldsymbol B}_{u}\,{\boldsymbol B}_{k_1}
\,
\dfrac{1 + \gamma^0}{2}\,\chi^{(m)}
\right]\\[2mm]
& = \begin{multlined}[t][160mm]\dfrac{1}{16}\,\sqrt{\dfrac{m^{\prime}_1\,m_1}{p^{\prime o}_1\,p^o_1}}\,\left[
\chi^{(m')\dagger}\,(1 + \gamma^0)\,(\hat k_{\rho/\lambda}^\prime\cdot\vec\sigma)\,(1 + \gamma^0)\,
{\boldsymbol B}_{k^\prime_1}^{-1}\,{\boldsymbol B}_{u^\prime}^{-1}\,{\boldsymbol B}_{p^\prime_1}
\,
{\boldsymbol B}_{p^\prime_1}^{-1}\,J\right.\\
\left.\times\,{\boldsymbol B}_{p_1}\,
{\boldsymbol B}_{p_1}^{-1}\,{\boldsymbol B}_{u}\,{\boldsymbol B}_{k_1}
\,(1 + \gamma^0)
\,
(1 + \gamma^0)\,\chi^{(m)}
\right]\\
\text{since}\quad(1 + \gamma^0)\,D({\boldsymbol R}_i) = D({\boldsymbol R}_i)\,(1 + \gamma^0)\end{multlined}\\[2mm]
& = \begin{multlined}[t][160mm]\dfrac{1}{8}\,\sqrt{\dfrac{m^{\prime}_1\,m_1}{p^{\prime o}_1\,p^o_1}}\,
\left[
\chi^{(m')\dagger}\,(1 + \gamma^0)\,(\hat k_{\rho/\lambda}^\prime\cdot\vec\sigma)\,(1 + \gamma^0)\,
{\boldsymbol B}_{k^\prime_1}^{-1}\,{\boldsymbol B}_{u^\prime}^{-1}\,J\,
{\boldsymbol B}_{u}\,{\boldsymbol B}_{k_1}
\,(1 + \gamma^0)
\,\chi^{(m)}
\right]\\ \text{since}\quad(1 + \gamma^0)^2 = 2\,(1 + \gamma^0)\end{multlined}\\[2mm]
& = \begin{multlined}[t][160mm]\dfrac{1}{8}\,\sqrt{\dfrac{m^{\prime}_1\,m_1}{p^{\prime o}_1\,p^o_1}}\,
\left[
\chi^{(m')\dagger}\,\gamma^0\,(1 + \gamma^0)\,(\hat k_{\rho/\lambda}^\prime\cdot\vec\sigma)\,(1 + \gamma^0)\,
{\boldsymbol B}_{k^\prime_1}^{-1}\,{\boldsymbol B}_{u^\prime}^{-1}\,J\,
{\boldsymbol B}_{u}\,{\boldsymbol B}_{k_1}
\,(1 + \gamma^0)
\,\chi^{(m)}
\right]\\ \text{since}\quad\gamma^0\,(1 + \gamma^0) = 1 + \gamma^0\end{multlined}\\[2mm]
& = \begin{multlined}[t][160mm]\dfrac{1}{8}\,\sqrt{\dfrac{m^{\prime}_1\,m_1}{p^{\prime o}_1\,p^o_1}}\,
\left[
\chi^{(m')\dagger}\,\gamma^0\,
{\boldsymbol B}_{u^{\prime}}^{-1}{\boldsymbol B}_{u^{\prime}}
\,(1 + \gamma^0)\,
{\boldsymbol B}_{u^{\prime}}^{-1}{\boldsymbol B}_{u^{\prime}}
\,
(\hat k_{\rho/\lambda}^\prime\cdot\vec\sigma)\,
{\boldsymbol B}_{u^{\prime}}^{-1}\right.\\
\times\left.{\boldsymbol B}_{u^{\prime}}\,
(1 + \gamma^0)\,
{\boldsymbol B}_{k^\prime_1}^{-1}\,{\boldsymbol B}_{u^\prime}^{-1}\,J\,
{\boldsymbol B}_{u}\,{\boldsymbol B}_{k_1}
\,(1 + \gamma^0)\,
{\boldsymbol B}_{u}^{-1}{\boldsymbol B}_{u}
\,\chi^{(m)}
\right]\\
\text{inserting in the right places $\ {\boldsymbol B}_{u^{(\prime)}}^{-1}{\boldsymbol B}_{u^{(\prime)}} = {\boldsymbol 1}$}
\end{multlined}\\[2mm]
& = \begin{multlined}[t][160mm]\dfrac{1}{8}\,\sqrt{\dfrac{m^{\prime}_1\,m_1}{p^{\prime o}_1\,p^o_1}}\,
\left[
\chi^{(m')\dagger}\,\gamma^0\,{\boldsymbol B}_{u^{\prime}}^{-1}\,
(1 + \fmslash{u}^{\prime})\,
\gamma_5\,\dfrac{\Bigl[\fmslash{p}_{\rho/\lambda}^{\prime} - (p_{\rho/\lambda}^{\prime}\cdot u^{\prime})\,\fmslash{u}^{\prime}\Bigr]}{\norm{\vec k_{\rho/\lambda}^{\prime}}}\right.\\
\times\left.
\,\dfrac{(1+\fmslash{u}^{\,\prime})(m^\prime_1 + \fmslash{p}^{\,\prime}_1)}{\sqrt{2\,m^\prime_1\,(m^\prime_1 + k^{\prime 0}_1)}}
\,J\,
\dfrac{(m_1 + \fmslash{p}_1)(1+\fmslash{u})}{\sqrt{2\,m_1\,(m_1 + k^0_1)}}\,
{\boldsymbol B}_{u}
\,\chi^{(m)}
\right]\\
\text{using the relations of the appendix \eqref{ann:boosts}}
\end{multlined}\\[2mm]
& = \begin{multlined}[t][160mm]\dfrac{1}{16}\,\dfrac{1}{\sqrt{p^{\prime o}_1\,p^o_1}}\,
\dfrac{1}{\sqrt{m^\prime_1 + k^{\prime 0}_1}}\,
\dfrac{1}{\sqrt{m_1 + k^{0}_1}}\,
\dfrac{1}{\norm{\vec k_{\rho/\lambda}^{\prime}}}\\
\times
\,\biggl[
\bar\chi^{(m')}\,{\boldsymbol B}_{u^{\prime}}^{-1}\,
(1 + \fmslash{u}^{\,\prime})\,
\gamma_5\,\Bigl[\fmslash{p}_{\rho/\lambda}^{\prime} - (p_{\rho/\lambda}^{\prime}\cdot u^{\prime})\,\fmslash{u}^{\prime}\Bigr]\,
(1+\fmslash{u}^{\prime})\,(m^\prime_1 + \fmslash{p}^{\prime}_1)
\,J\,
(m_1 + \fmslash{p}_1)\,(1+\fmslash{u})\,
{\boldsymbol B}_{u}
\,\chi^{(m)}
\biggr]
\end{multlined}
\end{align*}
In terms of the boosted spinors, we finally get:
\[
{
\begin{multlined}[160mm]
{\mathscr T}_1  = \dfrac{1}{4}\,\dfrac{1}{\sqrt{p^{\prime o}_1\,p^o_1}}\,
\dfrac{1}{\sqrt{m^\prime_1 + k^{\prime 0}_1}}\,
\dfrac{1}{\sqrt{m_1 + k^{0}_1}}\,
\dfrac{1}{\norm{\vec k_{\rho/\lambda}^{\prime}}}\\
\times\left[
\bar\chi^{(m')}_{u^{\prime}}\,
\gamma_5\,\Bigl[\fmslash{p}_{\rho/\lambda}^{\prime} - (p_{\rho/\lambda}^{\prime}\cdot u^{\prime})\,\fmslash{u}^{\prime}\Bigr]\,
(1+\fmslash{u}^{\prime})\,(m^\prime_1 + \fmslash{p}^{\prime}_1)
\,J\,
(m_1 + \fmslash{p}_1)\,
\chi^{(m)}_{u}
\right]
\end{multlined}
}
\]
\item{\underline{Calculation of the trace} :}\par
This trace has been already calculated previously in section~\ref{para:bracketUN} and was found to be:
\[
{
\begin{multlined}[170mm]
{\mathscr T}_2 = \dfrac{1}{\sqrt{m + k^{\prime o}_2}}\,\dfrac{1}{\sqrt{m + k^{o}_2}}\,
\dfrac{1}{\sqrt{m + k^{\prime o}_3}}\,\dfrac{1}{\sqrt{m + k^{o}_3}}\\[2mm]
\times
\left[
{
m^2(1 +  u\cdot u^\prime) + m( p_2 +  p_3)( u +  u^\prime)
+ ( p_2\cdot u)( p_3\cdot u^\prime) + ( p_3\cdot u)( p_2\cdot u^\prime)
+ ( p_2\cdot p_3)(1 -  u\cdot u^\prime)
}
\right]
\end{multlined}
}
\]
\item{\underline{Conclusion} :}\par
Gathering everything one obtains :
\begin{multline*}
    \brakket{\Psi^\prime,\,m^\prime}{J}{\Psi,\,m}_{(\rho/\lambda)} = \dfrac{1}{8\pi}\,\int\dfrac{\dd\vec p_2}{(2\pi)^3}\dfrac{1}{2\,p^{o}_2}\,\dfrac{\dd\vec p_3}{(2\pi)^3}\dfrac{1}{2\,p^{o}_3}\,
    \dfrac{\sqrt{u^{\prime o}\,u^o}}{ {p^{\prime o}_1} {p^{o}_1} }\,
	{\mathscr G}^{(\rho/\lambda)}\\[2mm]
    \times
    \psi^{*}(\norm{\vec k^\prime_\lambda}^2,\,\norm{\vec k^\prime_\rho}^2,\,{\vec k^\prime_\lambda}\cdot{\vec k^\prime_\rho})\,
    \varphi(\norm{\vec k_\lambda}^2,\,\norm{\vec k_\rho}^2,\,{\vec k_\lambda}\cdot{\vec k_\rho})
    \\[2mm]
    \times\,\sqrt{\dfrac{k^{o}_1}{m_1 + k^{o}_1} }\,
	\sqrt{\dfrac{k^{\prime o}_1}{m^\prime_1 + k^{\prime o}_1} }\,
	\sqrt{\dfrac{k^{o}_2}{m + k^{o}_2} }\,
	\sqrt{\dfrac{k^{\prime o}_2}{m + k^{\prime o}_2} }\,
	\sqrt{\dfrac{k^{o}_3}{m + k^{o}_3} }\,
	\sqrt{\dfrac{k^{\prime o}_3}{m + k^{\prime o}_3} }\,
	\dfrac{1}{\norm{\vec k_{\rho/\lambda}^{\prime}}}
	\\[2mm]
	\times
	\Bigl[
	{
	m^2(1 +  u\cdot u^\prime) + m( p_2 +  p_3)( u +  u^\prime)
	+ ( p_2\cdot u)( p_3\cdot u^\prime) + ( p_3\cdot u)( p_2\cdot u^\prime)
	+ ( p_2\cdot p_3)(1 -  u\cdot u^\prime)
	}
	\Bigr]\\[2mm]
	\times\biggl[
	\bar\chi^{(m')}_{ u^{\prime}}\,
	\gamma_5\,\Bigl[\fmslash{ p}_{\rho/\lambda}^{\prime} - ( p_{\rho/\lambda}^{\prime}\cdot u^{\prime})\,\fmslash{ u}^{\prime}\Bigr]\,
	(1+\fmslash{ u}^{\prime})\,(m^\prime_1 + \fmslash{ p}^{\prime}_1)
	\,J\,
	(m_1 + \fmslash{ p}_1)\,
	\chi^{(m)}_{ u}
	\biggr]
\end{multline*}
\end{maliste}
\subsubsection*{Limit of infinite mass}
Taking the ``infinite mass'' limit, one obtains:
\[
{
\begin{multlined}[\textwidth-10mm]
    \brakket{\Psi^\prime,\,m^\prime}{J}{\Psi,\,m}_{(\rho/\lambda)} = \dfrac{1}{8\pi}\,\dfrac{1}{\sqrt{v^{\prime o}\,v^o}}\,\int\dfrac{\dd\vec p_2}{(2\pi)^3}\dfrac{1}{2\,p^{o}_2}\,\dfrac{\dd\vec p_3}{(2\pi)^3}\dfrac{1}{2\,p^{o}_3}\,
	{\mathscr G}^{(\rho/\lambda)}\\[2mm]
    \times
    \psi^{*}(\norm{\vec k^\prime_\lambda}^2,\,\norm{\vec k^\prime_\rho}^2,\,{\vec k^\prime_\lambda}\cdot{\vec k^\prime_\rho})\,
    \varphi(\norm{\vec k_\lambda}^2,\,\norm{\vec k_\rho}^2,\,{\vec k_\lambda}\cdot{\vec k_\rho})
    \\[2mm]
    \times\,
	\sqrt{\dfrac{ p_2\cdot v}{m +  p_2\cdot v}}\,
    \sqrt{\dfrac{ p_2\cdot v^\prime}{m +  p_2\cdot v^\prime}}\,
    \sqrt{\dfrac{ p_3\cdot v}{m +  p_3\cdot v}}\,
    \sqrt{\dfrac{ p_3\cdot v^\prime}{m +  p_3\cdot v^\prime}}\,
	\dfrac{1}{\norm{\vec k_{\rho/\lambda}^{\prime}}}
	\\[2mm]
	\times
	\Bigl[
	{
	m^2(1 + w) + m( p_2 +  p_3)( v +  v^\prime)
	+ ( p_2\cdot v)( p_3\cdot v^\prime) + ( p_3\cdot v)( p_2\cdot v^\prime)
	+ ( p_2\cdot p_3)(1 - w)
	}
	\Bigr]\\[2mm]
	\times 2\,\biggl\{
	\bar\chi^{(m')}_{ v^{\prime}}\,
	\gamma_5\,\Bigl[\fmslash{ p}_{\rho/\lambda} - ( p_{\rho/\lambda}\cdot v^{\prime})\,\fmslash{ v}^{\prime}\Bigr]\,
	(1+\fmslash{ v}^{\prime})
	\,J\,
	\chi^{(m)}_{ v}
	\biggr\}
\end{multlined}
}
\]
Note that the ``prime''~on the boosted Jacobi momenta are removed since they are useless from now on: the internal variables are obtained by the boosts of velocities $ v$ and $ v'$. 
\subsubsection*{Covariance and its consequences}
This last expression of the transition amplitude is relativistically invariant (see equations~\eqref{eq:covaUN}, \eqref{eq:covaDEUX} and \eqref{eq:covaTROIS}).
\begin{maliste}
\item{\underline{Covariant structure} :}\par
By making the term within brackets explicit, we can write:
\[
\brakket{\Psi^\prime,\,m^\prime}{J}{\Psi,\,m}_{(\rho/\lambda)} = \dfrac{1}{4\pi}\,\dfrac{1}{\sqrt{v^{\prime o}\,v^o}}\,{\mathscr A}^\mu_{(\rho/\lambda)}\,
\biggl\{
\bar\chi^{(m')}_{ v^{\prime}}\,
\gamma_5\,\Bigl[\gamma_\mu -  v^{\prime}_\mu\,\fmslash{ v}^{\prime}\Bigr]\,
(1+\fmslash{ v}^{\prime})
\,J\,
\chi^{(m)}_{ v}
\biggr\}
\]
where ${\mathscr A}^\mu_{(\rho/\lambda)}$ corresponds to
\[
{
\begin{multlined}[\textwidth-10mm]
    {\mathscr A}^\mu_{(\rho/\lambda)} = \int\dfrac{\dd\vec p_2}{(2\pi)^3}\dfrac{1}{2\,p^{o}_2}\,\dfrac{\dd\vec p_3}{(2\pi)^3}\dfrac{1}{2\,p^{o}_3}\,
	{\mathscr G}^{(\rho/\lambda)}\,
    \psi^{*}(\norm{\vec k^\prime_\lambda}^2,\,\norm{\vec k^\prime_\rho}^2,\,{\vec k^\prime_\lambda}\cdot{\vec k^\prime_\rho})\,
    \varphi(\norm{\vec k_\lambda}^2,\,\norm{\vec k_\rho}^2,\,{\vec k_\lambda}\cdot{\vec k_\rho})
    \\[2mm]
    \times\,
	\sqrt{\dfrac{ p_2\cdot v}{m +  p_2\cdot v}}\,
    \sqrt{\dfrac{ p_2\cdot v^\prime}{m +  p_2\cdot v^\prime}}\,
    \sqrt{\dfrac{ p_3\cdot v}{m +  p_3\cdot v}}\,
    \sqrt{\dfrac{ p_3\cdot v^\prime}{m +  p_3\cdot v^\prime}}\,
	\dfrac{1}{\norm{\vec k_{\rho/\lambda}^{\prime}}}
	\\[2mm]
	\times
	\Bigl[
	{
	m^2(1 + w) + m( p_2 +  p_3)( v +  v^\prime)
	+ ( p_2\cdot v)( p_3\cdot v^\prime) + ( p_3\cdot v)( p_2\cdot v^\prime)
	+ ( p_2\cdot p_3)(1 - w)
	}
	\Bigr]\,p_{\rho/\lambda}^{\prime\mu}
\end{multlined}
}
\]
Since the initial amplitude is written in a covariant form, this means that this term ${\mathscr A}^\mu$ behaves as a four-vector under the Lorentz transformations. One knows therefore its structure
\[
{\mathscr A}^\mu\ =\ A(w)\,v^\mu\ +\ B(w)\,v^{\prime\,\mu}
\]
which is the unique way of writing a four-vector using the objects at our disposal (the two velocities and the metric tensor).\par
To obtain the coefficients $A(w)$ and $B(w)$, one can make the following contractions
\[
\left.
\begin{aligned}
{\mathscr A}^\mu\,v_\mu & = A(w)\ +\ w\,B(w)\\
{\mathscr A}^\mu\,v^{\prime}_\mu & = w\,A(w)\ +\ B(w)
\end{aligned}
\right\}
\qquad\Longrightarrow\qquad
\left\{
\begin{aligned}
A(w) &= \dfrac{1}{1 - w^2}\,{\mathscr A}^\mu(v_\mu - w\,v^{\prime}_\mu)\\[2mm]
B(w) &= \dfrac{1}{1 - w^2}\,{\mathscr A}^\mu(v_\mu^{\prime} - w\,v_\mu)
\end{aligned}
\right.
\]
\item{\underline{Consequences} :}\par
With the previous structure, the transition amplitude (the term in $B(w)$ vanishes) becomes:
\[
\brakket{\Psi^\prime,\,m^\prime}{J}{\Psi,\,m} = \dfrac{1}{4\pi}\,\dfrac{1}{\sqrt{v^{\prime o}\,v^o}}\,A(w)\,
\biggl\{
\bar\chi^{(m')}_{ v^{\prime}}\,
\gamma_5\,\Bigl[\fmslash{ v} - w\,\fmslash{ v}^{\prime}\Bigr]\,
(1+\fmslash{ v}^{\,\prime})
\,J\,
\chi^{(m)}_{ v}
\biggr\}
\]
The last term within brackets can be simplified :
\begin{multline*}
    \left\{
    \bar\chi^{(m')}_{ v^{\prime}}\,
    \gamma_5\,(\fmslash{ v} - w\,\fmslash{ v}^{\prime})
    \,(1+\fmslash{ v}^{\,\prime})
    \,J\,
    \chi^{(m)}_{ v}
    \right\}
    =
    \left\{
    \bar\chi^{(m')}_{ v^{\prime}}\,
    \gamma_5\,(\fmslash{ v} - w\,\fmslash{ v}^{\prime}
    +\fmslash{ v}\fmslash{ v}^{\,\prime} - w\,\fmslash{ v}^{\,\prime}\fmslash{ v}^{\,\prime})
    \,J\,
    \chi^{(m)}_{ v}
    \right\}\\
    =
    \left\{
    \bar\chi^{(m')}_{ v^{\prime}}\,
    \gamma_5\,(\fmslash{ v} - w\,\fmslash{ v}^{\prime}
    +2\,{ v}\!\cdot\!{ v}^{\,\prime} - \fmslash{ v}^{\,\prime}\fmslash{ v} - w\,{ v}^{\,\prime}\!\cdot\!{ v}^{\,\prime})
    \,J\,
    \chi^{(m)}_{ v}
    \right\} 
    =
    \left\{
    \bar\chi^{(m')}_{ v^{\prime}}\,
    \gamma_5\,(\fmslash{ v} - w\,\fmslash{ v}^{\prime}
    +2\,w - \fmslash{ v}^{\,\prime}\fmslash{ v} - w)
    \,J\,
    \chi^{(m)}_{ v}
    \right\}\\
     = \left\{
    \bar\chi^{(m')}_{ v^{\prime}}\,
    \gamma_5\,(\fmslash{ v} + w)
    \,J\,
    \chi^{(m)}_{ v}
    \right\} - \left\{
    \bar\chi^{(m')}_{ v^{\prime}}\,
    \gamma_5\,(\fmslash{ v}^{\,\prime}\fmslash{ v} + w\,\fmslash{ v}^{\prime})
    \,J\,
    \chi^{(m)}_{ v}
    \right\} \\
     = \left\{
    \bar\chi^{(m')}_{ v^{\prime}}\,
    \gamma_5\,(\fmslash{ v} + w)
    \,J\,
    \chi^{(m)}_{ v}
    \right\} + \left\{
    \bar\chi^{(m')}_{ v^{\prime}}\,
    \fmslash{ v}^{\,\prime}\,\gamma_5\,(\fmslash{ v} + w)
    \,J\,
    \chi^{(m)}_{ v}
    \right\}\\
    =\begin{multlined}[t][110mm]\left\{
    \bar\chi^{(m')}_{ v^{\prime}}\,
    \gamma_5\,(\fmslash{ v} + w)
    \,J\,
    \chi^{(m)}_{ v}
    \right\} + \left\{
    \bar\chi^{(m')}_{ v^{\prime}}\,
    \gamma_5\,(\fmslash{ v} + w)
    \,J\,
    \chi^{(m)}_{ v}
    \right\}
	\quad\text{since }\ \bar\chi^{(m')}_{ v^{\prime}}\fmslash{ v}^{\prime} = \bar\chi^{(m')}_{ v^{\prime}}
	\end{multlined}\\
     = 2\,\left\{
    \bar\chi^{(m')}_{ v^{\prime}}\,
    \gamma_5\,(\fmslash{ v} + w)
    \,J\,
    \chi^{(m)}_{ v}
    \right\}
\end{multline*}
Therefore, one arrives at the following expression: 
\[
{
\brakket{\Psi^\prime,\,m^\prime}{J}{\Psi,\,m}_{(\rho/\lambda)}\ =\ \dfrac{2}{1 - w^2}\,\dfrac{1}{4\pi}\,\dfrac{1}{\sqrt{v^{\prime o}\,v^o}}\,{\mathscr A}^\mu_{(\rho/\lambda)}(v_\mu - w\,v^{\prime}_\mu)
\left\{
\bar\chi^{(m')}_{ v^{\prime}}\,
\gamma_5\,(\fmslash{ v} + w)
\,J\,
\chi^{(m)}_{ v}
\right\}
}
\]
\item{\underline{Final expression of the transition amplitude} :}\par
Since
\[
\brakket{\Psi^\prime,\,m^\prime}{J}{\Psi,\,m} = \brakket{\Psi^\prime,\,m^\prime}{J}{\Psi,\,m}_{(\rho)} + \brakket{\Psi^\prime,\,m^\prime}{J}{\Psi,\,m}_{(\lambda)}
\]
one can then write 
\[
\brakket{\Psi^\prime,\,m^\prime}{J}{\Psi,\,m}^{\ell_\rho;\pm 1}\ =\ \dfrac{1}{2\pi}\,\dfrac{1}{1 - w^2}\,\dfrac{1}{\sqrt{v^{\prime o}\,v^o}}\,{\mathscr A}^\mu_{\ell_\rho;\pm 1}(v_\mu - w\,v^{\prime}_\mu)
\left\{
\bar\chi^{(m')}_{ v^{\prime}}\,
\gamma_5\,(\fmslash{ v} + w)
\,J\,
\chi^{(m)}_{ v}
\right\}
\qquad{({\text{even}\ \ell_\rho})}
\]
knowing that
\[
{\mathscr A}^\mu = {\mathscr A}^\mu_{(\rho)} + {\mathscr A}^\mu_{(\lambda)}
\]
which reads, explicitely 
\[
\begin{multlined}[\textwidth]
    {\mathscr A}^\mu_{\ell_\rho;\pm 1} = \int\dfrac{\dd\vec p_2}{(2\pi)^3}\dfrac{1}{2\,p^{o}_2}\,\dfrac{\dd\vec p_3}{(2\pi)^3}\dfrac{1}{2\,p^{o}_3}\,
    \psi^{*}_{\ell_\rho;\pm 1}(\norm{\vec k^\prime_\lambda}^2,\,\norm{\vec k^\prime_\rho}^2,\,{\vec k^\prime_\lambda}\cdot{\vec k^\prime_\rho})\,
    \varphi(\norm{\vec k_\lambda}^2,\,\norm{\vec k_\rho}^2,\,{\vec k_\lambda}\cdot{\vec k_\rho})
    \\[2mm]
    \times\,
	\sqrt{\dfrac{ p_2\cdot v}{m +  p_2\cdot v}}\,
    \sqrt{\dfrac{ p_2\cdot v^\prime}{m +  p_2\cdot v^\prime}}\,
    \sqrt{\dfrac{ p_3\cdot v}{m +  p_3\cdot v}}\,
    \sqrt{\dfrac{ p_3\cdot v^\prime}{m +  p_3\cdot v^\prime}}
	\\[2mm]
	\times
	\Bigl[
	{
	m^2(1 + w) + m( p_2 +  p_3)( v +  v^\prime)
	+ ( p_2\cdot v)( p_3\cdot v^\prime) + ( p_3\cdot v)( p_2\cdot v^\prime)
	+ ( p_2\cdot p_3)(1 - w)
	}
	\Bigr]\\[2mm]
	\times c_{\ell_\rho;\pm 1}\,\dfrac{1}{\norm{\hat k^\prime_\rho\wedge\hat k^\prime_\lambda}}\,\biggl[
	P_{\ell_\rho}^1\bigl(\hat k^\prime_\rho\cdot\hat k^\prime_\lambda\bigr)\,\dfrac{p_{\rho}^{\mu}}{\norm{\vec k_{\rho}^{\prime}}} - 
	P_{\ell_\rho\pm 1}^1\bigl(\hat k^\prime_\rho\cdot\hat k^\prime_\lambda\bigr)\,\dfrac{p_{\lambda}^{\mu}}{\norm{\vec k_{\lambda}^{\prime}}}
	\biggr]
\end{multlined}
\]
In this last expression, the covariance is also manifest thanks to the relations \eqref{eq:covaUN}, \eqref{eq:covaDEUX} and \eqref{eq:covaTROIS}, and if we recall that $\dfrac{1}{\sqrt{1-x^2}}\,P_{n}^1\bigl(x\bigr)$ is a polynomial in $x$.
\end{maliste}
\subsection{Term of the type $\boldsymbol{(ii)}$}
We are going to show that this type of terms do not contribute. The starting point is: 
\[
\left\{
\begin{aligned}
\Psi^{(m)}_{s_1,s_2,s_3}(\vec k_\lambda,\,\vec k_\rho)
&= \chi^{(m)}_{s_1}\,(\sigma_2)_{s_2,s_3}\,\varphi(\norm{\vec k_\lambda}^2,\,\norm{\vec k_\rho}^2,\,{\vec k_\lambda}\cdot{\vec k_\rho})\\[2mm]
\Psi^{\prime(m')}_{s_1^\prime,s_2^\prime,s_3^\prime}(\vec k^\prime_\lambda,\,\vec k^\prime_\rho) &=
\,\hat k^\prime_{\rho/\lambda}\cdot\bigl[
\vec\sigma\,\sigma_2\bigr]_{s_2^\prime,s_3^\prime}\,	
\chi^{(m')}_{s^\prime_1}\,\psi(\norm{\vec k^\prime_\lambda}^2,\,\norm{\vec k^\prime_\rho}^2,\,{\vec k^\prime_\lambda}\cdot{\vec k^\prime_\rho})
\end{aligned}
\right.
\]
Therefore
\begin{align*}
\Psi^{\prime*\,(m')}_{s_1^\prime,s_2^\prime,s_3^\prime}(\vec k^\prime_\lambda,\,\vec k^\prime_\rho) &=
\,\hat k^\prime_{\rho/\lambda}\cdot\bigl[
\vec\sigma\,\sigma_2\bigr]^{*}_{s_2^\prime,s_3^\prime}\,	
\chi^{*(m')}_{s^\prime_1}\,\psi^{*}(\norm{\vec k^\prime_\lambda}^2,\,\norm{\vec k^\prime_\rho}^2,\,{\vec k^\prime_\lambda}\cdot{\vec k^\prime_\rho})\\[2mm]
&=\,\hat k^\prime_{\rho/\lambda}\cdot\biggl({}^{t}\Bigl[\bigl(
\vec\sigma\,\sigma_2\bigr)^{\dagger}\Bigr]\biggr)_{s_2^\prime,s_3^\prime}\,	
{}^{t}\Bigl(\chi^{\dagger(m')}\Bigr)_{s^\prime_1}\,\psi^{*}(\norm{\vec k^\prime_\lambda}^2,\,\norm{\vec k^\prime_\rho}^2,\,{\vec k^\prime_\lambda}\cdot{\vec k^\prime_\rho})\\[2mm]
&=
\,\hat k^\prime_{\rho/\lambda}\cdot\bigl[
\sigma_2\,\vec\sigma\bigr]_{s_3^\prime,s_2^\prime}\,	
\chi^{\dagger(m')}_{s^\prime_1}\,\psi^{*}(\norm{\vec k^\prime_\lambda}^2,\,\norm{\vec k^\prime_\rho}^2,\,{\vec k^\prime_\lambda}\cdot{\vec k^\prime_\rho})
\qquad\text{ car }\sigma_i^\dagger = \sigma_i
\end{align*}
The transition amplitude to be computed is then given by
\begin{multline*}
    \brakket{\Psi^\prime,\,m^\prime}{J}{\Psi,\,m}_{(\rho/\lambda)} = \int\dfrac{\dd\vec p_2}{(2\pi)^3}\dfrac{\dd\vec p_3}{(2\pi)^3}\,
    \sqrt{u^{\prime o}\,u^o}\,
    \sqrt{\dfrac{ {k^{\prime o}_1} {k^{o}_1} }{ {p^{\prime o}_1} {p^{o}_1} }}\,
    \sqrt{\dfrac{ {k^{\prime o}_2} {k^{o}_2} }{ {p^{\prime o}_2} {p^{o}_2} }}\,
    \sqrt{\dfrac{ {k^{\prime o}_3} {k^{o}_3} }{ {p^{\prime o}_3} {p^{o}_3} }}\,\\[2mm]
	\times\,
    \psi^{*}(\norm{\vec k^\prime_\lambda}^2,\,\norm{\vec k^\prime_\rho}^2,\,{\vec k^\prime_\lambda}\cdot{\vec k^\prime_\rho})\,
    \varphi(\norm{\vec k_\lambda}^2,\,\norm{\vec k_\rho}^2,\,{\vec k_\lambda}\cdot{\vec k_\rho})
    \\[2mm]
    \times\sum\limits_{s_1^\prime,s_2^\prime,s_3^\prime}\sum\limits_{s_1,s_2,s_3}
    \chi^{(m)}_{s_1}\,(\sigma_2)_{s_2,s_3}\,
	\hat k^\prime_{\rho/\lambda}\cdot\bigl[
	\sigma_2\,\vec\sigma\bigr]_{s_3^\prime,s_2^\prime}\,	
	\chi^{\dagger(m')}_{s^\prime_1}\\[2mm]
    \times\,\left[D({\boldsymbol R}_1^{\prime -1})\,J(\vec p^{\,\prime}_1,\,\vec p_1)\,D({\boldsymbol R}_1)\right]_{s_1^\prime,s_1}\,
    \left[D({\boldsymbol R}_2^{\prime -1}{\boldsymbol R}_2)\right]_{s_2^\prime,s_2}\,
    \left[D({\boldsymbol R}_3^{\prime -1}{\boldsymbol R}_3)\right]_{s_3^\prime,s_3}
\end{multline*}
Using the techniques employed in the previous calculation, this amplitude can be written in the following form:
\begin{multline*}
    \brakket{\Psi^\prime,\,m^\prime}{J}{\Psi,\,m}_{(\rho/\lambda)} = \int\dfrac{\dd\vec p_2}{(2\pi)^3}\dfrac{\dd\vec p_3}{(2\pi)^3}\,
    \sqrt{u^{\prime o}\,u^o}\,
    \sqrt{\dfrac{ {k^{\prime o}_1} {k^{o}_1} }{ {p^{\prime o}_1} {p^{o}_1} }}\,
    \sqrt{\dfrac{ {k^{\prime o}_2} {k^{o}_2} }{ {p^{\prime o}_2} {p^{o}_2} }}\,
    \sqrt{\dfrac{ {k^{\prime o}_3} {k^{o}_3} }{ {p^{\prime o}_3} {p^{o}_3} }}\,\\[2mm]
    \times\,
    \psi^{*}(\norm{\vec k^\prime_\lambda}^2,\,\norm{\vec k^\prime_\rho}^2,\,{\vec k^\prime_\lambda}\cdot{\vec k^\prime_\rho})\,
    \varphi(\norm{\vec k_\lambda}^2,\,\norm{\vec k_\rho}^2,\,{\vec k_\lambda}\cdot{\vec k_\rho})
    \\[2mm]
    \times\,\left[
        \chi^{\dagger(m')}
        D({\boldsymbol R}_1^{\prime -1})\,J(\vec p^{\,\prime}_1,\,\vec p_1)\,D({\boldsymbol R}_1)
        \chi^{(m)}
        \right]\,
        \text{Tr}\left[
            \bigl[\hat k^\prime_{\rho/\lambda}\cdot\vec\sigma\bigr]\,
D({\boldsymbol R}_2^{\prime -1}{\boldsymbol R}_2)\,D({\boldsymbol R}_3^{-1}{\boldsymbol R}^\prime_3)\right]
\end{multline*}
\subsubsection*{Formulation in 4D}
\begin{maliste}
\item{\underline{Term in brackets} :}\par
The term to be computed is the following:
\[
    {\mathscr T}_1 = \left[
        \chi^{\dagger(m')}\,
        D({\boldsymbol R}_1^{\prime -1})\,J(\vec p^{\,\prime}_1,\,\vec p_1)\,D({\boldsymbol R}_1)
        \chi^{(m)}
        \right]
\]
It has been dealt with in section~\ref{para:bracketUN} and the result is:
\[
{
{\mathscr T}_1  = \dfrac{1}{2}\,\dfrac{1}{\sqrt{p^{\prime o}_1\,p^o_1}}\,
\dfrac{1}{\sqrt{m^\prime_1 + k^{\prime 0}_1}}\,
\dfrac{1}{\sqrt{m_1 + k^{0}_1}}\,
\left[
\bar\chi^{(m')}_{ u^{\prime}}\,
(m^\prime_1 + \fmslash{ p}^{\,\prime}_1)
\,J\,
(m_1 + \fmslash{ p}_1)\,
\chi^{(m)}_{ u}
\right]
}
\]
\item{\underline{Calculation of the trace} :}\par
\[
{\mathscr T}_2\ =\ \text{Tr}\left[
    \bigl[\hat k^\prime_{\rho/\lambda}\cdot\vec\sigma\bigr]\,
D({\boldsymbol R}_2^{\prime -1}{\boldsymbol R}_2)\,D({\boldsymbol R}_3^{-1}{\boldsymbol R}^\prime_3)\right]
\]
One uses the methods already presented for the calculation of traces terms:
\begin{align*}
{\mathscr T}_2\ &=\
    \text{Tr}
\left[
    \dfrac{1+\gamma^0}{2}\,
    \bigl[\hat k^\prime_{\rho/\lambda}\cdot\vec\sigma\bigr]\,
{\boldsymbol B}_{ k^\prime_2}^{-1}\,{\boldsymbol B}_{ u^\prime}^{-1}\,{\boldsymbol B}_{ p^\prime_2}\,
{\boldsymbol B}_{ p_2}^{-1}\,{\boldsymbol B}_{ u}\,{\boldsymbol B}_{ k_2}\,
{\boldsymbol B}_{ k_3}^{-1}\,{\boldsymbol B}_{ u}^{-1}\,{\boldsymbol B}_{ p_3}\,
{\boldsymbol B}_{ p^\prime_3}^{-1}\,{\boldsymbol B}_{ u^\prime}\,{\boldsymbol B}_{ k_3^\prime}
\right]\\[2mm]
&=\ \begin{multlined}[t][160mm]\dfrac{1}{16}\,
    \text{Tr}
\left[
    (1+\gamma_0)^4\,
    \bigl[\hat k^\prime_{\rho/\lambda}\cdot\vec\sigma\bigr]
{\boldsymbol B}_{ k^\prime_2}^{-1}\,{\boldsymbol B}_{ u^\prime}^{-1}\,{\boldsymbol B}_{ p^\prime_2}\,
{\boldsymbol B}_{ p_2}^{-1}\,{\boldsymbol B}_{ u}\,{\boldsymbol B}_{ k_2}\,
{\boldsymbol B}_{ k_3}^{-1}\,{\boldsymbol B}_{ u}^{-1}\,{\boldsymbol B}_{ p_3}\,
{\boldsymbol B}_{ p^\prime_3}^{-1}\,{\boldsymbol B}_{ u^\prime}\,{\boldsymbol B}_{ k_3^\prime}
\right]\\ \text{because}\quad(1 + \gamma^0)^4 = 8(1 + \gamma^0)\end{multlined}\\[2mm]
& = \begin{multlined}[t][160mm]\dfrac{1}{16}\,\text{Tr}\left[
    \bigl[\hat k^\prime_{\rho/\lambda}\cdot\vec\sigma\bigr]\,
    (1 + \gamma^0)\,
    {\boldsymbol B}_{ k^\prime_2}^{-1}\,{\boldsymbol B}_{ u^\prime}^{-1}\,{\boldsymbol B}_{ p^\prime_2}\,
    {\boldsymbol B}_{ p_2}^{-1}\,{\boldsymbol B}_{ u}\,{\boldsymbol B}_{ k_2}\,(1 + \gamma^0)^2\,
    {\boldsymbol B}_{ k_3}^{-1}\,{\boldsymbol B}_{ u}^{-1}\,{\boldsymbol B}_{ p_3}\,
    {\boldsymbol B}_{ p^\prime_3}^{-1}\,{\boldsymbol B}_{ u^\prime}\,{\boldsymbol B}_{ k_3^\prime}\,(1 + \gamma^0)
    \right]\\ \quad\text{car}\quad(1 + \gamma^0)\,D({\boldsymbol R}_i) = D({\boldsymbol R}_i)\,(1 + \gamma^0)\quad\text{et}\quad (1 + \gamma^0)\,\sigma^{i} = \sigma^{i}\,(1 + \gamma^0)\end{multlined}\\[2mm]
& = \begin{multlined}[t][160mm]\dfrac{1}{16}\,\text{Tr}\left[
    \bigl[\hat k^\prime_{\rho/\lambda}\cdot\vec\sigma\bigr]\,
(1 + \gamma^0)\,
{\boldsymbol B}_{ k^\prime_2}^{-1}\,{\boldsymbol B}_{ u^\prime}^{-1}\,
{\boldsymbol B}_{ u}\,{\boldsymbol B}_{ k_2}\,(1 + \gamma^0)^2\,
{\boldsymbol B}_{ k_3}^{-1}\,{\boldsymbol B}_{ u}^{-1}\,
{\boldsymbol B}_{ u^\prime}\,{\boldsymbol B}_{ k_3^\prime}\,(1 + \gamma^0)
\right]\\ \text{because $\ {\vec p}^{\,\prime}_2\ =\ {\vec p}_2\ $ et $\ {\vec p}^{\,\prime}_3\ =\ {\vec p}_3$}\end{multlined}\\[2mm]
& = \begin{multlined}[t][160mm]\dfrac{1}{16}\,\text{Tr}\left[{\boldsymbol B}_{ u^\prime}\,
    \bigl[\hat k^\prime_{\rho/\lambda}\cdot\vec\sigma\bigr]
    \,{\boldsymbol B}_{ u^\prime}^{-1}\,
{\boldsymbol B}_{ u^\prime}\,(1 + \gamma^0)\,
{\boldsymbol B}_{ k^\prime_2}^{-1}\,{\boldsymbol B}_{ u^\prime}^{-1}\,
{\boldsymbol B}_{ u}\,{\boldsymbol B}_{ k_2}\,(1 + \gamma^0)\,{\boldsymbol B}_{ u}^{-1}\right. \\ \times\,\left.{\boldsymbol B}_{ u}\,(1 + \gamma^0)\,
{\boldsymbol B}_{ k_3}^{-1}\,{\boldsymbol B}_{ u}^{-1}\,
{\boldsymbol B}_{ u^\prime}\,{\boldsymbol B}_{ k_3^\prime}\,(1 + \gamma^0)\,{\boldsymbol B}_{ u^\prime}^{-1}
\right] \quad\text{inserting in the right places $\ {\boldsymbol B}_{ u^{(\prime)}}^{-1}{\boldsymbol B}_{ u^{(\prime)}} = {\boldsymbol 1}$}\end{multlined}\\[2mm]
& = \begin{multlined}[t][160mm]\dfrac{1}{64}\dfrac{1}{m^2}\,
\dfrac{1}{\sqrt{m + k^{\prime o}_2}}\,
\dfrac{1}{\sqrt{m + k^{o}_2}}\,
\dfrac{1}{\sqrt{m + k^{\prime o}_3}}\,
\dfrac{1}{\sqrt{m + k^{o}_3}}\\ \times\,
\text{Tr}\left[{\boldsymbol B}_{ u^\prime}\,
\bigl[\hat k^\prime_{\rho/\lambda}\cdot\vec\sigma\bigr]
\,{\boldsymbol B}_{ u^\prime}^{-1}\,
(1+\fmslash{ u}^\prime)(m + \fmslash{ p}_2)
(m + \fmslash{ p}_2)(1+\fmslash{ u})
(1+\fmslash{ u})(m + \fmslash{ p}_3)
(m + \fmslash{ p}_3)(1+\fmslash{ u}^\prime)
\right]\\ \quad\text{using relations \eqref{ann:eq3}}\end{multlined}\\[2mm]
& = \begin{multlined}[t][160mm]\dfrac{1}{8}\,
\dfrac{1}{\sqrt{m + k^{\prime o}_2}}\,
\dfrac{1}{\sqrt{m + k^{o}_2}}\,
\dfrac{1}{\sqrt{m + k^{\prime o}_3}}\,
\dfrac{1}{\sqrt{m + k^{o}_3}}\\ \times\,
\text{Tr}\left[{\boldsymbol B}_{ u^\prime}\,
\bigl[\hat k^\prime_{\rho/\lambda}\cdot\vec\sigma\bigr]
\,{\boldsymbol B}_{ u^\prime}^{-1}\,
(1+\fmslash{ u}^\prime)(m + \fmslash{ p}_2)
(1+\fmslash{ u})(m + \fmslash{ p}_3)(1+\fmslash{ u}^\prime)
\right]\\
\text{since $\ (1+\fmslash{ u})^2 = 2(1+\fmslash{ u})\ $ et $\ (m+\fmslash{ p})^2 = 2\,m(m+\fmslash{ p})$}\end{multlined}\\[2mm]
& = \begin{multlined}[t][160mm]\dfrac{1}{8}\,
\dfrac{1}{\sqrt{m + k^{\prime o}_2}}\,
\dfrac{1}{\sqrt{m + k^{o}_2}}\,
\dfrac{1}{\sqrt{m + k^{\prime o}_3}}\,
\dfrac{1}{\sqrt{m + k^{o}_3}}\\ \times\,
\text{Tr}\left[\gamma_5\,\dfrac{\Bigl[\fmslash{ p}_{\rho/\lambda}^{\prime} - ( p_{\rho/\lambda}^{\prime}\cdot u^{\prime})\,\fmslash{ u}^{\prime}\Bigr]}{\norm{\vec k_{\rho/\lambda}^{\prime}}}\,
(1+\fmslash{ u}^\prime)(m + \fmslash{ p}_2)
(1+\fmslash{ u})(m + \fmslash{ p}_3)(1+\fmslash{ u}^\prime)
\right]\\
\text{using the properties of appendix \eqref{ann:boosts}}\end{multlined}
\end{align*}
And therefore :
\[
{
{\mathscr T}_2\ =\
\begin{multlined}[t][160mm]\dfrac{1}{8}\,
\dfrac{1}{\sqrt{m + k^{\prime o}_2}}\,
\dfrac{1}{\sqrt{m + k^{o}_2}}\,
\dfrac{1}{\sqrt{m + k^{\prime o}_3}}\,
\dfrac{1}{\sqrt{m + k^{o}_3}}\,
\dfrac{1}{\norm{\vec k_{\rho/\lambda}^{\prime}}}\\[2mm] \times\,
\text{Tr}\left[
\gamma_5\,\Bigl[\fmslash{ p}_{\rho/\lambda}^{\prime} - ({ p}_{\rho/\lambda}^{\prime}\cdot u^\prime)\,\fmslash{ u}^\prime\Bigr]\,
(1+\fmslash{ u}^\prime)(m + \fmslash{ p}_2)
(1+\fmslash{ u})(m + \fmslash{ p}_3)(1+\fmslash{ u}^\prime)
\right]
\end{multlined}
}
\]
\item{\underline{Conclusion} :}\par
Gathering everything one gets (one removes again the ``prime''~of ${ p}_{\rho/\lambda}^{\prime}$)
\begin{multline*}
    \brakket{\Psi^\prime,\,m^\prime}{J}{\Psi,\,m}_{(\rho/\lambda)} = \dfrac{1}{16}\,\int\dfrac{\dd\vec p_2}{(2\pi)^3}\dfrac{\dd\vec p_3}{(2\pi)^3}\,
    \sqrt{u^{\prime o}\,u^o}\,
    \sqrt{\dfrac{ {k^{\prime o}_1} {k^{o}_1} }{ {p^{\prime o}_1} {p^{o}_1} }}\,
    \sqrt{\dfrac{ {k^{\prime o}_2} {k^{o}_2} }{ {p^{\prime o}_2} {p^{o}_2} }}\,
    \sqrt{\dfrac{ {k^{\prime o}_3} {k^{o}_3} }{ {p^{\prime o}_3} {p^{o}_3} }}\,\\[2mm]
	\times\,
    \psi^{*}(\norm{\vec k^\prime_\lambda}^2,\,\norm{\vec k^\prime_\rho}^2,\,{\vec k^\prime_\lambda}\cdot{\vec k^\prime_\rho})\,
    \varphi(\norm{\vec k_\lambda}^2,\,\norm{\vec k_\rho}^2,\,{\vec k_\lambda}\cdot{\vec k_\rho})
    \\[2mm]
    \times\,
    \dfrac{1}{\sqrt{p^{\prime o}_1\,p^o_1}}\,
\dfrac{1}{\sqrt{m^\prime_1 + k^{\prime 0}_1}}\,
\dfrac{1}{\sqrt{m_1 + k^{0}_1}}\,\,
\dfrac{1}{\sqrt{m + k^{\prime o}_2}}\,
\dfrac{1}{\sqrt{m + k^{o}_2}}\,
\dfrac{1}{\sqrt{m + k^{\prime o}_3}}\,
\dfrac{1}{\sqrt{m + k^{o}_3}}\,
\dfrac{1}{\norm{\vec k_{\rho/\lambda}^{\prime}}}\\[2mm]
\times\,
\text{Tr}\left[
\gamma_5\,\Bigl[\fmslash{ p}_{\rho/\lambda} - ( p_{\rho/\lambda}\cdot u^\prime)\,\fmslash{ u}^\prime\Bigr]\,
(1+\fmslash{ u}^\prime)(m + \fmslash{ p}_2)
(1+\fmslash{ u})(m + \fmslash{ p}_3)(1+\fmslash{ u}^\prime)
\right]
\\[2mm]
\times\,
\left[
\bar\chi^{(m')}_{ u^{\prime}}\,
(m^\prime_1 + \fmslash{ p}^{\prime}_1)
\,J\,
(m_1 + \fmslash{ p}_1)\,
\chi^{(m)}_{ u}
\right]
\end{multline*}
\end{maliste}
\subsubsection*{Infinite mass limit}
After some algebra, we obtain: 
\[
    \begin{multlined}[160mm]
        \brakket{\Psi^\prime,\,m^\prime}{J}{\Psi,\,m}_{(\rho/\lambda)} = \dfrac{1}{8}\,
        \dfrac{1}{\sqrt{v^{\prime o}\,v^o}}\,
        \int\dfrac{\dd\vec p_2}{(2\pi)^3}\dfrac{\dd\vec p_3}{(2\pi)^3}\,
        \dfrac{1}{p^o_2}\,\dfrac{1}{p^o_3}\,
        \psi^{*}(\norm{\vec k^\prime_\lambda}^2,\,\norm{\vec k^\prime_\rho}^2,\,{\vec k^\prime_\lambda}\cdot{\vec k^\prime_\rho})\,
        \varphi(\norm{\vec k_\lambda}^2,\,\norm{\vec k_\rho}^2,\,{\vec k_\lambda}\cdot{\vec k_\rho})
        \\[2mm]
        \times\,\sqrt{\dfrac{ p_2\cdot v}{m +  p_2\cdot v}}\,
    \sqrt{\dfrac{ p_2\cdot v^\prime}{m +  p_2\cdot v^\prime}}\,
    \sqrt{\dfrac{ p_3\cdot v}{m +  p_3\cdot v}}\,
    \sqrt{\dfrac{ p_3\cdot v^\prime}{m +  p_3\cdot v^\prime}}\,
\dfrac{1}{\norm{\vec k_{\rho/\lambda}^{\prime}}}\,
    \left[
    \bar\chi^{(m')}_{ v^{\prime}}
    \,J\,
    \chi^{(m)}_{ v}
    \right]
    \\[2mm]\times\,
	\text{Tr}\left[
	\gamma_5\,\Bigl[\fmslash{ p}_{\rho/\lambda} - ( p_{\rho/\lambda}\cdot v^\prime)\,\fmslash{ v}^\prime\Bigr]\,
	(1+\fmslash{ v}^\prime)(m + \fmslash{ p}_2)
	(1+\fmslash{ v})(m + \fmslash{ p}_3)(1+\fmslash{ v}^\prime)
	\right]
   \end{multlined}
\]
One can expand the trace by using the expressions of $ p_{\rho/\lambda}$ and the properties of the Levi-Civita tensor which appears.
\begin{maliste}
\item{\underline{$p_{\rho}$ case} :}
\begin{align*}
	\text{Tr}\left[
	\gamma_5\,\Bigl[\fmslash{ p}_\rho - ( p_\rho\cdot v^\prime)\,\fmslash{ v}^\prime\Bigr]\,
	(1+\fmslash{ v}^\prime)(m + \fmslash{ p}_2)
	(1+\fmslash{ v})(m + \fmslash{ p}_3)(1+\fmslash{ v}^\prime)
	\right]
    &\ =\
    4\,i\,
	\Bigl[
	\bigl( p_2 -  p_3\bigr)\cdot v^\prime
	\Bigr]
	\,\epsilon^{\sigma\beta\delta\tau}\,{p_2}_\sigma\,{p_3}_\beta\,v_\delta\,v^\prime_\tau\\[2mm]
    &\ =\
    8\,i\,\bigl( p_\rho\cdot v^\prime \bigr)\,\epsilon^{\sigma\beta\delta\tau}\,{p_2}_\sigma\,{p_3}_\beta\,v_\delta\,v^\prime_\tau
\end{align*}
\item{\underline{$p_{\lambda}$ case} :}
\begin{align*}
	\text{Tr}\left[
	\gamma_5\,\Bigl[\fmslash{ p}_\lambda - ( p_\lambda\cdot v^\prime)\,\fmslash{ v}^\prime\Bigr]\,
	(1+\fmslash{ v}^\prime)(m + \fmslash{ p}_2)
	(1+\fmslash{ v})(m + \fmslash{ p}_3)(1+\fmslash{ v}^\prime)
	\right]
    &\ =\
    4\,i\,
	\Bigl[
	\bigl( p_2 +  p_3\bigr)\cdot v^\prime + 2\,m
	\Bigr]\,\epsilon^{\sigma\beta\delta\tau}\,{p_2}_\sigma\,{p_3}_\beta\,v_\delta\,v^\prime_\tau\\[2mm]
    &\ =\
    8\,i\,\bigl(m +  p_\lambda\cdot v^\prime\bigr)\,\epsilon^{\sigma\beta\delta\tau}\,{p_2}_\sigma\,{p_3}_\beta\,v_\delta\,v^\prime_\tau
\end{align*}
\item{\underline{Summary} :}\par
Comparing both expressions, one sees that the trace takes the form:
\[
	\text{Tr}\left[
	\gamma_5\,\Bigl[\fmslash{ p}_{\rho/\lambda} - ( p_{\rho/\lambda}\cdot v^\prime)\,\fmslash{ v}^\prime\Bigr]\,
	(1+\fmslash{ v}^\prime)(m + \fmslash{ p}_2)
	(1+\fmslash{ v})(m + \fmslash{ p}_3)(1+\fmslash{ v}^\prime)
	\right]\ =\ 8\,i\,F_{(\rho/\lambda)}\,\epsilon^{\sigma\beta\delta\tau}\,{p_2}_\sigma\,{p_3}_\beta\,v_\delta\,v^\prime_\tau\,
\]
where $F_{(\rho/\lambda)}$ is a scalar Lorentz invariant depending on $ p_{\rho/\lambda}\cdot v^\prime$ and maybe on $m$.\par
Therefore
    \begin{multline*}
        \brakket{\Psi^\prime,\,m^\prime}{J}{\Psi,\,m}_{(\rho/\lambda)} = 
        \dfrac{i}{\sqrt{v^{\prime o}\,v^o}}\,
        \int\dfrac{\dd\vec p_2}{(2\pi)^3}\dfrac{\dd\vec p_3}{(2\pi)^3}\,
        \dfrac{1}{p^o_2}\,\dfrac{1}{p^o_3}\,
		\psi^{*}(\norm{\vec k^\prime_\lambda}^2,\,\norm{\vec k^\prime_\rho}^2,\,{\vec k^\prime_\lambda}\cdot{\vec k^\prime_\rho})\,
		\varphi(\norm{\vec k_\lambda}^2,\,\norm{\vec k_\rho}^2,\,{\vec k_\lambda}\cdot{\vec k_\rho})
        \\[2mm]
        \times\,\sqrt{\dfrac{ p_2\cdot v}{m +  p_2\cdot v}}\,
    \sqrt{\dfrac{ p_2\cdot v^\prime}{m +  p_2\cdot v^\prime}}\,
    \sqrt{\dfrac{ p_3\cdot v}{m +  p_3\cdot v}}\,
    \sqrt{\dfrac{ p_3\cdot v^\prime}{m +  p_3\cdot v^\prime}}\,
\dfrac{1}{\norm{\vec k_{\rho/\lambda}^{\prime}}}
    \\[2mm]\times\,F_{(\rho/\lambda)}\,
	\epsilon^{\sigma\beta\delta\tau}\,{p_2}_\sigma\,{p_3}_\beta\,v_\delta\,v^\prime_\tau\,
    \,\left[
        \bar\chi^{(m')}_{ v^{\prime}}
        \,J\,
        \chi^{(m)}_{ v}
        \right]
   \end{multline*}
\end{maliste}
\subsubsection*{Using the covariance}
In the previous relation appears the following integral:
\begin{equation*}
    {\mathscr I}_{\alpha\beta}^{(\rho/\lambda)} \ =\
    \begin{multlined}[t][150mm]
        \int\dfrac{\dd\vec p_2}{(2\pi)^3}\dfrac{\dd\vec p_3}{(2\pi)^3}\,
        \dfrac{1}{p^o_2}\,\dfrac{1}{p^o_3}\,
		\psi^{*}(\norm{\vec k^\prime_\lambda}^2,\,\norm{\vec k^\prime_\rho}^2,\,{\vec k^\prime_\lambda}\cdot{\vec k^\prime_\rho})\,
		\varphi(\norm{\vec k_\lambda}^2,\,\norm{\vec k_\rho}^2,\,{\vec k_\lambda}\cdot{\vec k_\rho})\\[2mm]
        \times\,\sqrt{\dfrac{ p_2\cdot v}{m +  p_2\cdot v}}\,
        \sqrt{\dfrac{ p_2\cdot v^\prime}{m +  p_2\cdot v^\prime}}\,
        \sqrt{\dfrac{ p_3\cdot v}{m +  p_3\cdot v}}\,
        \sqrt{\dfrac{ p_3\cdot v^\prime}{m +  p_3\cdot v^\prime}}\,
\dfrac{1}{\norm{\vec k_{\rho/\lambda}^{\prime}}}\,
       F_{(\rho/\lambda)}\,{p_2}_\alpha\,{p_3}_\beta
    \end{multlined}\\[2mm]
\end{equation*}
which is manifestly covariant because of~\eqref{eq:covaUN}, \eqref{eq:covaDEUX} and \eqref{eq:covaTROIS}.\par
Thus, the original transition amplitude simply is:
\begin{equation*}
    \brakket{\Psi^\prime,\,m^\prime}{J}{\Psi,\,m}_{(\rho/\lambda)} = 
        \dfrac{i}{\sqrt{v^{\prime o}\,v^o}}\,
		\epsilon^{\alpha\beta\delta\tau}\,{\mathscr I}_{\alpha\beta}^{(\rho/\lambda)}\,
		v_\delta\,v^\prime_\tau
\,\left[
        \bar\chi^{(m')}_{ v^{\prime}}
        \,J\,
        \chi^{(m)}_{ v}
        \right]
\end{equation*}
However, because of its covariance, the previous integral ${\mathscr I}_{\alpha\beta}^{(\rho/\lambda)}$ can be put in the form:
\begin{equation*}
    {\mathscr I}_{\alpha\beta}^{(\rho/\lambda)} =
    A_{(\rho/\lambda)}(w)\,g_{\alpha\beta} +  B_{(\rho/\lambda)}(w)\,v_\alpha\,v_\beta +  C_{(\rho/\lambda)}(w)\,v^{\prime}_{\alpha}\,v^{\prime}_{\beta} +  D_{(\rho/\lambda)}(w)\,v_{\alpha}\,v^{\prime}_{\beta} +  E_{(\rho/\lambda)}(w)\,v^{\prime}_{\alpha}\,v_{\beta}
\end{equation*}
A direct calculation then gives:
\begin{equation*}
	\epsilon^{\alpha\beta\delta\tau}\,{\mathscr I}_{\alpha\beta}^{(\rho/\lambda)}\,
	v_\delta\,v^\prime_\tau
     \ =\ 0
\end{equation*}
Therefore, we obtain:
\[
    \brakket{\Psi^\prime,\,m^\prime}{J}{\Psi,\,m} = 0
\]
Hence, there are no contributions of terms of the type ${(ii)}$, that is no contributions of the terms ${j=0}$ with ${\ell_\rho}$ odd to all orders.
\subsection{Term of the type $\boldsymbol{(iii)}$}
The starting point is now 
\[
\left\{
\begin{aligned}
\Psi^{(m)}_{s_1,s_2,s_3}(\vec k_\lambda,\,\vec k_\rho)
&= \dfrac{1}{4\pi}\dfrac{i}{\sqrt{2}}\,\chi^{(m)}_{s_1}\,(\sigma_2)_{s_2,s_3}\,\varphi(\norm{\vec k_\lambda}^2,\,\norm{\vec k_\rho}^2,\,{\vec k_\lambda}\cdot{\vec k_\rho})\\[2mm]
\Psi^{\prime(m')}_{s_1^\prime,s_2^\prime,s_3^\prime}(\vec k^\prime_\lambda,\,\vec k^\prime_\rho) &= \dfrac{1}{2}\,{\mathscr G}^{(\rho/\lambda)}
\,\biggl[
\bigl(\vec\sigma\,\sigma_2\bigr)_{s^\prime_2,s^\prime_3}\,\wedge\,\hat k^\prime_{\rho/\lambda}
\biggr]
\cdot\Bigl[\vec\sigma\,\chi^{(m^\prime)}\Bigr]_{s_1^\prime}
\,\psi(\norm{\vec k^\prime_\lambda}^2,\,\norm{\vec k^\prime_\rho}^2,\,{\vec k^\prime_\lambda}\cdot{\vec k^\prime_\rho})
\end{aligned}
\right.
\]
Therefore
\begin{align*}
\Psi^{\prime*(m')}_{s_1^\prime,s_2^\prime,s_3^\prime}(\vec k^\prime_\lambda,\,\vec k^\prime_\rho) &=
\dfrac{1}{2}\,{\mathscr G}^{(\rho/\lambda)}
\,\biggl[
\bigl(\vec\sigma\,\sigma_2\bigr)_{s^\prime_2,s^\prime_3}\,\wedge\,\hat k^\prime_{\rho/\lambda}
\biggr]^{*}
\cdot\Bigl[\vec\sigma\,\chi^{(m^\prime)}\Bigr]^{*}_{s_1^\prime}
\,\psi^{*}(\norm{\vec k^\prime_\lambda}^2,\,\norm{\vec k^\prime_\rho}^2,\,{\vec k^\prime_\lambda}\cdot{\vec k^\prime_\rho})\\[2mm]
&=
\dfrac{1}{2}\,{\mathscr G}^{(\rho/\lambda)}
\,\biggl[
{}^{t}\bigl(\vec\sigma\,\sigma_2\bigr)^\dagger_{s^\prime_2,s^\prime_3}\,\wedge\,\hat k^\prime_{\rho/\lambda}
\biggr]
\cdot{}^{t}\Bigl[\vec\sigma\,\chi^{(m^\prime)}\Bigr]^\dagger_{s_1^\prime}
\,\psi^{*}(\norm{\vec k^\prime_\lambda}^2,\,\norm{\vec k^\prime_\rho}^2,\,{\vec k^\prime_\lambda}\cdot{\vec k^\prime_\rho})\\[2mm]
&=
\dfrac{1}{2}\,{\mathscr G}^{(\rho/\lambda)}
\,\biggl[
\bigl(\sigma_2\,\vec\sigma\bigr)_{s^\prime_3,s^\prime_2}\,\wedge\,\hat k^\prime_{\rho/\lambda}
\biggr]
\cdot\Bigl[\chi^{\dagger(m^\prime)}\,\vec\sigma\Bigr]_{s_1^\prime}
\,\psi^{*}(\norm{\vec k^\prime_\lambda}^2,\,\norm{\vec k^\prime_\rho}^2,\,{\vec k^\prime_\lambda}\cdot{\vec k^\prime_\rho})\\[2mm]
&=
\dfrac{1}{2}\,{\mathscr G}^{(\rho/\lambda)}
\,\biggl[
\hat k^\prime_{\rho/\lambda}\wedge\chi^{\dagger(m^\prime)}\,\vec\sigma
\biggr]_{s_1^\prime}\cdot
\bigl(\sigma_2\,\vec\sigma\bigr)_{s^\prime_3,s^\prime_2}
\,\psi^{*}(\norm{\vec k^\prime_\lambda}^2,\,\norm{\vec k^\prime_\rho}^2,\,{\vec k^\prime_\lambda}\cdot{\vec k^\prime_\rho})
\\[2mm]
&=
\dfrac{1}{2}\,{\mathscr G}^{(\rho/\lambda)}
\,\biggl[
\chi^{\dagger(m')}\,\bigl(\hat k^\prime_{\rho/\lambda}\wedge\vec\sigma\bigr)
\biggr]_{s_1^\prime}\cdot
\bigl(\sigma_2\,\vec\sigma\bigr)_{s^\prime_3,s^\prime_2}
\,\psi^{*}(\norm{\vec k^\prime_\lambda}^2,\,\norm{\vec k^\prime_\rho}^2,\,{\vec k^\prime_\lambda}\cdot{\vec k^\prime_\rho})
\end{align*}
and the transition amplitude to be computed becomes 
\begin{align*}
    \brakket{\Psi^\prime,\,m^\prime}{J}{\Psi,\,m}_{(\rho/\lambda)} &=\begin{multlined}[t][140mm] \dfrac{1}{8\pi}\,\dfrac{i}{\sqrt{2}}\,\int\dfrac{\dd\vec p_2}{(2\pi)^3}\dfrac{\dd\vec p_3}{(2\pi)^3}\,
    \sqrt{u^{\prime o}\,u^o}\,
    \sqrt{\dfrac{ {k^{\prime o}_1} {k^{o}_1} }{ {p^{\prime o}_1} {p^{o}_1} }}\,
    \sqrt{\dfrac{ {k^{\prime o}_2} {k^{o}_2} }{ {p^{\prime o}_2} {p^{o}_2} }}\,
    \sqrt{\dfrac{ {k^{\prime o}_3} {k^{o}_3} }{ {p^{\prime o}_3} {p^{o}_3} }}\,
    \\[2mm]
	\times\psi^{*}(\norm{\vec k^\prime_\lambda}^2,\,\norm{\vec k^\prime_\rho}^2,\,{\vec k^\prime_\lambda}\cdot{\vec k^\prime_\rho})\,
\varphi(\norm{\vec k_\lambda}^2,\,\norm{\vec k_\rho}^2,\,{\vec k_\lambda}\cdot{\vec k_\rho})\\[2mm]
    \times{\mathscr G}^{(\rho/\lambda)}\,\sum\limits_{s_1^\prime,s_2^\prime,s_3^\prime}\sum\limits_{s_1,s_2,s_3}
    \chi^{(m)}_{s_1}\,(\sigma_2)_{s_2,s_3}\,
    \bigl[\sigma_2\,\vec\sigma\bigr]_{s_3^\prime,s_2^\prime}
    \cdot
    \Bigl[\chi^{\dagger(m')}\,\bigl(\hat k^\prime_{\rho/\lambda}\wedge\vec\sigma\bigr)\Bigr]_{s_1^\prime}\\[2mm]
    \times\,\left[D({\boldsymbol R}_1^{\prime -1})\,J(\vec p^{\,\prime}_1,\,\vec p_1)\,D({\boldsymbol R}_1)\right]_{s_1^\prime,s_1}\,
    \left[D({\boldsymbol R}_2^{\prime -1}{\boldsymbol R}_2)\right]_{s_2^\prime,s_2}\,
    \left[D({\boldsymbol R}_3^{\prime -1}{\boldsymbol R}_3)\right]_{s_3^\prime,s_3}\end{multlined}\\[2mm]
&=\begin{multlined}[t][140mm] \dfrac{1}{8\pi}\,\dfrac{i}{\sqrt{2}}\,\int\dfrac{\dd\vec p_2}{(2\pi)^3}\dfrac{\dd\vec p_3}{(2\pi)^3}\,
    \sqrt{u^{\prime o}\,u^o}\,
    \sqrt{\dfrac{ {k^{\prime o}_1} {k^{o}_1} }{ {p^{\prime o}_1} {p^{o}_1} }}\,
    \sqrt{\dfrac{ {k^{\prime o}_2} {k^{o}_2} }{ {p^{\prime o}_2} {p^{o}_2} }}\,
    \sqrt{\dfrac{ {k^{\prime o}_3} {k^{o}_3} }{ {p^{\prime o}_3} {p^{o}_3} }}\,
	\\[2mm]
	\times\psi^{*}(\norm{\vec k^\prime_\lambda}^2,\,\norm{\vec k^\prime_\rho}^2,\,{\vec k^\prime_\lambda}\cdot{\vec k^\prime_\rho})\,
\varphi(\norm{\vec k_\lambda}^2,\,\norm{\vec k_\rho}^2,\,{\vec k_\lambda}\cdot{\vec k_\rho})
    \\[2mm]
    \times\,\dfrac{1}{\norm{\vec k^\prime_{\rho/\lambda}}}\,{\mathscr G}^{(\rho/\lambda)}\,
	        \text{Tr}\left[
	            \vec\sigma\,
	D({\boldsymbol R}_2^{\prime -1}{\boldsymbol R}_2)\,
	D({\boldsymbol R}_3^{-1}{\boldsymbol R}^\prime_3)
	\right]\\[2mm]\cdot
	\left[
	        \chi^{\dagger(m')}\,\bigl(\vec k^\prime_{\rho/\lambda}\wedge\vec\sigma\bigr)\,
	        D({\boldsymbol R}_1^{\prime -1})\,J(\vec p^{\,\prime}_1,\,\vec p_1)\,D({\boldsymbol R}_1)
	        \chi^{(m)}
	        \right]\end{multlined}
\end{align*}

\subsubsection*{Formulation in 4D}
\begin{maliste}
\item{\underline{Making the structure covariant} :}\par
One needs to consider a term of the form:
\[
\Bigl[\vec k\wedge\vec\sigma\Bigr]\cdot\inv{\vec\sigma}
\ =\ -\,\Bigl[\inv{\vec\sigma}\wedge\vec\sigma\Bigr]\cdot\vec k
\ =\ -\epsilon_{ijk}\,\inv{\sigma^j}\,\sigma^k\,k^i
\]
where we have shaded the $\sigma$ matrix which enters the trace.\\
Therefore, using the relations shown in appendix~\eqref{ann:boosts}, the action of the boosts of the type ${\boldsymbol B}_{ u^\prime}$ gives the structure:
\begin{equation}\label{eq:typeIII}
    \Bigl[\vec k\wedge\vec\sigma\Bigr]\cdot\inv{\vec\sigma}\ \leadsto\
-\,\epsilon_{\mu\nu\alpha\beta}\,u^{\prime\mu}\,p^{\nu}\,\inv{\gamma_5\,\fmslash{ u}^\prime\,\gamma^\alpha}\,\gamma_5\,\gamma^\beta\,\fmslash{ u}^\prime
\end{equation}
\item{\underline{Calculation of the trace} :}\par
Proceeding in the usual way (still keeping the $\sigma^j$ term as such):
\[
    {\mathscr T}_1\ \leadsto\
        \text{Tr}
    \left[
        \dfrac{1+\gamma^0}{2}\,\inv{\sigma^j}\,
    {\boldsymbol B}_{ k^\prime_2}^{-1}\,{\boldsymbol B}_{ u^\prime}^{-1}\,{\boldsymbol B}_{ p^\prime_2}\,
    {\boldsymbol B}_{ p_2}^{-1}\,{\boldsymbol B}_{ u}\,{\boldsymbol B}_{ k_2}\,
    {\boldsymbol B}_{ k_3}^{-1}\,{\boldsymbol B}_{ u}^{-1}\,{\boldsymbol B}_{ p_3}\,
    {\boldsymbol B}_{ p^\prime_3}^{-1}\,{\boldsymbol B}_{ u^\prime}\,{\boldsymbol B}_{ k_3^\prime}
    \right]
\]
Therefore, it follows:
\begin{align*}
{\mathscr T}_1 & = \begin{multlined}[t][160mm]\dfrac{1}{32}\text{Tr}\left[
(1 + \gamma^0)^5\,\inv{\sigma^j}\,
{\boldsymbol B}_{ k^\prime_2}^{-1}\,{\boldsymbol B}_{ u^\prime}^{-1}\,{\boldsymbol B}_{ p^\prime_2}\,
{\boldsymbol B}_{ p_2}^{-1}\,{\boldsymbol B}_{ u}\,{\boldsymbol B}_{ k_2}\,
{\boldsymbol B}_{ k_3}^{-1}\,{\boldsymbol B}_{ u}^{-1}\,{\boldsymbol B}_{ p_3}\,
{\boldsymbol B}_{ p^\prime_3}^{-1}\,{\boldsymbol B}_{ u^\prime}\,{\boldsymbol B}_{ k_3^\prime}
\right]\\ \quad\text{since}\ \ (1 + \gamma^0)^5 = 16(1 + \gamma^0)\end{multlined}\\[2mm]
& = \begin{multlined}[t][160mm]\dfrac{1}{32}\text{Tr}\left[
(1 + \gamma^0)\,\inv{\sigma^j}\,(1 + \gamma^0)\,
{\boldsymbol B}_{ k^\prime_2}^{-1}\,{\boldsymbol B}_{ u^\prime}^{-1}\,{\boldsymbol B}_{ p^\prime_2}\,
{\boldsymbol B}_{ p_2}^{-1}\,{\boldsymbol B}_{ u}\,{\boldsymbol B}_{ k_2}\,(1 + \gamma^0)^2\,
{\boldsymbol B}_{ k_3}^{-1}\,{\boldsymbol B}_{ u}^{-1}\,{\boldsymbol B}_{ p_3}\,
{\boldsymbol B}_{ p^\prime_3}^{-1}\,{\boldsymbol B}_{ u^\prime}\,{\boldsymbol B}_{ k_3^\prime}\,(1 + \gamma^0)
\right]\\ \quad\text{because}\quad(1 + \gamma^0)\,D({\boldsymbol R}_i) = D({\boldsymbol R}_i)\,(1 + \gamma^0)\end{multlined}\\[2mm]
& = \begin{multlined}[t][160mm]\dfrac{1}{32}\text{Tr}\left[
(1 + \gamma^0)\,\inv{\sigma^j}\,(1 + \gamma^0)\,
{\boldsymbol B}_{ k^\prime_2}^{-1}\,{\boldsymbol B}_{ u^\prime}^{-1}\,
{\boldsymbol B}_{ u}\,{\boldsymbol B}_{ k_2}\,(1 + \gamma^0)^2\,
{\boldsymbol B}_{ k_3}^{-1}\,{\boldsymbol B}_{ u}^{-1}\,
{\boldsymbol B}_{ u^\prime}\,{\boldsymbol B}_{ k_3^\prime}\,(1 + \gamma^0)
\right]\\ \text{because $ {\vec p}^{\,\prime}_2\,=\,{\vec p}_2\ $ and $\ {\vec p}^{\,\prime}_3\,=\,{\vec p}_3$}\end{multlined}\\[2mm]
& = \begin{multlined}[t][160mm]\dfrac{1}{32}\text{Tr}\left[
{\boldsymbol B}_{ u^\prime}\,(1 + \gamma^0)\,
{\boldsymbol B}_{ u^\prime}^{-1}\,{\boldsymbol B}_{ u^\prime}\,
\inv{\sigma^j}\,
{\boldsymbol B}_{ u^\prime}^{-1}\,{\boldsymbol B}_{ u^\prime}\,(1 + \gamma^0)\,
{\boldsymbol B}_{ k^\prime_2}^{-1}\,{\boldsymbol B}_{ u^\prime}^{-1}\right.\\
\left.\times\,
{\boldsymbol B}_{ u}\,{\boldsymbol B}_{ k_2}\,(1 + \gamma^0)\,{\boldsymbol B}_{ u}^{-1}{\boldsymbol B}_{ u}\,(1 + \gamma^0)\,
{\boldsymbol B}_{ k_3}^{-1}\,{\boldsymbol B}_{ u}^{-1}\,
{\boldsymbol B}_{ u^\prime}\,{\boldsymbol B}_{ k_3^\prime}\,(1 + \gamma^0)\,{\boldsymbol B}_{ u^\prime}^{-1}
\right]\\ \quad\text{inserting $\ {\boldsymbol B}_{ u^{(\prime)}}^{-1}{\boldsymbol B}_{ u^{(\prime)}} = {\boldsymbol 1}$ in the right places}\end{multlined}\\[2mm]
& = \begin{multlined}[t][160mm]\dfrac{1}{128}\dfrac{1}{m^2}\,
\dfrac{1}{\sqrt{m + k^{\prime o}_2}}\,
\dfrac{1}{\sqrt{m + k^{o}_2}}\,
\dfrac{1}{\sqrt{m + k^{\prime o}_3}}\,
\dfrac{1}{\sqrt{m + k^{o}_3}}\\ \times
\text{Tr}\left[
(1+\fmslash{ u}^\prime)\,{\boldsymbol B}_{ u^\prime}\,
\inv{\sigma^j}\,
{\boldsymbol B}_{ u^\prime}^{-1}\,
(1+\fmslash{ u}^\prime)(m + \fmslash{ p}_2)
(m + \fmslash{ p}_2)(1+\fmslash{ u})
(1+\fmslash{ u})(m + \fmslash{ p}_3)
(m + \fmslash{ p}_3)(1+\fmslash{ u}^\prime)
\right]\\ \quad\text{using the appendix \eqref{ann:boosts}}
\end{multlined}\\[2mm]
& = \begin{multlined}[t][160mm]\dfrac{1}{8}\,
\dfrac{1}{\sqrt{m + k^{\prime o}_2}}\,
\dfrac{1}{\sqrt{m + k^{o}_2}}\,
\dfrac{1}{\sqrt{m + k^{\prime o}_3}}\,
\dfrac{1}{\sqrt{m + k^{o}_3}}\\
\times
\text{Tr}\left[
{\boldsymbol B}_{ u^\prime}\,
\inv{\sigma^j}\,
{\boldsymbol B}_{ u^\prime}^{-1}\,
(1+\fmslash{ u}^\prime)(m + \fmslash{ p}_2)
(1+\fmslash{ u})(m + \fmslash{ p}_3)
(1+\fmslash{ u}^\prime)
\right]\\
\quad\text{since $\ (1+\fmslash{ u})^2 = 2(1+\fmslash{ u})\ $ and $\ (m+\fmslash{ p})^2 = 2\,m(m+\fmslash{ p})$}\end{multlined}
\end{align*}
One then replaces the boosted $\sigma^j$ term by using~\eqref{eq:typeIII} and one gets:
\begin{align*}
    {\mathscr T}_1  &=\ \dfrac{1}{8}\,
\dfrac{1}{\sqrt{m + k^{\prime o}_2}}\,
\dfrac{1}{\sqrt{m + k^{o}_2}}\,
\dfrac{1}{\sqrt{m + k^{\prime o}_3}}\,
\dfrac{1}{\sqrt{m + k^{o}_3}}\,
\text{Tr}\left[
\gamma_5\,\fmslash{ u}^\prime\,\gamma^\alpha\,
(1+\fmslash{ u}^\prime)(m + \fmslash{ p}_2)
(1+\fmslash{ u})(m + \fmslash{ p}_3)
(1+\fmslash{ u}^\prime)
\right]\\[2mm]
&=\ -\,\dfrac{1}{8}\,
\dfrac{1}{\sqrt{m + k^{\prime o}_2}}\,
\dfrac{1}{\sqrt{m + k^{o}_2}}\,
\dfrac{1}{\sqrt{m + k^{\prime o}_3}}\,
\dfrac{1}{\sqrt{m + k^{o}_3}}\,
\text{Tr}\left[
\fmslash{ u}^\prime\,\gamma_5\,\gamma^\alpha\,
(1+\fmslash{ u}^\prime)(m + \fmslash{ p}_2)
(1+\fmslash{ u})(m + \fmslash{ p}_3)
(1+\fmslash{ u}^\prime)
\right]\\[2mm]
&=\ -\,\dfrac{1}{8}\,
\dfrac{1}{\sqrt{m + k^{\prime o}_2}}\,
\dfrac{1}{\sqrt{m + k^{o}_2}}\,
\dfrac{1}{\sqrt{m + k^{\prime o}_3}}\,
\dfrac{1}{\sqrt{m + k^{o}_3}}\,
\text{Tr}\left[
\gamma_5\,\gamma^\alpha\,
(1+\fmslash{ u}^\prime)(m + \fmslash{ p}_2)
(1+\fmslash{ u})(m + \fmslash{ p}_3)
(1+\fmslash{ u}^\prime)\,\fmslash{ u}^\prime
\right]
\end{align*}
Therefore, we finally obtain:
\[
{\mathscr T}_1 \ =\  -\,\dfrac{1}{8}\,
\dfrac{1}{\sqrt{m + k^{\prime o}_2}}\,
\dfrac{1}{\sqrt{m + k^{o}_2}}\,
\dfrac{1}{\sqrt{m + k^{\prime o}_3}}\,
\dfrac{1}{\sqrt{m + k^{o}_3}}\,
\text{Tr}\Bigl[
\gamma_5\,\gamma^\alpha\,
(1+\fmslash{ u}^\prime)(m + \fmslash{ p}_2)
(1+\fmslash{ u})(m + \fmslash{ p}_3)
(1+\fmslash{ u}^\prime)
\Bigr]
\]
\item{\underline{Term within brackets} :}\par
One deals with a term of the form:
\[
        {\mathscr T}_{2}\ =\ \Bigl[
        \chi^{\dagger(m')}\,
        \sigma^k\,
        D({\boldsymbol R}_1^{\prime -1})\,J(\vec p^{\,\prime}_1,\,\vec p_1)\,D({\boldsymbol R}_1)
        \chi^{(m)}
        \Bigr]
\]
With the help of the techniques recalled in the appendix, one proceeds as follows (leaving aside the term $\sigma^k$ for the time being):
\begin{align*}
    {\mathscr T}_{2} \
    &\leadsto\ \sqrt{\dfrac{m^{\prime}_1\,m_1}{p^{\prime o}_1\,p^o_1}}\,\left[
    \chi^{(m')\dagger}\,\dfrac{1 + \gamma^0}{2}\,
    \sigma^k\,
    D({\boldsymbol R}_1^{\prime -1})\,
    \dfrac{1 + \gamma^0}{2}\,{\boldsymbol B}_{ p^\prime_1}^{-1}\,J\,{\boldsymbol B}_{ p_1}\,
    \dfrac{1 + \gamma^0}{2}
    \,D({\boldsymbol R}_1)\,
    \dfrac{1 + \gamma^0}{2}\,\chi^{(m)}
    \right]\\[2mm]
    &=\ \dfrac{1}{16}\,\sqrt{\dfrac{m^{\prime}_1\,m_1}{p^{\prime o}_1\,p^o_1}}\,\left[
        \chi^{(m')\dagger}\,(1 + \gamma^0)\,
        \sigma^k\,
        {\boldsymbol B}_{ k^\prime_1}^{-1}\,{\boldsymbol B}_{ u^\prime}^{-1}\,{\boldsymbol B}_{ p^\prime_1}\,
        (1 + \gamma^0)\,{\boldsymbol B}_{ p^\prime_1}^{-1}\,J\,{\boldsymbol B}_{ p_1}\,
        (1 + \gamma^0)
        \,
        {\boldsymbol B}_{ p_1}^{-1}\,{\boldsymbol B}_{ u}\,{\boldsymbol B}_{ k_1}
        \,
        (1 + \gamma^0)\,\chi^{(m)}
        \right]\\[2mm]
    &=\ \dfrac{1}{16}\,\sqrt{\dfrac{m^{\prime}_1\,m_1}{p^{\prime o}_1\,p^o_1}}\,\left[
            \chi^{(m')\dagger}\,\gamma^0\,(1 + \gamma^0)\,
            \sigma^k\,
            {\boldsymbol B}_{ k^\prime_1}^{-1}\,{\boldsymbol B}_{ u^\prime}^{-1}\,{\boldsymbol B}_{ p^\prime_1}\,
            (1 + \gamma^0)\,{\boldsymbol B}_{ p^\prime_1}^{-1}\,J\,{\boldsymbol B}_{ p_1}\,
            (1 + \gamma^0)
            \,
            {\boldsymbol B}_{ p_1}^{-1}\,{\boldsymbol B}_{ u}\,{\boldsymbol B}_{ k_1}
            \,
            (1 + \gamma^0)\,\chi^{(m)}
            \right]\\[2mm]
    &=\ \dfrac{1}{16}\,\sqrt{\dfrac{m^{\prime}_1\,m_1}{p^{\prime o}_1\,p^o_1}}\,\left[
        \chi^{(m')\dagger}\,\gamma^0\,
        \sigma^k\,
        (1 + \gamma^0)\,(1 + \gamma^0)\,
        {\boldsymbol B}_{ k^\prime_1}^{-1}\,{\boldsymbol B}_{ u^\prime}^{-1}\,{\boldsymbol B}_{ p^\prime_1}\,
        {\boldsymbol B}_{ p^\prime_1}^{-1}\,J\,{\boldsymbol B}_{ p_1}\,
        \,
        {\boldsymbol B}_{ p_1}^{-1}\,{\boldsymbol B}_{ u}\,{\boldsymbol B}_{ k_1}
        \,
        (1 + \gamma^0)
        (1 + \gamma^0)\,\chi^{(m)}
        \right]\\[2mm]
    &=\ \dfrac14\,\sqrt{\dfrac{m^{\prime}_1\,m_1}{p^{\prime o}_1\,p^o_1}}\,\left[
            \bar\chi^{(m')}\,
            {\boldsymbol B}_{ u^{\prime}}^{-1}{\boldsymbol B}_{ u^{\prime}}\,
            \sigma^k\,
            {\boldsymbol B}_{ u^{\prime}}^{-1}{\boldsymbol B}_{ u^{\prime}}\,
            (1 + \gamma^0)\,
            {\boldsymbol B}_{ k^\prime_1}^{-1}\,
            {\boldsymbol B}_{ u^{\prime}}^{-1}{\boldsymbol B}_{ u^{\prime}}\,
            {\boldsymbol B}_{ u^\prime}^{-1}\,J\,
            {\boldsymbol B}_{ u}\,{\boldsymbol B}_{ k_1}
            \,(1 + \gamma^0)
            {\boldsymbol B}_{ u}^{-1}{\boldsymbol B}_{ u}\,
            \,\chi^{(m)}
            \right]\\[2mm]
&=\ \dfrac14\,\sqrt{\dfrac{m^{\prime}_1\,m_1}{p^{\prime o}_1\,p^o_1}}\,\left[
                \bar\chi^{(m')}_{ u^{\prime}}\,
                {\boldsymbol B}_{ u^{\prime}}\,
                \sigma^k\,
                {\boldsymbol B}_{ u^{\prime}}^{-1}{\boldsymbol B}_{ u^{\prime}}\,
                (1 + \gamma^0)\,
                {\boldsymbol B}_{ k^\prime_1}^{-1}\,
                {\boldsymbol B}_{ u^{\prime}}^{-1}\,J\,
                {\boldsymbol B}_{ u}\,{\boldsymbol B}_{ k_1}
                \,(1 + \gamma^0)
                {\boldsymbol B}_{ u}^{-1}\,
                \,\chi^{(m)}_{ u}
                \right]\\[2mm]
&=\ \dfrac18\,\dfrac{1}{\sqrt{p^{\prime o}_1\,p^o_1}}\,
\dfrac{1}{\sqrt{m^\prime_1 + k^{\prime 0}_1}}\,
\dfrac{1}{\sqrt{m_1 + k^{0}_1}}\,
\,\left[
    \bar\chi^{(m')}_{ u^{\prime}}\,
    {\boldsymbol B}_{ u^{\prime}}\,
    \sigma^k\,
    {\boldsymbol B}_{ u^{\prime}}^{-1}\,
    (1+\fmslash{ u}^{\,\prime})\,
    (m^\prime_1 + \fmslash{ p}^{\,\prime}_1)
    \,J\,
    (m_1 + \fmslash{ p}_1)\,(1+\fmslash{ u})\,
    \,\chi^{(m)}_{ u}
    \right]
\end{align*}
We substitute now the boosted $\sigma^k$ term by its corresponding expression given by the relation~\eqref{eq:typeIII}:
\[
{\mathscr T}_{2} =\ \dfrac18\,\dfrac{1}{\sqrt{p^{\prime o}_1\,p^o_1}}\,
\dfrac{1}{\sqrt{m^\prime_1 + k^{\prime 0}_1}}\,
\dfrac{1}{\sqrt{m_1 + k^{0}_1}}\,
\,\left[
    \bar\chi^{(m')}_{ u^{\prime}}\,
	\Bigl(\gamma_5\,\gamma^\beta\,\fmslash{ u}^\prime\Bigr)\,
    (1+\fmslash{ u}^{\,\prime})\,
    (m^\prime_1 + \fmslash{ p}^{\,\prime}_1)
    \,J\,
    (m_1 + \fmslash{ p}_1)\,(1+\fmslash{ u})\,
    \,\chi^{(m)}_{ u}
    \right]
\]
which finally leads to:
\[
{\mathscr T}_{2} =\ \dfrac18\,\dfrac{1}{\sqrt{p^{\prime o}_1\,p^o_1}}\,
	\dfrac{1}{\sqrt{m^\prime_1 + k^{\prime 0}_1}}\,
	\dfrac{1}{\sqrt{m_1 + k^{0}_1}}
	\,\left[
	    \bar\chi^{(m')}_{ u^{\prime}}\,
		\gamma_5\,\gamma^\beta\,
	    (1+\fmslash{ u}^{\,\prime})\,
	    (m^\prime_1 + \fmslash{ p}^{\,\prime}_1)
	    \,J\,
	    (m_1 + \fmslash{ p}_1)\,(1+\fmslash{ u})\,
	    \,\chi^{(m)}_{ u}
	    \right]
\]
\item{\underline{Conclusion} :}\par
Gathering everything, we obtain the 4D formulation of the transition amplitude:
\begin{multline*}
    \brakket{\Psi^\prime,\,m^\prime}{J}{\Psi,\,m}_{(\rho/\lambda)} = \dfrac{1}{8\pi}\,\dfrac{i}{\sqrt{2}}\,\dfrac{1}{64}\,
	\int\dfrac{\dd\vec p_2}{(2\pi)^3}\dfrac{\dd\vec p_3}{(2\pi)^3}\,
    \sqrt{u^{\prime o}\,u^o}\,
    \sqrt{\dfrac{ {k^{\prime o}_1} {k^{o}_1} }{ {p^{\prime o}_1} {p^{o}_1} }}\,
    \sqrt{\dfrac{ {k^{\prime o}_2} {k^{o}_2} }{ {p^{\prime o}_2} {p^{o}_2} }}\,
    \sqrt{\dfrac{ {k^{\prime o}_3} {k^{o}_3} }{ {p^{\prime o}_3} {p^{o}_3} }}\,
	\\[2mm]
	\times\psi^{*}(\norm{\vec k^\prime_\lambda}^2,\,\norm{\vec k^\prime_\rho}^2,\,{\vec k^\prime_\lambda}\cdot{\vec k^\prime_\rho})\,
	\varphi(\norm{\vec k_\lambda}^2,\,\norm{\vec k_\rho}^2,\,{\vec k_\lambda}\cdot{\vec k_\rho})
    \\[2mm]    
\times\,
\dfrac{1}{\sqrt{p^{\prime o}_1\,p^o_1}}\,
\dfrac{1}{\sqrt{m^\prime_1 + k^{\prime 0}_1}}\,
\dfrac{1}{\sqrt{m_1 + k^{0}_1}}\,\,
\dfrac{1}{\sqrt{m + k^{\prime o}_2}}\,
\dfrac{1}{\sqrt{m + k^{o}_2}}\,
\dfrac{1}{\sqrt{m + k^{\prime o}_3}}\,
\dfrac{1}{\sqrt{m + k^{o}_3}}\,{\mathscr G}^{(\rho/\lambda)}\,
\dfrac{1}{\norm{\vec k_{\rho/\lambda}^{\prime}}}\\[2mm]
\times\,\epsilon_{\mu\nu\alpha\beta}\,u^{\prime\mu}\,p^{\prime\nu}_{(\rho/\lambda)}\,\text{Tr}\Bigl[
\gamma_5\,\gamma^\alpha\,
(1+\fmslash{ u}^\prime)(m + \fmslash{ p}_2)
(1+\fmslash{ u})(m + \fmslash{ p}_3)
(1+\fmslash{ u}^\prime)
\Bigr]\\[2mm]
    \times\left[
	    \bar\chi^{(m')}_{ u^{\prime}}\,
		\gamma_5\,\gamma^\beta\,
	    (1+\fmslash{ u}^{\,\prime})\,
	    (m^\prime_1 + \fmslash{ p}^{\,\prime}_1)
	    \,J\,
	    (m_1 + \fmslash{ p}_1)\,(1+\fmslash{ u})\,
	    \,\chi^{(m)}_{ u}
	    \right]
\end{multline*}
\end{maliste}
\subsubsection*{Infinite mass limit}
We proceed exactly in the same way as for the two previous other types and we get:
\begin{multline*}
\brakket{\Psi^\prime,\,m^\prime}{J}{\Psi,\,m}_{(\rho/\lambda)} =	\dfrac{1}{32\pi}\,\dfrac{i}{\sqrt{2}}
\,\dfrac{1}{\sqrt{v^{\prime o}\,v^o}}\,\int\dfrac{\dd\vec p_2}{(2\pi)^3}\dfrac{1}{2\,p^{o}_2}\,\dfrac{\dd\vec p_3}{(2\pi)^3}\dfrac{1}{2\,p^{o}_3}\,
	{\mathscr G}^{(\rho/\lambda)}\,
	\\[2mm]
	\times\psi^{*}(\norm{\vec k^\prime_\lambda}^2,\,\norm{\vec k^\prime_\rho}^2,\,{\vec k^\prime_\lambda}\cdot{\vec k^\prime_\rho})\,
\varphi(\norm{\vec k_\lambda}^2,\,\norm{\vec k_\rho}^2,\,{\vec k_\lambda}\cdot{\vec k_\rho})
    \\[2mm]
    \times\,
	\sqrt{\dfrac{ p_2\cdot v}{m +  p_2\cdot v}}\,
    \sqrt{\dfrac{ p_2\cdot v^\prime}{m +  p_2\cdot v^\prime}}\,
    \sqrt{\dfrac{ p_3\cdot v}{m +  p_3\cdot v}}\,
    \sqrt{\dfrac{ p_3\cdot v^\prime}{m +  p_3\cdot v^\prime}}\,
	\dfrac{1}{\norm{\vec k_{\rho/\lambda}^{\prime}}}
	\\[2mm]
\times\,\epsilon_{\mu\nu\alpha\beta}\,v^{\prime\mu}\,p^{\prime\nu}_{(\rho/\lambda)}\,\text{Tr}\Bigl[
\gamma_5\,\gamma^\alpha\,
(1+\fmslash{ v}^\prime)(m + \fmslash{ p}_2)
(1+\fmslash{ v})(m + \fmslash{ p}_3)
(1+\fmslash{ v}^\prime)
\Bigr]\,\left[
	    \bar\chi^{(m')}_{ v^{\prime}}\,
		\gamma_5\,\gamma^\beta\,
	    (1+\fmslash{ v}^{\,\prime})
	    \,J\,
	    \chi^{(m)}_{ v}
	    \right]
\end{multline*}
\subsubsection*{Calculation of the trace}
The trace can be directly computed:
\[
\text{Tr}\Bigl[
\gamma_5\,\gamma_\alpha\,
(1+\fmslash{ v}^\prime)(m + \fmslash{ p}_2)
(1+\fmslash{ v})(m + \fmslash{ p}_3)
(1+\fmslash{ v}^\prime)
\Bigr]\ =\ -\,8\,i\,\epsilon_{\alpha\tau\sigma\delta}\,v^{\prime\delta}\,\Bigl[
        (m +  p_3\cdot v^\prime)\,p_2^\tau\,v^\sigma
        - (m +  p_2\cdot v^\prime)\,p_3^\tau\,v^\sigma
        + (1-w)\,\,p_2^\tau\,p_3^\sigma
        \Bigr]
\]
so that the transition amplitude now becomes:
\begin{multline*}
\brakket{\Psi^\prime,\,m^\prime}{J}{\Psi,\,m}_{(\rho/\lambda)} =	\dfrac{1}{4\pi}\,\dfrac{1}{\sqrt{2}}
\,\dfrac{1}{\sqrt{v^{\prime o}\,v^o}}\,\int\dfrac{\dd\vec p_2}{(2\pi)^3}\dfrac{1}{2\,p^{o}_2}\,\dfrac{\dd\vec p_3}{(2\pi)^3}\dfrac{1}{2\,p^{o}_3}\,
	{\mathscr G}^{(\rho/\lambda)}\\[2mm]
	\times\psi^{*}(\norm{\vec k^\prime_\lambda}^2,\,\norm{\vec k^\prime_\rho}^2,\,{\vec k^\prime_\lambda}\cdot{\vec k^\prime_\rho})\,
	\varphi(\norm{\vec k_\lambda}^2,\,\norm{\vec k_\rho}^2,\,{\vec k_\lambda}\cdot{\vec k_\rho})
    \\[2mm]
    \times\,
	\sqrt{\dfrac{ p_2\cdot v}{m +  p_2\cdot v}}\,
    \sqrt{\dfrac{ p_2\cdot v^\prime}{m +  p_2\cdot v^\prime}}\,
    \sqrt{\dfrac{ p_3\cdot v}{m +  p_3\cdot v}}\,
    \sqrt{\dfrac{ p_3\cdot v^\prime}{m +  p_3\cdot v^\prime}}\,
	\dfrac{1}{\norm{\vec k_{\rho/\lambda}^{\prime}}}
	\\[2mm]
\qquad\times\,\epsilon_{\mu\nu\alpha\beta}\,\epsilon^{\alpha}_{\ \tau\sigma\delta}\,v^{\prime\mu}\,v^{\prime\delta}\,p^{\prime\nu}_{(\rho/\lambda)}\,
\Bigl[
        (m +  p_3\cdot v^\prime)\,p_2^\tau\,v^\sigma
        - (m +  p_2\cdot v^\prime)\,p_3^\tau\,v^\sigma
        + (1-w)\,\,p_2^\tau\,p_3^\sigma
        \Bigr]
\,\left[
	    \bar\chi^{(m')}_{ v^{\prime}}\,
		\gamma_5\,\gamma^\beta\,
	    (1+\fmslash{ v}^{\,\prime})
	    \,J\,
	    \chi^{(m)}_{ v}
	    \right]
\end{multline*}
In order to recover in the following the usual notations, we will change the names of the Lorentz indices and start with the equivalent expression:
\begin{multline*}
\brakket{\Psi^\prime,\,m^\prime}{J}{\Psi,\,m}_{(\rho/\lambda)} =	\dfrac{1}{4\pi}\,\dfrac{1}{\sqrt{2}}
\,\dfrac{1}{\sqrt{v^{\prime o}\,v^o}}\,\int\dfrac{\dd\vec p_2}{(2\pi)^3}\dfrac{1}{2\,p^{o}_2}\,\dfrac{\dd\vec p_3}{(2\pi)^3}\dfrac{1}{2\,p^{o}_3}\,
	{\mathscr G}^{(\rho/\lambda)}\,
	\\[2mm]
	\times\psi^{*}(\norm{\vec k^\prime_\lambda}^2,\,\norm{\vec k^\prime_\rho}^2,\,{\vec k^\prime_\lambda}\cdot{\vec k^\prime_\rho})\,
\varphi(\norm{\vec k_\lambda}^2,\,\norm{\vec k_\rho}^2,\,{\vec k_\lambda}\cdot{\vec k_\rho})
    \\[2mm]
    \times\,
	\sqrt{\dfrac{ p_2\cdot v}{m +  p_2\cdot v}}\,
    \sqrt{\dfrac{ p_2\cdot v^\prime}{m +  p_2\cdot v^\prime}}\,
    \sqrt{\dfrac{ p_3\cdot v}{m +  p_3\cdot v}}\,
    \sqrt{\dfrac{ p_3\cdot v^\prime}{m +  p_3\cdot v^\prime}}\,
	\dfrac{1}{\norm{\vec k_{\rho/\lambda}^{\prime}}}
	\\[2mm]
\qquad\times\,\epsilon_{\tau\mu\alpha\beta}\,\epsilon^{\alpha}_{\ \nu\sigma\delta}\,v^{\prime\tau}\,v^{\prime\delta}\,p^{\prime\mu}_{(\rho/\lambda)}\,
\Bigl[
        (m +  p_3\cdot v^\prime)\,p_2^\nu\,v^\sigma
        - (m +  p_2\cdot v^\prime)\,p_3^\nu\,v^\sigma
        + (1-w)\,\,p_2^\nu\,p_3^\sigma
        \Bigr]
\,\left[
	    \bar\chi^{(m')}_{ v^{\prime}}\,
		\gamma_5\,\gamma^\beta\,
	    (1+\fmslash{ v}^{\,\prime})
	    \,J\,
	    \chi^{(m)}_{ v}
	    \right]
\end{multline*}
This expression is manifestly covariant thanks to the relations~\eqref{eq:covaUN}, \eqref{eq:covaDEUX} and \eqref{eq:covaTROIS}. 
\subsubsection*{Using the covariance}
 One isolates now the following two manifestly covariant integrals:
    \begin{align*}
        {\mathscr I}_{(1,\,\rho/\lambda)}^{\mu\nu} &\ =\
        \begin{multlined}[t][150mm]
            \int\dfrac{\dd\vec p_2}{(2\pi)^3}\dfrac{1}{2\,p^{o}_2}\,\dfrac{\dd\vec p_3}{(2\pi)^3}\dfrac{1}{2\,p^{o}_3}\,
			{\mathscr G}^{(\rho/\lambda)}\,
			\psi^{*}(\norm{\vec k^\prime_\lambda}^2,\,\norm{\vec k^\prime_\rho}^2,\,{\vec k^\prime_\lambda}\cdot{\vec k^\prime_\rho})\,
			\varphi(\norm{\vec k_\lambda}^2,\,\norm{\vec k_\rho}^2,\,{\vec k_\lambda}\cdot{\vec k_\rho})\\[2mm]
            \times\,\sqrt{\dfrac{ p_2\cdot v}{m +  p_2\cdot v}}\,
            \sqrt{\dfrac{ p_2\cdot v^\prime}{m +  p_2\cdot v^\prime}}\,
            \sqrt{\dfrac{ p_3\cdot v}{m +  p_3\cdot v}}\,
            \sqrt{\dfrac{ p_3\cdot v^\prime}{m +  p_3\cdot v^\prime}}
			\,\dfrac{1}{\norm{\vec k_{\rho/\lambda}^{\prime}}}\\[2mm]
			\times\,\Bigl[\bigl(m +  p_3\cdot v^\prime\bigr)\,p_2^\nu\ -\ \bigl(m +  p_2\cdot v^\prime\bigr)\,p_3^\nu\Bigr]\,{p^{\,\prime\mu}_{(\rho/\lambda)}}
        \end{multlined}\\[2mm]
        {\mathscr I}_{(2,\,\rho/\lambda)}^{\mu\nu\sigma} &\ =\
        \begin{multlined}[t][150mm]
            \int\dfrac{\dd\vec p_2}{(2\pi)^3}\dfrac{1}{2\,p^{o}_2}\,\dfrac{\dd\vec p_3}{(2\pi)^3}\dfrac{1}{2\,p^{o}_3}\,
			{\mathscr G}^{(\rho/\lambda)}\,
			\psi^{*}(\norm{\vec k^\prime_\lambda}^2,\,\norm{\vec k^\prime_\rho}^2,\,{\vec k^\prime_\lambda}\cdot{\vec k^\prime_\rho})\,
			\varphi(\norm{\vec k_\lambda}^2,\,\norm{\vec k_\rho}^2,\,{\vec k_\lambda}\cdot{\vec k_\rho})\\[2mm]
            \times\,\sqrt{\dfrac{ p_2\cdot v}{m +  p_2\cdot v}}\,
            \sqrt{\dfrac{ p_2\cdot v^\prime}{m +  p_2\cdot v^\prime}}\,
            \sqrt{\dfrac{ p_3\cdot v}{m +  p_3\cdot v}}\,
            \sqrt{\dfrac{ p_3\cdot v^\prime}{m +  p_3\cdot v^\prime}}\,\dfrac{1}{\norm{\vec k_{\rho/\lambda}^{\prime}}}\,
			p_2^\nu\,p_3^\sigma\,{\hat p^{\,\prime\mu}_{(\rho/\lambda)}}
        \end{multlined}
    \end{align*}
which allow the transition amplitude to be rewritten as:
    \begin{multline*}
        \brakket{\Psi^\prime,\,m^\prime}{J}{\Psi,\,m}_{(\rho/\lambda)} = \dfrac{1}{4\pi}\,\dfrac{1}{\sqrt{2}}\,\dfrac{1}{\sqrt{v^o\,v^{\prime o}}}\,\left\{
        {\mathscr I}_{(1,\,\rho/\lambda)}^{\mu\nu}\,
        \epsilon_{\tau\mu\alpha\beta}\,{\epsilon^\alpha}_{\nu\sigma\delta}
        \,v^{\prime\tau}\,v^{\prime\delta}\,v^\sigma
                \ +\ (1-w)\,{\mathscr I}_{(2,\,\rho/\lambda)}^{\mu\nu\sigma}\,\epsilon_{\tau\mu\alpha\beta}\,{\epsilon^\alpha}_{\nu\sigma\delta}
                \,v^{\prime\tau}\,v^{\prime\delta}
        \right\}\\ \times\,\left[
            \bar\chi^{(m')}_{ v^{\prime}}\,
            \gamma_5\,\gamma^\beta
            \,
            (1+\fmslash{ v}^{\,\prime})
            \,J
            \,\chi^{(m)}_{ v}
            \right]
    \end{multline*}
However, because of their covariance, the former two integrals can be expressed as a linear combination of Lorentz covariant terms:
\begin{align*}
    {\mathscr I}_{(1,\,\rho/\lambda)}^{\mu\nu} &= \sum\limits_id_i^{(\rho/\lambda)}(w)\,\Lambda_i^{\mu\nu}\\[2mm]
    &= d_1^{(\rho/\lambda)}(w)\,v^\mu\,v^\nu + d_2^{(\rho/\lambda)}(w)\,v^{\prime\mu}\,v^{\nu} + d_3^{(\rho/\lambda)}(w)\,v^{\mu}\,v^{\prime\nu} + d_4^{(\rho/\lambda)}(w)\,v^{\prime\mu}\,v^{\prime\nu} + d_5^{(\rho/\lambda)}(w)\,g^{\mu\nu}
\end{align*}
as well as:
\begin{align*}
    {\mathscr I}_{(2,\,\rho/\lambda)}^{\mu\nu\sigma} &= \sum\limits_ic_i^{(\rho/\lambda)}(w)\,\Lambda_i^{\mu\nu\sigma}\\[2mm]
	&=\begin{multlined}[t][145mm]
	c_1^{(\rho/\lambda)}(w)\,v^\mu\,v^\nu\,v^\sigma + c_2^{(\rho/\lambda)}(w)\,v'^\mu\,v^\nu\,v^\sigma + c_3^{(\rho/\lambda)}(w)\,v^\mu\,v'^\nu\,v^\sigma + c_4^{(\rho/\lambda)}(w)\,v'^\mu\,v'^\nu\,v^\sigma +c_5^{(\rho/\lambda)}(w)\,v^\mu\,v^\nu\,v'^\sigma\\[2mm]
+  c_6^{(\rho/\lambda)}(w)\,v'^\mu\,v^\nu\,v'^\sigma + c_7^{(\rho/\lambda)}(w)\,v^\mu\,v'^\nu\,v'^\sigma + c_8^{(\rho/\lambda)}(w)\,v'^\mu\,v'^\nu\,v'^\sigma + c_9^{(\rho/\lambda)}(w)\,g^{\mu\nu}v^\sigma + c_{10}^{(\rho/\lambda)}(w)\,g^{\mu\nu}\,v^{\prime\sigma}\\[2mm]
+ c_{11}^{(\rho/\lambda)}(w)\,g^{\mu\sigma}v^\nu + c_{12}^{(\rho/\lambda)}(w)\,g^{\mu\sigma}\,v^{\prime\nu} + c_{13}^{(\rho/\lambda)}(w)\,g^{\nu\sigma}v^\mu + c_{14}^{(\rho/\lambda)}(w)\,g^{\nu\sigma}\,v^{\prime\mu}
\end{multlined}
\end{align*}
Then, when performing the contractions in the expression of $\brakket{\Psi^\prime,\,m^\prime}{J}{\Psi,\,m}$, only certain terms contribute. Indeed, a straightforward calculation (see the appendix~\ref{ann:projecteur}) allows us to obtain:
\[
\left\{
\begin{aligned}
    \ {\mathscr I}_{(1,\,\rho/\lambda)}^{\mu\nu}\,\epsilon_{\tau\mu\alpha\beta}\,{\epsilon^\alpha}_{\nu\sigma\delta}
    \,v^{\prime\tau}\,v^{\prime\delta}\,v^\sigma & =\ -\,2\,d_5^{(\rho/\lambda)}\,(v_\beta - w\,v^\prime_\beta)\\[2mm]
    {\mathscr I}_{(2,\,\rho/\lambda)}^{\mu\nu\sigma}\,\epsilon_{\tau\mu\alpha\beta}\,{\epsilon^\alpha}_{\nu\sigma\delta}\,v^{\prime\tau}\,v^{\prime\delta} & =\ -\,2\,\bigl(c_9^{(\rho/\lambda)} - c_{11}^{(\rho/\lambda)}\bigr)\,(v_\beta - w\,v^\prime_\beta)
\end{aligned}
\right.
\]
We have therefore:
\begin{multline*}
    \brakket{\Psi^\prime,\,m^\prime}{J}{\Psi,\,m}_{(\rho/\lambda)} = -\,\dfrac{1}{2\pi}\,\dfrac{1}{\sqrt{2}}\,\dfrac{1}{\sqrt{v^o\,v^{\prime o}}}\,\biggl[
    d_5^{(\rho/\lambda)}(w)
            \ +\ (1-w)\,\Bigl(c_9^{(\rho/\lambda)}(w) - c_{11}^{(\rho/\lambda)}(w)\Bigr) 
    \biggr]\\[2mm]\times\,(v_\beta - w\,v^\prime_\beta)\left\{
        \bar\chi^{(m')}_{ v^{\prime}}\,
        \gamma_5\,\gamma^\beta
        \,
        (1+\fmslash{ v}^{\,\prime})
        \,J
        \,\chi^{(m)}_{ v}
        \right\}
\end{multline*}
The last contraction has been already found in the study of the terms of type (i):
\[
(v_\beta - w\,v^\prime_\beta)\left[
        \bar\chi^{(m')}_{ v^{\prime}}\,
        \gamma_5\,\gamma^\beta
        \,
        (1+\fmslash{ v}^{\,\prime})
        \,J
        \,\chi^{(m)}_{ v}
        \right]\ =\ 
	    \left[
	    \bar\chi^{(m')}_{ v^{\prime}}\,
	    \gamma_5\,(\fmslash{ v} - w\,\fmslash{ v}^{\prime})
	    \,(1+\fmslash{ v}^{\,\prime})
	    \,J\,
	    \chi^{(m)}_{ v}
	    \right]\ =\ 2\,\left[
    \bar\chi^{(m')}_{ v^{\prime}}\,
    \gamma_5\,(\fmslash{ v} + w)
    \,J\,
    \chi^{(m)}_{ v}
    \right]
\]
Gathering everything one gets:
\[
{
    \brakket{\Psi^\prime,\,m^\prime}{J}{\Psi,\,m}_{(\rho/\lambda)} = -\,\dfrac{1}{\pi\sqrt{2}}\,\dfrac{1}{\sqrt{v^o\,v^{\prime o}}}\,\biggl[
        d_5^{(\rho/\lambda)}(w)\ +\ (1-w)\,\Bigl(c_9^{(\rho/\lambda)}(w) - c_{11}^{(\rho/\lambda)}(w)\Bigr)
    \biggr]\left\{
        \bar\chi^{(m')}_{ v^{\prime}}\,
        \gamma_5\,(\fmslash{ v} + w)
        \,J\,
        \chi^{(m)}_{ v}
        \right\}
}
\]
We then need the expression of the coefficients $d_5^{(\rho/\lambda)}$, $c_9^{(\rho/\lambda)}$ and $c_{11}^{(\rho/\lambda)}$ in terms of the initial integrals ${\mathscr I}_{(1,\,\rho/\lambda)}^{\mu\nu}$ and ${\mathscr I}_{(2,\,\rho/\lambda)}^{\mu\nu\sigma}$ (see appendix~\ref{ann:projecteur}): 
\[
{
   \left\{\
\begin{aligned}
&d_5^{(\rho/\lambda)}(w) = -\,\dfrac12\,{\mathscr I}_{(1,\,\rho/\lambda)}^{\mu\nu}\,\biggl[\dfrac{1}{1-w^2}\bigl(v_\mu\,v_\nu + v^\prime_\mu\,v^\prime_\nu\bigr) - \dfrac{w}{1-w^2}
\bigl(v_\mu\,v^{\prime}_\nu + v^{\prime}_\mu\,v_{\nu}\bigr) - g_{\mu\nu}\biggr] \\[2mm]
&c_9^{(\rho/\lambda)}(w) - c_{11}^{(\rho/\lambda)}(w)=
    -\,\dfrac{1}{2}\,\dfrac{1}{1-w^2}\,
    {\mathscr I}_{(2,\,\rho/\lambda)}^{\mu\nu\sigma}\,
    \biggl[
    v_\nu\,\bigl(g_{\mu\sigma} - v^\prime_\mu\,v^\prime_\sigma\bigr) - v_\sigma\,\bigl(g_{\mu\nu} - v^\prime_\mu\,v^\prime_\nu\bigr)
    + w\,\bigl(g_{\mu\nu}v^\prime_\sigma - g_{\mu\sigma}v^\prime_\nu\bigr)
    \biggr]
\end{aligned}
\right.
}
\]
Finally, the complete transition amplitude is:
\begin{multline*}
    \brakket{\Psi^\prime,\,m^\prime}{J}{\Psi,\,m}^{\ell_\rho;\pm 1} = \dfrac{1}{2\pi}\,\dfrac{1}{\sqrt{2}}\,\dfrac{1}{\sqrt{v^o\,v^{\prime o}}}\,\Biggl[
{\mathscr B}_{\ell_\rho;\pm 1}^{\mu\nu}\,\biggl[\dfrac{1}{1-w^2}\bigl(v_\mu\,v_\nu + v^\prime_\mu\,v^\prime_\nu\bigr) - \dfrac{w}{1-w^2}
\bigl(v_\mu\,v^{\prime}_\nu + v^{\prime}_\mu\,v_{\nu}\bigr) - g_{\mu\nu}\biggr]\Biggr.\\[2mm]
\Biggl.
+\ \dfrac{1}{1+w}\,
    {\mathscr C}_{\ell_\rho;\pm 1}^{\mu\nu\sigma}\,
    \biggl[
    v_\nu\,\bigl(g_{\mu\sigma} - v^\prime_\mu\,v^\prime_\sigma\bigr) - v_\sigma\,\bigl(g_{\mu\nu} - v^\prime_\mu\,v^\prime_\nu\bigr)
    + w\,\bigl(g_{\mu\nu}v^\prime_\sigma - g_{\mu\sigma}v^\prime_\nu\bigr)
    \biggr]
\Biggr]\\[2mm]
\times\left\{
        \bar\chi^{(m')}_{ v^{\prime}}\,
        \gamma_5\,(\fmslash{ v} + w)
        \,J\,
        \chi^{(m)}_{ v}
        \right\}
\qquad\qquad{({\text{for odd}\ \ell_\rho})}\qquad
\end{multline*}
with the covariant integrals given by:
\[
{\mathscr B}_{\ell_\rho;\pm 1}^{\mu\nu}\ =\ {\mathscr I}_{(1,\,\rho)}^{\mu\nu} + {\mathscr I}_{(1,\,\lambda)}^{\mu\nu}
\qquad\text{as well as}\qquad
{\mathscr C}_{\ell_\rho;\pm 1}^{\mu\nu\sigma}\ =\ {\mathscr I}_{(2,\,\rho)}^{\mu\nu\sigma} + {\mathscr I}_{(2,\,\lambda)}^{\mu\nu\sigma}
\]
Explicitely, removing the ``prime''~of the boosted momenta ${ p}_{\rho/\lambda}^{\prime}$:
\[
        {\mathscr B}_{\ell_\rho;\pm 1}^{\mu\nu} \ =\
        \begin{multlined}[t][140mm]
            \int\dfrac{\dd\vec p_2}{(2\pi)^3}\dfrac{1}{2\,p^{o}_2}\,\dfrac{\dd\vec p_3}{(2\pi)^3}\dfrac{1}{2\,p^{o}_3}\,
            \psi^{*}_{\ell_\rho;\pm 1}(\norm{\vec k^\prime_\lambda}^2,\,\norm{\vec k^\prime_\rho}^2,\,{\vec k^\prime_\lambda}\cdot{\vec k^\prime_\rho})\,
            \varphi(\norm{\vec k_\lambda}^2,\,\norm{\vec k_\rho}^2,\,{\vec k_\lambda}\cdot{\vec k_\rho})\\[2mm]
            \times\,\sqrt{\dfrac{ p_2\cdot v}{m +  p_2\cdot v}}\,
            \sqrt{\dfrac{ p_2\cdot v^\prime}{m +  p_2\cdot v^\prime}}\,
            \sqrt{\dfrac{ p_3\cdot v}{m +  p_3\cdot v}}\,
            \sqrt{\dfrac{ p_3\cdot v^\prime}{m +  p_3\cdot v^\prime}}\\[2mm]
			\times\,\Bigl[\bigl(m +  p_3\cdot v^\prime\bigr)\,p_2^\nu\ -\ \bigl(m +  p_2\cdot v^\prime\bigr)\,p_3^\nu\Bigr]\\[2mm]
	\times c_{\ell_\rho;\pm 1}\,\dfrac{1}{\norm{\hat k^\prime_\rho\wedge\hat k^\prime_\lambda}}\,\biggl[
	P_{\ell_\rho}^1\bigl(\hat k^\prime_\rho\cdot\hat k^\prime_\lambda\bigr)\,\dfrac{p_{\rho}^{\mu}}{\norm{\vec k_{\rho}^{\prime}}} - 
	P_{\ell_\rho\pm 1}^1\bigl(\hat k^\prime_\rho\cdot\hat k^\prime_\lambda\bigr)\,\dfrac{p_{\lambda}^{\mu}}{\norm{\vec k_{\lambda}^{\prime}}}
	\biggr]
        \end{multlined}
\]
and
\[
        {\mathscr C}_{\ell_\rho;\pm 1}^{\mu\nu\sigma} \ =\
        \begin{multlined}[t][140mm]
            \int\dfrac{\dd\vec p_2}{(2\pi)^3}\dfrac{1}{2\,p^{o}_2}\,\dfrac{\dd\vec p_3}{(2\pi)^3}\dfrac{1}{2\,p^{o}_3}\,
            \psi^{*}_{\ell_\rho;\pm 1}(\norm{\vec k^\prime_\lambda}^2,\,\norm{\vec k^\prime_\rho}^2,\,{\vec k^\prime_\lambda}\cdot{\vec k^\prime_\rho})\,
            \varphi(\norm{\vec k_\lambda}^2,\,\norm{\vec k_\rho}^2,\,{\vec k_\lambda}\cdot{\vec k_\rho})\\[2mm]
            \times\,\sqrt{\dfrac{ p_2\cdot v}{m +  p_2\cdot v}}\,
            \sqrt{\dfrac{ p_2\cdot v^\prime}{m +  p_2\cdot v^\prime}}\,
            \sqrt{\dfrac{ p_3\cdot v}{m +  p_3\cdot v}}\,
            \sqrt{\dfrac{ p_3\cdot v^\prime}{m +  p_3\cdot v^\prime}}\\[2mm]
	\times c_{\ell_\rho;\pm 1}\,\dfrac{1}{\norm{\hat k^\prime_\rho\wedge\hat k^\prime_\lambda}}\,\biggl[
	P_{\ell_\rho}^1\bigl(\hat k^\prime_\rho\cdot\hat k^\prime_\lambda\bigr)\,\dfrac{p_{\rho}^{\mu}}{\norm{\vec k_{\rho}^{\prime}}} - 
	P_{\ell_\rho\pm 1}^1\bigl(\hat k^\prime_\rho\cdot\hat k^\prime_\lambda\bigr)\,\dfrac{p_{\lambda}^{\mu}}{\norm{\vec k_{\lambda}^{\prime}}}
	\biggr]\,p_2^\nu\,p_3^\sigma
        \end{multlined}
\]
The covariance is still manifest because of the relations~\eqref{eq:covaUN}, \eqref{eq:covaDEUX} and \eqref{eq:covaTROIS}.
\subsection{The Isgur-Wise function $\boldsymbol{\sigma_\Lambda(w)}$}
\subsubsection{Definition}
The definition of the Isgur-Wise function $\sigma_\Lambda(w)$ is, see equation~\eqref{eq:IW} :
\begin{equation*}
{
\brakket{\Psi^\prime,\,m^\prime}{J}{\Psi,\,m}\ =\ \dfrac{1}{2\sqrt{3}}\,\dfrac{1}{\sqrt{v^o\,v^{\prime o}}}\,\sigma_\Lambda(w)
\,\left\{
\bar\chi^{(m')}_{ v^{\prime}}\,
\gamma_5\,(\fmslash{ v} + w)
\,J\,
\chi^{(m)}_{ v}
\right\}
}
\end{equation*}
\subsubsection{General expression}
With the notations previously introduced, one gets:
\begin{center}
\begin{tabular}{|Sc|Sc|}
\hline
\multicolumn{2}{|>{}c|}{\bfseries Modes $(\boldsymbol{\ell_\rho\,;\,\ell_\lambda={\ell_\rho\pm1}})$}\\
\hline
$\boldsymbol{\text{\bfseries even}\ \ell_\rho}$&$\sigma_{\ell_\rho;\pm 1}(w)\ =\ \dfrac{\sqrt{3}}{\pi}\,\dfrac{1}{1 - w^2}\,{\mathscr A}^\mu_{\ell_\rho;\pm 1}(v_\mu - w\,v^{\prime}_\mu)$\\
\hline
{$\boldsymbol{\text{\bfseries odd}\ \ell_\rho}$}&$
\begin{multlined}[100mm]
    \sigma_{\ell_\rho;\pm 1}(w) = \dfrac{1}{\pi}\,\sqrt{\dfrac{3}{2}}\,\Biggl\{
{\mathscr B}_{\ell_\rho;\pm 1}^{\mu\nu}\,\biggl[\dfrac{1}{1-w^2}\bigl(v_\mu\,v_\nu + v^\prime_\mu\,v^\prime_\nu\bigr) - \dfrac{w}{1-w^2}
\bigl(v_\mu\,v^{\prime}_\nu + v^{\prime}_\mu\,v_{\nu}\bigr) - g_{\mu\nu}\biggr]\Biggr.\\[-4mm]
\Biggl.
+\ \dfrac{1}{1+w}\,
    {\mathscr C}_{\ell_\rho;\pm 1}^{\mu\nu\sigma}\,
    \biggl[
    v_\nu\,\bigl(g_{\mu\sigma} - v^\prime_\mu\,v^\prime_\sigma\bigr) - v_\sigma\,\bigl(g_{\mu\nu} - v^\prime_\mu\,v^\prime_\nu\bigr)
    + w\,\bigl(g_{\mu\nu}v^\prime_\sigma - g_{\mu\sigma}v^\prime_\nu\bigr)
    \biggr]
\Biggr\}
\end{multlined}
$\\
\hline
\multicolumn{2}{|>{}c|}{\bfseries so that $\qquad\boldsymbol{\sigma_\Lambda(w) = \sum\limits_{\ell_{\rho}=0}^{+\infty}\sigma_{\ell_\rho;\pm 1}(w)}$}\\
\hline
\end{tabular}
\end{center}
\section{Conclusions and outlook}
This paper presents a technical and rather detailed account of the Bakamjian-Thomas formalism applied to baryons, and especially to the transitions of the type 
$\Lambda_b \to \Lambda_c\left({1 \over 2}^\pm \right)$ in the heavy quark mass limit for the $b$ and $c$ quarks, which are used for instance in the study of the semileptonic decays ${\Lambda_b \to \Lambda_c\left({1 \over 2}^\pm \right) \ell \overline{\nu}}\,$.\par
The BT model was built in order to provide a relativistic description of bound states with a finite number of constituents. Therefore, it is obvioulsy not a field theory. However, besides the relativistic features of the bound states, we have shown that the model provides, in the heavy quark mass limit, transition amplitudes which are covariant and which exhibit the Isgur-Wise scaling.\par
As emphasized above, the present three-body case $Qqq$ is rather involved as can be easily seen by using the natural Jacobi relative momenta ${\vec p}_\rho$ and ${\vec p}_\lambda$ because one has a relative orbital angular momentum ${\vec\ell}_\rho$ between the two spectator light quarks, and a relative angular momentum $\vec{\ell}_\lambda$ between the center-of-mass of the two light quarks and the heavy quark. This feature significantly complicates the structure of the matrix elements. That is why, with the aim of familiarizing the reader with the formalism, and since the calculations are quite complex, we have decided to write the present paper in a very detailed manner.\par
In the end, we have been able to compute the IW function $\xi_\Lambda(w)$ used in the transition ${\Lambda_b \to \Lambda_c\left({1 \over 2}^+ \right) \ell \overline{\nu}}$, i.e. $L = 0 \to L= 0$, and also the IW function $\sigma_\Lambda(w)$ used in the transition ${\Lambda_b \to \Lambda_c\left({1 \over 2}^- \right) \ell \overline{\nu}}$, i.e. $L = 0 \to L= 1$, where $L$ denotes the total orbital angular momentum which is a combination of the two relative orbital angular momenta ${\vec L} = \vec{\ell_\rho} + \vec{\ell_\lambda}$. It follows that $\ell_\rho$ and $\ell_\lambda$ can take an infinite number of values for both types of transitions $L = 0 \to L = 0, 1$.\par\medskip
The logical continuation of this work is as follows. On the theoretical side, one should try to prove the Bjorken sum rule, which requires the excited IW functions at zero recoil $\sigma^{(n)}_\Lambda(1)$ ($n$ denoting the radial excitation of the internal wave functions), which can be obtained directly from the present work, and the corresponding closure relations. We have performed this study in detail in the case of mesons~\cite{LOR-1,MORENAS-1} and in the case of baryons with a point-like spectator diquark~\cite{DIQ}. 
On the phenomenological side, we have studied the semileptonic decays ${\Lambda_b \to \Lambda_c\left({1 \over 2}^+ \right) \ell \overline{\nu}}$ and ${\Lambda_b \to \Lambda_c\left({1 \over 2}^- \right) \ell \overline{\nu}}$~\cite{DIQ}, although we used sometimes simplified models for the baryons, such as for example by considering a bound state of a heavy quark and a point-like spectator diquark. It will be very interesting to go further by using the more realistic present approach with three-body states (which could be described by the mass operator mentioned in the introduction). However, in view of the results of the present paper, this seems to be a rather formidable task.
\section{Acknowledgements}
We are indebted to Damir Bečirević for interesting discussions in the early stages of this work.
\bigskip
\appendix
\renewcommand{\thesubsubsection}{\Alph{section}.\arabic{subsection}.\alph{subsubsection}}
\section{Appendices}
\renewcommand{\theequation}{A.\arabic{equation}}
\subsection{From the 2D formalism to the 4D formalism}\label{para:DDform}
The goal is to extend the 2D objects to 4D objects inserting zeros to complete where it is necessary. This implies in particular that 
\begin{description}
\item{$\vartriangleright$ \underline{column matrices ${2\times 1}$}: }calling $(\chi)_{_{\text{2D}}}$ a matrix with two rows and one column, the formulation in 4 dimensions is done through the following procedure:
\[
(\chi)_{_{\text{2D}}}\qquad\leadsto\qquad {\dfrac{1 + \gamma^0}{2}(\chi)_{_{\text{4D}}}} = \begin{pmatrix}(\chi)_{_{\text{2D}}}\\ 0\end{pmatrix}
\]
where $(\chi)_{_{\text{4D}}}$ is a matrix with 4 rows and 1 column, containing in the first two rows $(\chi)_{_{\text{2D}}}$ and, in the last two, anything arbitrary. 
\item{$\vartriangleright$ \underline{squared matrices ${2\times 2}$}: }it is here more useful to consider the specific case of the Pauli matrices $\sigma_i = \sigma^i$. To go to 4D, one would like something of the form:
\[
\begin{pmatrix} \sigma^i&0\\ 0&0\end{pmatrix}\qquad\Longrightarrow\qquad
\sigma^i\qquad\leadsto\qquad {
\gamma_5\,\gamma^0\,\gamma^i\,\dfrac{1 + \gamma^0}{2}
}\qquad\text{car}\qquad
\gamma_5\,\gamma^0\,\gamma^i = \begin{pmatrix} \sigma^i&0\\ 0&\sigma^i\end{pmatrix}
\]
For any $2\times 2$ matrix $(A)$, one can write for example:
\[
\begin{pmatrix} A&0\\ 0&0\end{pmatrix}\qquad\Longrightarrow\qquad
(A)_{_{\text{2D}}}\qquad\leadsto\qquad {\dfrac{1 + \gamma^0}{2}\,
(A)_{_{\text{4D}}}\,\dfrac{1 + \gamma^0}{2}
}\qquad\text{with}\qquad
(A)_{_{\text{4D}}} = \begin{pmatrix} A&B\\ C&D\end{pmatrix}
\]
where the submatrices $(B)$, $(C)$ and $(D)$ are arbitrary.
\end{description}
\subsection{Representation of boosts for the four-vectors}
In a general way, for $p = {\boldsymbol{B}}_{ u}\, k$, one has:
\[
\left\{
\
\begin{aligned}
p^0&\ =\ k^0\,\cosh\zeta\ +\ \bigl(\hat u\cdot\vec k\bigr)\,\sinh\zeta\\
\vec p&\ =\ \Bigl[\vec k - \bigl(\hat u\cdot\vec k\bigr)\,\hat u\Bigr] + \cosh\zeta\,\bigl(\hat u\cdot\vec k\bigr)\,\hat u + k^0\,\sinh\zeta\,\hat u
\end{aligned}
\right.
\]
with
\[
\cosh\zeta = \dfrac{u^0}{m} = \gamma
\qquad ; \qquad
\sinh\zeta = \dfrac{\norm{\vec u}}{m} = \gamma\,\norm{\vec\beta}
\qquad ; \qquad
\vec u = \norm{\vec u}\,\hat u
\qquad ; \qquad
{\vec u}^2 = {u^0}^2 - m^2
\]
Therefore: 
\[
{
    \left\{
        \
        \begin{aligned}
        &p^0\ =\ \dfrac{1}{m}\Bigl[u^0\,k^0\ +\ \vec u\cdot\vec k\Bigr]\\[2mm]
        &\vec p\ =\ \Bigl[\vec k - \bigl(\vec k\cdot\hat u\bigr)\,\hat u\Bigr] + \dfrac{u^0}{m}\,\bigl(\vec k\cdot\hat u\bigr)\,\hat u + \dfrac{k^0}{m}\,\vec u
        \end{aligned}
        \right.
}
\]
If one chooses for $u$ the four-vector velocity $u = (u^0,\,\vec u) = \gamma (1,\,\vec v)$, these relations are obvious.
\subsection{Representation of boosts for spinors}
\subsubsection{Expression in 4 dimensions}
For an arbitrary four-vector $k$:
\[
{
    {\boldsymbol{B}}_{ k}\ =\ \dfrac{m + \fmslash{ k}\,\gamma^0}{\sqrt{2m\,(m + k^o)}}
}
\qquad\qquad\text{where}\qquad
    \gamma^0 = \begin{pmatrix}\mathbb{1}&0\\ 0&-\mathbb{1} \end{pmatrix}\qquad\text{et}\qquad
    \gamma^i = \begin{pmatrix}0&\sigma^i\\ -\sigma^i&0 \end{pmatrix}
\]
We can deduce the explicit matrix elements of this boost:
\[
    {\boldsymbol{B}}_{ k}\ =\ \sqrt{\dfrac{m + k^o}{2m}}
    \begin{pmatrix}
        \mathbb{1} & \dfrac{\vec k\cdot\vec \sigma}{m + k^o}\\
        \dfrac{\vec k\cdot\vec \sigma}{m + k^o} & \mathbb{1}
    \end{pmatrix}
\]
\subsubsection{Inverse of the boost in 4 dimensions}
One obtains in this case:
\[
{
    {\boldsymbol{B}}_{ k}^{-1}\ =\ \gamma^0\,{\boldsymbol{B}}_{ k}^\dagger\,\gamma^0\ =\ \dfrac{m + \gamma^0\,\fmslash{ k}}{\sqrt{2m\,(m + k^o)}}
}
\]
The corresponding matrix elements are given as follows:
\[
    {\boldsymbol{B}}_{ k}^{-1}\ =\ \sqrt{\dfrac{m + k^o}{2m}}
    \begin{pmatrix}
        \mathbb{1} & -\,\dfrac{\vec k\cdot\vec \sigma}{m + k^o}\\
        -\,\dfrac{\vec k\cdot\vec \sigma}{m + k^o} & \mathbb{1}
    \end{pmatrix}\qquad\qquad(\vec k\,\leadsto\,-\,\vec k)
\]
\subsection{Wigner rotations}\label{ann:wigner}
In the transition amplitudes written in the BT formalism, the term $D({\boldsymbol R}_j)_{s_i,s_i^\prime}$ is the $(s_i,s_i^\prime)$ matrix element of the Wigner rotation of a spin $1/2$ particle. It reflects the behaviour of spins when going from an arbitrary reference frame (momentum $p_j$) to the internal reference frame (momentum $k_j$). It can be expressed using the boosts introduced earlier as follows:
\[
    {\boldsymbol R}_j = {\boldsymbol B}^{-1}_{p_j}{\boldsymbol B}_{u}{\boldsymbol B}_{k_j}
    \qquad\qquad\text{with}\qquad M_o\,u = \sum\limits_j p_j\quad\text{and}\quad
    k_j = {\boldsymbol B}^{-1}_{u}\,p_j
\]
where $M_o = \sqrt{\sum(p_j)^2}$ is a mass term/operator of the baryon.
\subsection{Regarding the boosts Dirac matrices}\label{ann:boosts}
When going to the 4D formalism, one sometimes needs to calculate:
\[
{\boldsymbol B}_{ u}\,\gamma^\mu\,{\boldsymbol B}_{ u}^{-1}
\]
Taking the expressions of the boosts
\[
{\boldsymbol B}_{ u} = \boostu{u}\qquad\text{ainsi que}\qquad{\boldsymbol B}_{ u}^{-1} = \recboostu{u}
\]
one can obtain in a straightforward way:
\begin{equation}\label{eq:boost}
{
    \Gamma^\mu_{(u)} = {\boldsymbol B}_{ u}\,\gamma^\mu\,{\boldsymbol B}_{ u}^{-1}\ =\ \gamma^\mu\ -\ \dfrac{u^\mu}{1+u^o}\bigl(\gamma_0 + \fmslash{u}\bigr)\ +\ 
g^{0\mu}\,\biggl[\dfrac{1 + 2\,u^o}{1+u^o}\,\fmslash{u} - \dfrac{\gamma^0}{1+u^o}\biggr]
}
\end{equation}
In other words:
\[
{
\ \left\{
\begin{aligned}
    \Gamma^0_{(u)} = {\boldsymbol B}_{ u}\,\gamma^0\,{\boldsymbol B}_{ u}^{-1}&\ =\ \fmslash{u}\\[2mm]
    \Gamma^i_{(u)} = {\boldsymbol B}_{ u}\,\gamma^i\,{\boldsymbol B}_{ u}^{-1}&\ =\ \gamma^i\ -\ \dfrac{u^i}{1+u^o}\bigl(\gamma_0 + \fmslash{u}\bigr)\quad
\end{aligned}
\right.
}
\]
One can also verify that:
\[
    {
        \Gamma^\mu_{(u)}\,{\Gamma_{(u)}}_\mu = \gamma^\mu\,\gamma_\mu
        }\qquad\text{and}\qquad
{
{\boldsymbol B}_{ u}\,\gamma^5 = \gamma^5\,{\boldsymbol B}_{ u}
}\qquad\text{as well as}\qquad
{
{\boldsymbol B}_{ u}^{-1}\,\gamma^5 = \gamma^5\,{\boldsymbol B}_{ u}^{-1}
}
\]
Finally, for any four-vector $q$, one has the general property
\[
{{\boldsymbol B}_{ u}\,\fmslash{ q}\,{\boldsymbol B}_{ u}^{-1} = \gamma_\mu\left({\boldsymbol B}_{ u}\, q\right)^\mu}
\]
One can also show the two following useful relations:
\begin{equation}\label{ann:eq3}
{
{\boldsymbol B}_{ u}\,{\boldsymbol B}_{ k}\,(1+\gamma^0)\,{\boldsymbol B}_{ u}^{-1}\ =\
\dfrac{(m + \fmslash{p})(1+\fmslash{u})}{\sqrt{2\,m\,(m + k^0)}}
}
\qquad\text{and}\qquad
{
{\boldsymbol B}_{ u}\,(1+\gamma^0)\,{\boldsymbol B}_{ k}^{-1}\,{\boldsymbol B}_{ u}^{-1}\ =\
\dfrac{(1+\fmslash{u})(m + \fmslash{p})}{\sqrt{2\,m\,(m + k^0)}}
}
\end{equation}
\[
\text{provided that}\qquad{ k = {\boldsymbol B}_{ u}^{-1}\, p}
\]
\subsubsection*{Consequences}
\begin{maliste}
\item{\underline{First consequence} :}\par
Let us consider the boost of the following term 
\begin{align*}
\Bigl[\inv{\vec\sigma}\wedge\vec\sigma\Bigr]_i\ =\ \epsilon_{ijk}\,\inv{\sigma^j}\,\sigma^k
&\quad\leadsto\quad
-\,\epsilon_{\mu\nu\rho\sigma}\,n^{\mu}\,\bigl(\inv{\gamma_5\,\gamma_0\,\gamma^\rho}\bigr)\,\bigl(\gamma_5\,\gamma_0\,\gamma^\sigma\bigr)\\[2mm]
&\quad\leadsto\quad
\epsilon_{\mu\nu\rho\sigma}\,n^{\mu}\,\bigl(\inv{\gamma_5\,\gamma_0\,\gamma^\rho}\bigr)\,\bigl(\gamma_5\,\gamma^\sigma\,\gamma_0\bigr)
\qquad\text{because $\sigma\neq0$}
\end{align*}
The vector product does not vanish since the $\vec\sigma$ are not identical (they appear in different structures and do not talk to each other).\par
Then
\begin{align*}
{\boldsymbol B}_{ u}\,\Bigl[\inv{\vec\sigma}\wedge\vec\sigma\Bigr]_i\,{\boldsymbol B}_{ u}^{-1}
&\ =\ 
\epsilon_{\mu\nu\rho\sigma}\,u^{\mu}\,\inv{\gamma_5}\,({\boldsymbol B}_{ u}\,\inv{\gamma_0}\,{\boldsymbol B}_{ u}^{-1})
\,({\boldsymbol B}_{ u}\,\inv{\gamma^\rho}\,{\boldsymbol B}_{ u}^{-1})\,\gamma_5
\,({\boldsymbol B}_{ u}\,\gamma^\sigma\,{\boldsymbol B}_{ u}^{-1})
\,({\boldsymbol B}_{ u}\,\gamma_0\,{\boldsymbol B}_{ u}^{-1})
\end{align*}
Looking at the relation~\eqref{eq:boost}, one realizes that:
\begin{malistealph}
	\item the terms containing $u^\sigma$ and $u^\rho$ will not contribute because of the contraction with $\epsilon_{\mu\nu\rho\sigma}\,u^{\mu}$
	\item the terms in $g^{0\sigma}$ and in $g^{0\rho}$, in the contraction with $\epsilon_{\mu\nu\rho\sigma}\,u^{\mu}$, will imply the values 1, 2 and 3 for $\mu$. Therefore, the coordinates $u^{\mu}$ will be the coordinates of the three-velocity $\vec u$ which, in the infinite mass limit, becomes the baryon velocity $\vec v$. Since the relations are covariant, one can work in any reference frame and if one takes the rest frame, these terms are clearly vanishing. 
\end{malistealph}
It follows that the non-vanishing contributions are such that 
\[
{
{\boldsymbol B}_{ u}\,\Bigl[\inv{\vec\sigma}\wedge\vec\sigma\Bigr]_i\,{\boldsymbol B}_{ u}^{-1}
\ =\ 
\epsilon_{\mu\nu\rho\sigma}\,u^{\mu}\,\inv{\gamma_5\,\fmslash{u}\,\gamma^\rho}\,\gamma_5
\,\gamma^\sigma\,\fmslash{u}}
\]
\item{\underline{Second consequence} :}\par
Let us consider now a term of the form:
\begin{align*}
\vec k\cdot\vec\sigma = k_i\,\sigma_i
\quad\leadsto\quad
-\gamma_5\,\gamma_0\,\gamma_i\,k^i &= \gamma_5\,\gamma_i\,k^i\,\gamma_0\quad(i\neq0)\\[2mm]
&=\gamma_5\,\bigl(\fmslash{k} - {k^0}\,\gamma_0\bigr)\,\gamma_0\\[2mm]
&=\gamma_5\,\bigl(\fmslash{k} - (k\cdot n)\,\fmslash{n}\bigr)\,\gamma_0
\end{align*}
Since we have
\[
{\boldsymbol B}_{ u}\,\fmslash{k}\,{\boldsymbol B}_{ u}^{-1} = \gamma^\nu\bigl({\boldsymbol B}_{ u}\,k\bigr)_\nu = \fmslash{p}
\qquad\text{and also}\qquad
{\boldsymbol B}_{ u}\,\fmslash{n}\,{\boldsymbol B}_{ u}^{-1} = \gamma^\nu\bigl({\boldsymbol B}_{ u}\,n\bigr)_\nu = \fmslash{u}
\]
and because of the invariance of the scalar product of four-vectors 
\[
k\cdot n  = p\cdot u 
\]
one gets
\[
{
{\boldsymbol B}_{ u}\,\bigl(\vec k\cdot\vec\sigma\bigr)\,{\boldsymbol B}_{ u}^{-1}\ =\ 
\gamma_5\,\bigl[\fmslash{p} - (p\cdot u)\,\fmslash{u}\bigr]\,\fmslash{u}
}
\]
\item{\underline{Third consequence} :}\par
Let us consider the following term: 
\begin{align*}
\bigl(\vec k_1\wedge\vec k_2\bigr)\cdot\vec\sigma = \epsilon_{ijk}\,k_1^j\,k_2^k\,\sigma^i
\end{align*}
It is of the previous form with $\vec k = \vec k_1\wedge\vec k_2$.\par
Therefore:
\begin{align*}
k_i = \epsilon^{ijk}\,{k_1}_j\,{k_2}_k\quad\leadsto\quad
k_\nu = -\,n^{\mu}\,\epsilon_{\mu\nu\rho\sigma}\,k_1^\rho\,k_2^\sigma
\end{align*}
and thus:
\[
{\boldsymbol B}_{ u}\,\fmslash{k}\,{\boldsymbol B}_{ u}^{-1} = \gamma^\nu\bigl({\boldsymbol B}_{ u}\,k\bigr)_\nu
= -\,\epsilon_{\mu\nu\rho\sigma}\,u^{\mu}\,p_1^\rho\,p_2^\sigma\,\gamma^\nu
\qquad\text{and also}\qquad
p\cdot u = k\cdot n = 0
\]
Finally, we have:
\[
{
{\boldsymbol B}_{ u}\,\Bigl[\bigl(\vec k_1\wedge\vec k_2\bigr)\cdot\vec\sigma\Bigr]\,{\boldsymbol B}_{ u}^{-1}\ =\ 
-\,\epsilon_{\mu\nu\rho\sigma}\,u^{\mu}\,p_1^\rho\,p_2^\sigma\,\gamma_5\,\gamma^\nu\,\fmslash{u}
}
\]
\end{maliste}
\subsubsection*{Internal variables and Lorentz boosts}
Introducing the internal momentum $k_i$:
\[
{ k_i = {\boldsymbol{B}}^{-1}_{ u} p_i}
\]
one gets:
\begin{align*}
\vec k_1\cdot\vec k_2 = \llvec{{\boldsymbol{B}}^{-1}_{ u} p_1}\cdot\llvec{{\boldsymbol{B}}^{-1}_{ u} p_2}
&= \Bigl({\boldsymbol{B}}^{-1}_{ u} p_1\Bigr)^{\!o}
\Bigl({\boldsymbol{B}}^{-1}_{ u} p_2\Bigr)^{\!o}
- \Bigl({\boldsymbol{B}}^{-1}_{ u} p_1\Bigr)\cdot\Bigl({\boldsymbol{B}}^{-1}_{ u} p_2\Bigr)\\
&= \Bigl({\boldsymbol{B}}^{-1}_{ u} p_1\Bigr)^{\!o}
\Bigl({\boldsymbol{B}}^{-1}_{ u} p_2\Bigr)^{\!o}
 -  p_1\cdot p_2\qquad\text{(scalar product invariance)}\\
&= \Bigl(\bigl({\boldsymbol{B}}^{-1}_{ u} p_1\bigr)\cdot n\Bigr)
\Bigl(\bigl({\boldsymbol{B}}^{-1}_{ u} p_2\bigr)\cdot n\Bigr)
-  p_1\cdot p_2\qquad\text{(setting $ n = (1,\,\vec 0)$)}\\
&= \Bigl( p_1\cdot\bigl({\boldsymbol{B}}_{ u}\, n\bigr)\Bigr)
\Bigl( p_2\cdot\bigl({\boldsymbol{B}}_{ u}\, n\bigr)\Bigr)
-  p_1\cdot p_2\qquad\text{(scalar product invariance)}\\
&= ( p_1\cdot u)( p_2\cdot u) -  p_1\cdot p_2\qquad\text{(definition of the boost ${\boldsymbol{B}}_{ u}$)}
\end{align*}
In summary:
\begin{equation}\label{eq:internes}
{\vec k_1\cdot\vec k_2 = \llvec{{\boldsymbol{B}}^{-1}_{ u} p_1}\cdot\llvec{{\boldsymbol{B}}^{-1}_{ u} p_2} = ( p_1\cdot u)( p_2\cdot u) -  p_1\cdot p_2}\qquad\Longrightarrow\qquad
{
\vec k^{\,2} = \norm{\llvec{{\boldsymbol{B}}^{-1}_{ u} p}}^2 = ( p\cdot u)^2 -  p\cdot p
}
\end{equation}
\subsection{The $\boldsymbol{J(\vec p^{\,\prime},\vec p)}$ current term}\label{ann:current}
In the BT formalism, it is necessary to have a four-dimensional expression in terms of boosts of the current term ${J(\vec p^{\,\prime},\vec p)}$ at the origin of the transition $b\to c$ in the baryon decays considered. Typically, this current is given by:
\[
    J(\vec p^{\,\prime},\vec p) = \bar u_{s'}(p')\,J\,u_s(p)
\]
where $u_s(p)$ is the free Dirac spinor for a particle having momentum $p$ and spin projection $s$. This spinor can be obtained by boosting the spinor of a particle with zero momentum, as follows:
\[
    u_s(p) = \sqrt{\dfrac{m}{p^0}}\,{\boldsymbol B}_{p}\begin{pmatrix}\chi_s\\ 0\end{pmatrix}
\]
where $\chi_s$ is the Pauli spinor of the spin $1/2$ heavy quark and ${\boldsymbol B}_{p}$ is the Lorentz boost $(\sqrt{p^2},\vec 0)\ \leadsto\ p$ which reads in the Dirac notation:
\[
    {\boldsymbol{B}}_{p}\ =\ \dfrac{m + \fmslash{p}\,\gamma^0}{\sqrt{2m\,(m + p^0)}}
\]
Finally, in order to achieve the full ``covariantization'' of this current, $\bigl(\begin{smallmatrix}\chi_s\\ 0\end{smallmatrix}\bigr)$ must be transformed into a $4\times 4$ matrix following the procedure outlined in the paragraph~\ref{para:DDform}. This amounts to performing the substitution:
\[
    \begin{pmatrix}\chi_s\\ 0\end{pmatrix}\quad\leadsto\quad\dfrac{1+\gamma^0}{2}
\]
As a result:
\[
    J(\vec p^{\,\prime},\vec p) = \sqrt{\dfrac{m'}{p^{\prime 0}}\dfrac{m}{p^0}}\,\dfrac{1+\gamma^0}{2}\,{\boldsymbol B}^\dagger_{p'}\,\gamma^0\,J\,{\boldsymbol B}_{p}\,\dfrac{1+\gamma^0}{2}
    \qquad\qquad(\bar u = u^\dagger\,\gamma^0)
\]
which finally leads to:
\[
    J(\vec p^{\,\prime},\vec p) = \sqrt{\dfrac{m'}{p^{\prime 0}}\dfrac{m}{p^0}}\,\dfrac{1+\gamma^0}{2}\,{\boldsymbol B}^{-1}_{p'}\,J\,{\boldsymbol B}_{p}\,\dfrac{1+\gamma^0}{2}
    \qquad\qquad\text{since}\quad {\boldsymbol B}^\dagger_{p} = \gamma^0\,{\boldsymbol B}^{-1}_{p}\,\gamma^0
\]
\subsection{Basis and standard coordinates}\label{ann:standardbasis}
The standard basis (expressed in terms of the cartesian coordinates) is defined by:
\[
{
\vec e^{\,(0)} = \vec e_3\quad;\quad \vec e^{\,(+1)} = -\,\dfrac{1}{\sqrt{2}}(\vec e_1 - i\,\vec e_2)\quad;\quad
\vec e^{\,(-1)} = \dfrac{1}{\sqrt{2}}(\vec e_1 + i\,\vec e_2)
}
\]
In this basis, the components of a tri-vector $\vec a$ reads:
\[
{\vec a = \sum\limits_{\mathclap{n=-1,0,1}}a^{(n)}\vec e^{\,(n)}}\qquad\text{where}\qquad{
a^{\,(0)} = a_3\quad;\quad a^{(+1)} = -\,\dfrac{1}{\sqrt{2}}(a_1 + i\,a_2)\quad;\quad
a^{(-1)} = \dfrac{1}{\sqrt{2}}(a_1 - i\,a_2)
}
\]
and they satisfy (if the $a_i$ are real numbers) the following property:
\[
{a^{\,(n)}{}^{*} = (-)^na^{\,(-n)}}
\]
The scalar product is defined by:
\[
{
\scalprod{b}{\vec a} = \vec a\cdot\vec b^*\ =\ \sum\limits_{i=1}^3 a_i\,b_i^*\ =\ \sum\limits_{\mathclap{n=-1,0,1}}a^{(n)}\,b^{(n)*}\ =\ \sum\limits_{\mathclap{n=-1,0,1}} (-)^n a^{(n)}\,b^{(-n)}
}
\]
where the last equality holds if the $b_i$ are real.\par
One also verifies that the basis is orthonormal:
\[
\scalprod{e^{\,(m)}}{\vec e^{\,(n)}} = \vec e^{\,(n)}\cdot\vec e^{\,(m)}{}^{*} = \delta^{nm}\qquad\text{with}\qquad\vec e^{\,(m)}{}^{*} = (-)^m\vec e^{\,(-m)}
\]
and that
\[
{
a^{\,(n)} = \scalprod{e^{\,(n)}}{\vec a} = \vec a\cdot\vec e^{\,(n)}{}^{*}
}
\]
\subsection{Ladder operators on the Pauli spinors}\label{ann:ladder}
The relations given in this paragraph are used extensively when computing the spin and orbital momentum parts of the heavy baryon wave functions in terms of the Legendre polynomial.\par
One introduces indeed the usual ladder operators:
\[
{\sigma^{(+)} = \dfrac12(\sigma_1 + i\,\sigma_2)}\qquad\text{and also}\qquad
{\sigma^{(-)} = \dfrac12(\sigma_1 - i\,\sigma_2)}
\]
which act on the Pauli spinors $\chi^{(m)}$ according to:
\[
{
\left\{
\begin{aligned}
\ &\sigma^{(+)}\,\chi^{(1/2)} = 0\\
\ &\sigma^{(+)}\,\chi^{(-1/2)} = \chi^{(1/2)}
\end{aligned}
\right.
\qquad\text{and}\qquad
\left\{
\begin{aligned}
\ &\sigma^{(-)}\,\chi^{(1/2)} = \chi^{(-1/2)}\\
\ &\sigma^{(-)}\,\chi^{(-1/2)} = 0
\end{aligned}
\right.
}
\]
There is also the third Pauli matrix, which is not a ladder operator, but which satisfies:
\[
{\sigma^{(0)} = \sigma_3}\qquad\text{with}\qquad{
\left\{
\begin{aligned}
\ &\sigma^{(0)}\,\chi^{(1/2)} = \chi^{(1/2)}\\
\ &\sigma^{(0)}\,\chi^{(-1/2)} = -\chi^{(-1/2)}
\end{aligned}
\right.
}
\]
Finally, it is noted that:
\[
{\sigma^{(+)}\ =\ -\dfrac{1}{\sqrt{2}}\,\sigma^{(+1)}}
\qquad\text{and also}\qquad
{\sigma^{(-)}\ =\ \dfrac{1}{\sqrt{2}}\,\sigma^{(-1)}}
\]
where $\sigma^{(\pm 1)}$ are the standard coordinates of $\vec\sigma$.
\subsection{Double factorial}\label{ann:facto}
Some properties satisfied by the double factorial $n!!$ are:
\begin{align*}
&n!\ =\ n!!\,(n-1)!!
&&(2n)!!\ =\ 2^n\,n!\\[2mm]	
&(2n+1)!!\ =\ \dfrac{(2n+1)!}{(2n)!!}\ =\ \dfrac{(2n+1)!}{2^n\,n!}
&&(2n-1)!!\ =\ \dfrac{(2n)!}{(2n)!!}\ =\ \dfrac{(2n)!}{2^n\,n!}\\[2mm]	
\end{align*}
\subsection{Generalized Legendre polynomial $\boldsymbol{P_\ell^1(x)}$}\label{ann:legendre}
An important piece of information for the rest is the expression of the coefficient of the term of order $\ell-1$ in the generalized Legendre polynomial $P_\ell^1(x)$.\par
A possible generic expression for the $P_\ell^1(x)$ is the following:
\[
P_\ell^1(x) = -\,2^\ell\,\sqrt{1-x^2}\,\sum\limits_{k=1}^\ell\dfrac{1}{(k-1)!}\,\dfrac{1}{(\ell-k)!}\,
\dfrac{\Gamma\Biggl[\dfrac{k+\ell+1}{2}\Biggr]}{\Gamma\Bigl[\dfrac{k-\ell+1}{2}\Bigr]}\,x^{k-1}
\]
Therefore the coefficient of the term $x^{\ell-1}$ reads:
\[
t_\ell = -\,2^\ell\,\dfrac{1}{(\ell-1)!}\,\dfrac{1}{0!}\,
\dfrac{\Gamma\Biggl[\dfrac{2\ell+1}{2}\Biggr]}{\Gamma\Bigl[\dfrac{1}{2}\Bigr]}
\]
Using the following property of the function $\Gamma(x)$
\[
\Gamma[n+1/2]\ =\ \dfrac{(2n-1)!!}{2^n}\,\sqrt{\pi}\ =\ \dfrac{(2n)!}{n!\,4^n}\,\sqrt{\pi}
\]
one finds the expression of the coefficient $t_\ell$ we are looking for
\[
{
t_\ell\ =\ -\,\dfrac{1}{2^{\ell}}\,\dfrac{(2\ell)!}{\ell!\,(\ell - 1)!}\ =\ 
-\,\dfrac{(2\ell -1)!!}{(\ell-1)!}
}
\]
\subsection{Some useful Clebsch-Gordan coefficients}\label{ann:clebsch}
One can check by evaluating each sides of the equalities the following identities:
\[
    \CG{1/2,\, s_1}{1/2,\, s_2}{0,\, m} = \dfrac{i}{\sqrt{2}}\,\bigl(\sigma_2\bigr)_{s_1s_2}
    \qquad\text{and}\qquad
    \CG{1/2,\, s_1}{1/2,\, s_2}{1,\, m} = \dfrac{i}{\sqrt{2}}\,\Bigl[\bigl({\vec e}^{\,(m)}\cdot\vec\sigma\bigr)\,\sigma_2\Bigr]_{s_1s_2}
\]
where ${\vec e}^{\,(m)}$ is the $m$ component of the standard basis given in~\ref{ann:standardbasis} and $s_i$ is the spin projection with the following convention:
\begin{center}
\begin{tabular}{|c|c|c|}
\hline
&Clebsch-Gordan coefficient&Pauli matrix element\\
\hline
\multirow{ 2}{*}{$s_i$}&1/2&1\\
&-1/2&2\\
\hline
\end{tabular}
\end{center}
In addition, explicit expressions of other Clebsch-Gordan coefficients which are useful in the calculations are given below:
\begin{align*}
\CG{n,\, n }{1,\, 0}{n+1,\, n}&\ =\ \dfrac{1}{\sqrt{n+1}}\\[2mm]
\CG{n,\, n-1 }{1,\, 1}{n+1,\, n}&\ =\ \sqrt{\dfrac{n}{n+1}}\\[2mm]
\CG{n,\, n }{1,\, -1}{n+1,\, n-1}&\ =\ \dfrac{1}{\sqrt{(n+1)(2n+1)}}\\[2mm]
\CG{n,\, n-2 }{1,\, 1}{n+1,\, n-1}&\ =\ \sqrt{\dfrac{n(2n-1)}{(n+1)(2n+1)}}\\[2mm]
\CG{n,\, n - 1}{n - 1,\, 1 - n}{1,\, 0}&\ =\ \sqrt{\dfrac{3}{n\,(2n+1)}}\\[2mm]
\CG{n,\, n}{n +1,\, - n}{1,\, 0}&\ =\ \sqrt{\dfrac{3}{(n+1)\,(2n+3)}}\\[2mm]
\CG{n,\, n - 1}{n +1,\, 1 - n}{1,\, 0}&\ =\ -\,2\,\sqrt{\dfrac{3n}{(n+1)(2n+1)(2n+3)}}\\[2mm]
\CG{n,\, n - 2}{n -1,\, 2 - n}{1,\, 0}&\ =\ -\,2\,\sqrt{\dfrac{3(n-1)}{n(2n-1)(2n+1)}}
\end{align*}
\subsection{Explicit example: the $\boldsymbol{(\ell_\rho,\ell_\lambda) = (0,1)}$ mode of the $\boldsymbol{L=1}$ wave function}\label{ann:example}
We want to compute in this section the spin and orbital angular momentum contribution to the $L=1$ wave function of the mode ${(\ell_\rho,\ell_\lambda) = (0,1)}$ with the small component $j=1$.\\
The starting point is the decomposition~\eqref{eq:pair}:
\[
\begin{split}
        {\mathscr T} = \sum\limits_{m,m^\prime}
        \CG{1/2,\,s_2}{1/2,\,s_3}{0,\,0}\,
        \underbracket[0.5pt]{\CG{0,\,0}{1,\,m'}{1,\,m'}}_{=1}\,
        \underbracket[0.5pt]{\CG{0,\, m}{1,\, (m^\prime - m)}{1,\, m^\prime}}_{=\,\delta_{m0}}\qquad\qquad\qquad
        \\
        \qquad\qquad\qquad\times
        \CG{1,\,m'}{1/2,\,(M - m')}{1/2,\,M}\,
        \underbracket[0.5pt]{Y_{0}^{(m)}(\hat k_\rho)}_{=1/\sqrt{4\pi}}\,
        Y_{1}^{(m'-m)}(\hat k_\lambda)\,\chi^{(M-m')}
\end{split}
\]
which reads, using the appendix~\ref{ann:clebsch}:
\[
    {\mathscr T} = \dfrac{i}{\sqrt{2}}\,\bigl(\sigma_2\bigr)_{s_2s_3}\,{\mathscr A}^{M}
    \quad\text{with}\quad
    {\mathscr A}^{M} = \dfrac{1}{\sqrt{4\pi}}\,\sum\limits_\mu\CG{1,\,M-\mu}{1/2,\,\mu}{1/2,\,M}\,Y_{1}^{(M-\mu)}(\hat k_\lambda)\,\chi^{(\mu)}
\]
The Clebsch-Gordan can be expressed using a recursion relation~\cite{clebsch}:
\[
    \CG{1,\,M-\mu}{1/2,\,\mu}{1/2,\,M} = (-)^{1/2+\mu}\sqrt{\dfrac23}\,\CG{1/2,\,-\mu}{1/2,\,M}{1,\,M-\mu} = (-)^{1/2+\mu}\dfrac{i}{\sqrt{3}}\,\,\Bigl[\bigl({\vec e}^{\,(M-\mu)}\cdot\vec\sigma\bigr)\,\sigma_2\Bigr]_{-\mu,M}
\]
where we have used the appendix~\ref{ann:clebsch} again.\\
From the expression of the spherical harmonics given in section~\ref{sec:harmo}, we obtain: 
\[
    {\mathscr A}^{M} = \dfrac{i}{{4\pi}}\,\sum\limits_\mu(-)^{1/2+\mu}\Bigl({\vec e}^{\,(M-\mu)}\cdot\,\hat k_\lambda\Bigr)\,\Bigl[\bigl({\vec e}^{\,(M-\mu)}\cdot\vec\sigma\bigr)\,\sigma_2\Bigr]_{-\mu,M}\,\chi^{(\mu)}
    \qquad\text{where}\quad\mu=\pm1/2
\]
\begin{maliste}
\item{\underline{$M=1/2$ case}:} in this situation, $M-\mu$ can take the values 0 and 1.\\
So the contribution we need to compute reads:
\[
    {\mathscr A}^{1/2} = \dfrac{i}{{4\pi}}\,\sum\limits_{n=0}^1(-)^{1-n}\Bigl({\vec e}^{\,(n)}\cdot\,\hat k_\lambda\Bigr)\,\Bigl[\bigl({\vec e}^{\,(n)}\cdot\vec\sigma\bigr)\,\sigma_2\Bigr]_{n-1/2,1/2}\,\chi^{(1/2-n)}
    \qquad(n=M-\mu)
\]
\begin{description}
    \item{$\vartriangleright$ {$n=0$ term}: }this first term is
\begin{align*}
    -\,\dfrac{i}{{4\pi}}\,\Bigl({\vec e}^{\,(0)}\cdot\,\hat k_\lambda\Bigr)\,\Bigl[\bigl({\vec e}^{\,(0)}\cdot\vec\sigma\bigr)\,\sigma_2\Bigr]_{-1/2,1/2}\,\chi^{(1/2)}&= -\,\dfrac{i}{{4\pi}}\,\hat k^{(0)}_\lambda\,\bigl(\sigma_3\,\sigma_2\bigr)_{-1/2,1/2}\,\chi^{(1/2)}\qquad\text{see appendix~\ref{ann:standardbasis}}\\[2mm]
    &= -\,\dfrac{i}{{4\pi}}\,\hat k^{(0)}_\lambda\,\bigl(\sigma_3\,\sigma_2\bigr)_{2,1}\,\chi^{(1/2)}\qquad\text{see table in appendix~\ref{ann:clebsch}}\\[2mm]
    &= -\,\dfrac{1}{{4\pi}}\,\hat k^{(0)}_\lambda\,\bigl(\sigma_1\bigr)_{2,1}\,\chi^{(1/2)}\qquad\text{since $\sigma_3\,\sigma_2=-i\sigma_1$}\\[2mm]
    &= -\,\dfrac{1}{{4\pi}}\,\hat k^{(0)}_\lambda\,\chi^{(1/2)}\qquad\text{because $\sigma_1=\bigl(\begin{smallmatrix}0&1\\1&0\end{smallmatrix}\bigr)$}
\end{align*}
    \item{$\vartriangleright$ {$n=1$ term}: }this second term now is
\begin{align*}
     \dfrac{i}{{4\pi}}\,\Bigl({\vec e}^{\,(+1)}\cdot\,\hat k_\lambda\Bigr)\,\Bigl[\bigl({\vec e}^{\,(+1)}\cdot\vec\sigma\bigr)\,\sigma_2\Bigr]_{1/2,1/2}\,\chi^{(-1/2)}&= \dfrac{i}{{4\pi}}\,\hat k^{(+1)}_\lambda\,\Bigl[\bigl({\vec e}^{\,(+1)}\cdot\vec\sigma\bigr)\,\sigma_2\Bigr]_{1,1}\,\chi^{(-1/2)}
\end{align*}
However:
\begin{multline*}
    \bigl({\vec e}^{\,(+1)}\cdot\vec\sigma\bigr)\,\sigma_2 = -\dfrac{1}{\sqrt{2}}\bigl(\sigma_1 + i\,\sigma_2\bigr)\sigma_2 = -\dfrac{1}{\sqrt{2}}\bigl(\sigma_1\sigma_2+ i\,\sigma_2^2\bigr) = -\dfrac{i}{\sqrt{2}}\bigl(\sigma_3+ \mathbf{1}\bigr) = -\dfrac{i}{\sqrt{2}}\Bigl(\begin{smallmatrix}2&0\\[1mm]0&0\end{smallmatrix}\Bigr)\\[2mm]
    \Longrightarrow\qquad\Bigl[\bigl({\vec e}^{\,(+1)}\cdot\vec\sigma\bigr)\,\sigma_2\Bigr]_{1,1} = -i\,\sqrt{2}
\end{multline*}
which leads to:
\[
    \dfrac{i}{{4\pi}}\,\Bigl({\vec e}^{\,(+1)}\cdot\,\hat k_\lambda\Bigr)\,\Bigl[\bigl({\vec e}^{\,(+1)}\cdot\vec\sigma\bigr)\,\sigma_2\Bigr]_{1/2,1/2}\,\chi^{(-1/2)} = \dfrac{1}{{4\pi}}\,\sqrt{2}\,\hat k^{(+1)}_\lambda\,\chi^{(-1/2)}
\]
    \item{$\vartriangleright$ {Summary}: }the case $M=1/2$ involves both $\chi^{(1/2)}$ and $\chi^{(-1/2)}$. It would be more appropriate to have only $\chi^{(1/2)}$. In order to achieve this, we will use the ladder operators described in the appendix~\ref{ann:ladder}. We have indeed:
\[
    \chi^{(-1/2)} = \sigma^{(-)}\chi^{(1/2)}\qquad\text{as well as}\qquad\chi^{(1/2)} = \sigma^{(0)}\chi^{(1/2)}
\]
To summarize:
\begin{gather*}
\text{$M=1/2$, $n=0$ term:}\qquad-\,\dfrac{1}{{4\pi}}\,\hat k^{(0)}_\lambda\,\sigma^{(0)}\,\chi^{(1/2)}\\[2mm]
\text{$M=1/2$, $n=1$ term:}\qquad\dfrac{1}{{4\pi}}\,\sqrt{2}\,\hat k^{(+1)}_\lambda\,\sigma^{(-)}\,\chi^{(1/2)}
\end{gather*}
so that we finally obtain:
\[
    {\mathscr A}^{1/2} = -\,\dfrac{1}{{4\pi}}\,\Bigl[\hat k^{(0)}_\lambda\,\sigma^{(0)} - \sqrt{2}\,\hat k^{(+1)}_\lambda\,\sigma^{(-)}\Bigr]\,\chi^{(1/2)}
\]
\end{description}
\item{\underline{$M=-1/2$ case}:} in this second situation, $M-\mu$ can take the values 0 and $-1$.\\
So the new contribution we need to compute reads (setting now $n=\mu - M$):
\[
    {\mathscr A}^{-1/2} = \dfrac{i}{{4\pi}}\,\sum\limits_{n=0}^1(-)^{n}\Bigl({\vec e}^{\,(-n)}\cdot\,\hat k_\lambda\Bigr)\,\Bigl[\bigl({\vec e}^{\,(-n)}\cdot\vec\sigma\bigr)\,\sigma_2\Bigr]_{1/2-n,-1/2}\,\chi^{(n-1/2)}
\]
By following the same line of reasoning as in the previous case, we obtain:
\[
    {\mathscr A}^{-1/2} = -\,\dfrac{1}{{4\pi}}\,\Bigl[\hat k^{(0)}_\lambda\,\sigma^{(0)} + \sqrt{2}\,\hat k^{(-1)}_\lambda\,\sigma^{(+)}\Bigr]\,\chi^{(-1/2)}
\]
where the ladder operator $\sigma^{(+)}$ is used.
\item{\underline{Conclusion}:} the ladder operators also satisfy
\[
    \sigma^{(+)}\chi^{(1/2)} = 0\qquad\text{and}\qquad\sigma^{(-)}\chi^{(-1/2)} = 0
\]
Hence, the expressions of ${\mathscr A}^{1/2}$ and ${\mathscr A}^{-1/2}$ can be combined into a single one:
\[
    {\mathscr A}^{M} = -\,\dfrac{1}{{4\pi}}\,\Bigl[\hat k^{(0)}_\lambda\,\sigma^{(0)} + \sqrt{2}\,\hat k^{(-1)}_\lambda\,\sigma^{(+)} - \sqrt{2}\,\hat k^{(+1)}_\lambda\,\sigma^{(-)}\Bigr]\,\chi^{(M)}\qquad\quad\text{with}\quad M=\pm1/2
\]
We now needs to simplify the term in brackets. Recalling the following definitions
\[
\left\{
\begin{aligned}
\ &\hat k^{(0)} = \hat k_3\qquad;\qquad\hat k^{(+1)} = -\dfrac{1}{\sqrt{2}}\bigl(\hat k_1 + i\,\hat k_2\bigr)\qquad;\qquad\hat k^{(-1)} = \dfrac{1}{\sqrt{2}}\bigl(\hat k_1 - i\,\hat k_2\bigr)\qquad\text{(see appendix~\ref{ann:standardbasis})}\\[2mm]
\ &\sigma^{(0)} = \sigma_3\qquad;\qquad\sigma^{(+)} = \dfrac{1}{{2}}\bigl(\sigma_1 + i\,\sigma_2\bigr)\qquad;\qquad\sigma^{(-)} = \dfrac{1}{{2}}\bigl(\sigma_1 - i\,\sigma_2\bigr)\qquad\text{(see appendix~\ref{ann:ladder})}
\end{aligned}
\right.
\]
we get by simply expanding the bracket term:
\[
    \hat k^{(0)}_\lambda\,\sigma^{(0)} + \sqrt{2}\,\hat k^{(-1)}_\lambda\,\sigma^{(+)} - \sqrt{2}\,\hat k^{(+1)}_\lambda\,\sigma^{(-)} = \sum\limits_{i=1}^3 (\hat k_\lambda)_i\,\sigma_i = \hat k_\lambda\cdot\vec\sigma
\]
Finally, the desired spin and orbital angular momentum part is written as:
\[
{\mathscr T} = -\dfrac{1}{{4\pi}}\,\hat k_\lambda\cdot\vec\sigma\,\dfrac{i}{\sqrt{2}}\,\bigl(\sigma_2\bigr)_{s_2s_3}\,\chi^{(M)}
\]
which is the term found in the table~\ref{fig:pair}.
\end{maliste}
\subsection{Coefficients $\boldsymbol{c_{n;\pm 1}}$}\label{ann:lesCl}
This appendix descibes the method used to compute the coefficients ${c_{n;\pm 1}}$ defined in section~\ref{para:lesCl}.
\subsubsection{Method basics}
To determine the $c_{n;\pm 1}$, we will consider for example the modes $\rho^n\lambda^{n\pm 1}$ with $n$ even (and therefore $j=1$). The idea is to work on the $(0)$ component (the one proportional to $\sigma^{(0)}$) and to locate the coefficients of the terms in $x^{n-1}$ where $x$ is the scalar product $\hat k_\rho\cdot\hat k_\lambda$. It will then be enough to relate these coefficients to the corresponding~$t_n$.\par
This locating can be done in the following way: one starts from~\eqref{eq:pair} choosing $M=1/2$. The term proportional to $\sigma^{(0)}$ is then obtained by taking $m'=0$\footnote{The other possible value $m'=1$ generates a term of the type $\chi^{(-1/2)}$, proportional to $\sigma^{(-1)}$ because $\sigma^{(-1)}\chi^{(1/2)} = \sqrt{2}\,\chi^{(-1/2)}$} :
\begin{multline*}
\underbracket[0.5pt]{\CG{1,\,0}{1/2,\,1/2}{1/2,\,1/2}}_{-\frac{1}{\sqrt{3}}}\,
\underbracket[0.5pt]{\CG{1/2,\,s_2}{1/2,\,s_3}{0,\,0}}_{\frac{i}{\sqrt{2}}\,(\sigma_2)_{s_2,s_3}}\,
\sum\limits_{m}
\underbracket[0.5pt]{\CG{0,\,0}{1,\,0}{1,\,0}}_{1}\,\\[2mm]
\times\,\CG{n,\, m}{n\pm 1,\, - m}{1,\,0}
\,Y_{n}^{(m)}(\hat k_\rho)\,
Y_{n\pm 1}^{(-m)}(\hat k_\lambda)\,\chi^{(1/2)}
\end{multline*}
because $\sigma^{(0)}\chi^{(1/2)} = \chi^{(1/2)}$. The initial structure to handle is then:
\[
{\mathscr T}_{n;\pm 1}\ =\ 
-\dfrac{1}{\sqrt{3}}\,\sum\limits_{m}\CG{n,\, m}{n\pm 1,\, - m}{1,\,0}
\,Y_{n}^{(m)}(\hat k_\rho)\,
Y_{n\pm 1}^{(-m)}(\hat k_\lambda)
\]
where the spherical harmonics are expressed using Clebsch-Gordan coefficients, the normalization factors $N^{(n)}$ and the standard coordinates ${\hat k^{(n)}}$. The Clebsch-Gordan coefficients which enter the $Y^{(m)}_n$ are all of the form:
\[
\CGY{n_1,\,m_1}{1,\,m_2}{n_1+1,\,\mu}
\]
and satisfy the symmetry properties (often used in the following calculations):
\begin{align*}
\CGY{n_1,\,m_1}{1,\,m_2}{n_1+1,\,\mu}&\ =\ \CGY{1,\,m_2}{n_1,\,m_1}{n_1+1,\,\mu}\\
&\ =\ \CGY{n_1,\,-m_1}{1,\,-m_2}{n_1+1,\,-\mu}
\end{align*}
The index $Y$ indicates that this coefficient originates from the construction of the spherical harmonics $Y_n^{(m)}$.\par
In ${\mathscr T}_{n;\pm 1}$, one has to exhibit a first polynomial in $\hat k_\rho\cdot\hat k_\lambda$ that mutiplies ${\hat k_\rho^{(0)}}$ and a second polynomial in $\hat k_\rho\cdot\hat k_\lambda$ that multiplies ${\hat k_\lambda^{(0)}}$. These terms are given by a particular value of $m$.\par\bigskip
As an explicit example, let us consider the mode $\rho^2\lambda^3$ (${\mathscr T}_{2;+1}$). This case involves, in total, two momenta $\hat k_\rho$ and three momenta $\hat k_\lambda$. Therefore, the polynomial proportional to ${\hat k_\rho^{(0)}}$ has its higher order in $(\hat k_\rho\cdot\hat k_\lambda)(\hat k_\lambda\cdot\hat k_\lambda)$. But since the scalar products read:
\[
\left\{
\begin{aligned}
\ \hat k_\rho\cdot\hat k_\lambda&\ =\ \hat k_\rho^{(0)}\hat k_\lambda^{(0)} - \hat k_\rho^{(-1)}\hat k_\lambda^{(+1)} - \hat k_\rho^{(+1)}\hat k_\lambda^{(-1)}\\[2mm]
\ \hat k_\lambda\cdot\hat k_\lambda&\ =\ \hat k_\rho^{(0)}\hat k_\lambda^{(0)} - 2\,\hat k_\lambda^{(-1)}\hat k_\lambda^{(+1)}\ (=\ 1)
\end{aligned}
\right.
\]
a possible discriminating term to isolate is $\hat k_\rho^{(0)}\hat k_\rho^{(+1)}\hat k_\lambda^{(+1)}\bigl(\hat k_\lambda^{(-1)}\bigr)^2$, which can be obtained with $m=1$. Explicitly one extracts from ${\mathscr T}_{2;+1}$ the term:
\begin{multline*}
-\dfrac{1}{\sqrt{3}}\,N^{(2)}\,N^{(3)}\,\CG{2,\,1}{3,\,-1}{1,\,0}\,\Bigl[
\CGY{1,\,0}{1,\,1}{2,\,1} + \CGY{1,\,1}{1,\,0}{2,\,1}
\Bigr]\\
\times
\Bigl[
\CGY{2,\,2}{1,\,-1}{3,\,1} + \CGY{2,\,0}{1,\,1}{3,\,1}\,\Bigl(
\CGY{1,\,1}{1,\,-1}{2,\,0} + \CGY{1,\,-1}{1,\,1}{2,\,0}
\Bigr)
\Bigr](-1)\left(-\dfrac12\right)
\end{multline*}
where the coefficients $(-1)$ and $(-1/2)$ come from the fact that one $\hat k_\rho^{(+1)}\hat k_\lambda^{(-1)}$ identifies one $-(\hat k_\rho\cdot\hat k_\lambda)$ according to the expression of the scalar products, and one $\hat k_\lambda^{(+1)}\hat k_\lambda^{(-1)}$ identifies $-(\hat k_\lambda\cdot\hat k_\lambda)/2$.\par
In a similar way, the polynomial proportional to ${\hat k_\lambda^{(0)}}$ has its highest order in $(\hat k_\rho\cdot\hat k_\lambda)^2$. In this case, the discriminating coefficient to guess is $\hat k_\lambda^{(0)}\bigl(\hat k_\lambda^{(-1)}\bigr)^2\bigl(\hat k_\rho^{(+1)}\bigr)^2$ which is obtained from the term $m=2$. We then obtain:
\begin{multline*}
-\dfrac{1}{\sqrt{3}}\,N^{(2)}\,N^{(3)}\,\CG{2,\,2}{3,\,-2}{1,\,0}\\
\times
\Bigl[
\CGY{2,\,2}{1,\,0}{3,\,2} + \CGY{2,\,1}{1,\,1}{3,\,2}\,\Bigl(
\CGY{1,\,1}{1,\,0}{2,\,1} + \CGY{1,\,0}{1,\,1}{2,\,1}
\Bigr)
\Bigr](-1)^2
\end{multline*}
\subsubsection{Recurrence relations}
Let us focus here only on the relations between Clebsch-Gordan coefficients coming from the spherical harmonics for the modes $\rho^n\lambda^{n\pm 1}$ with $n$ even. We denote by $\alpha_{n;\pm 1}$ the coefficient of the higher power of $(\hat k_\lambda\cdot\hat k_\lambda)$ for the ${\hat k_\rho^{(0)}}$ term and by $\beta_{n;\pm 1}$ the one for the ${\hat k_\lambda^{(0)}}$ term. By applying the previous method to the explicit expressions of the modes $\rho^n\lambda^{\pm 1}$, we obtain the relations which give us $\alpha_{n;\pm 1}$ and $\beta_{n;\pm 1}$ for $n = 0,2,4$.\par 
Moreover, by comparing the structures calculated for even and odd $n$, we notice that these coefficients satisfy the additional properties:
\[
\alpha_{n;-1}\ =\ \beta_{n-1;+1}
\qquad\text{and also}\qquad
\beta_{n;-1}\ =\ \alpha_{n-1;+1}
\]
Therefore, we can group the relations into the following two blocks:
\begin{align*}
\ \beta_{0;+1}&\ =\ 1\\[2mm]
\ \beta_{1;+1}&\ =\ \CGY{1,\,1}{1,\,0}{2,\,1} + \CGY{1,\,0}{1,\,1}{2,\,1}\,\beta_{0;+1}\\[2mm]
\ \beta_{2;+1}&\ =\ \CGY{2,\,2}{1,\,0}{3,\,2} + \CGY{2,\,1}{1,\,1}{3,\,2}\,\beta_{1;+1}\\[2mm]
\ \beta_{3;+1}&\ =\ \CGY{3,\,3}{1,\,0}{4,\,3} + \CGY{3,\,2}{1,\,1}{4,\,3}\,\beta_{2;+1}
\end{align*}
and
\begin{align*}
\ \alpha_{0;+1}&\ =\ 0\\[2mm]
\ \alpha_{1;+1}&\ =\ \CGY{1,\,-1}{1,\,1}{2,\,0} + \CGY{1,\,1}{1,\,-1}{2,\,0}\\[2mm]
\ \alpha_{2;+1}&\ =\ \dfrac{\beta_{1;+1}}{\beta_{0;+1}}\Bigl[\CGY{2,\,2}{1,\,-1}{3,\,1}\,\beta_{0;+1} + \CGY{2,\,0}{1,\,1}{3,\,1}\,\alpha_{1;+1}\Bigr]\\[2mm]
\ \alpha_{3;+1}&\ =\ \dfrac{\beta_{2;+1}}{\beta_{1;+1}}\Bigl[\CGY{3,\,3}{1,\,-1}{4,\,2}\,\beta_{1;+1} + \CGY{3,\,1}{1,\,1}{4,\,2}\,\alpha_{2;+1}\Bigr]
\end{align*}
Two sequences then emerge:
\[
\left\{
\begin{aligned}
\beta_{n;+1}&\ =\ \CG{n,\,n}{1,\,0}{n+1,\,n} + \CG{n,\,n-1}{1,\,1}{n+1,\,n}\,\beta_{n-1;+1}\\[2mm]
\alpha_{n;+1}&\ =\ \dfrac{\beta_{n-1;+1}}{\beta_{n-2;+1}}\,\Bigl[\CG{n,\,n}{1,\,-1}{n+1,\,n-1}\,\beta_{n-2;+1} + \CG{n,\,n-2}{1,\,1}{n+1,\,n-1}\,\alpha_{n-1;+1}\Bigr]
\end{aligned}
\right.
\]
which, using the explicit expressions of the Clebsch-Gordan coefficients, can be rewritten as:
\[
\left\{
\begin{aligned}
\beta_{n;+1}&\ =\ \dfrac{1}{\sqrt{n+1}}\ +\ \sqrt{\dfrac{n}{n+1}}\,\beta_{n-1;+1}
\qquad\text{with}\qquad\beta_{0;+1}=1\\[2mm]
\alpha_{n;+1}&\ =\ \dfrac{\beta_{n-1;+1}}{\beta_{n-2;+1}}\,\left[\dfrac{1}{\sqrt{(n+1)(2n+1)}}\,\beta_{n-2;+1}\ +\  \sqrt{\dfrac{n(2n-1)}{(n+1)(2n+1)}}\,\alpha_{n-1;+1}\right]
\qquad\text{with}\qquad\alpha_{1;+1}=\sqrt{\dfrac23}
\end{aligned}
\right.
\]
The generic expression of these coefficients is then deduced by recurrence:
\[
{\beta_{n;+1}\ =\ \sqrt{n+1}}
\qquad\text{and}\qquad
{\alpha_{n;+1}\ =\ \sqrt{\dfrac{n(n+1)}{2n+1}}}
\]
while the remaining coefficients are given by:
\[
{\beta_{n;-1}\ =\ \alpha_{n-1;+1}\ =\ \sqrt{\dfrac{n(n-1)}{2n-1}}}
\qquad\text{as well as}\qquad
{\alpha_{n;-1}\ =\ \beta_{n-1;+1}\ =\ \sqrt{n}}
\]
\subsubsection{Extraction of the complete coefficients}
One must begin by taking into account the multiplicative factors $(-1)$ et $(-1/2)$. Looking at the structures gathered in the tables, one can notice that:
\begin{order}
\item{\underline{$n$ even}:} in the case $[n;-1]$, there is a factor $(-)^{n-1}$ for the $\alpha_{n;-1}$ and a factor $(-)^{n-1}/2$ for the coefficients $\beta_{n;-1}$ ; in the case $[n;+1]$, the factor for $\alpha_{n;+1}$ is $(-)^n/2$ while the one for $\beta_{n;+1}$ has the value $(-)^{n}$.
\item{\underline{$n$ odd}:} we have exactly the same situation
\end{order}
As a consequence, starting from the definition of ${\mathscr T}_{n;\pm 1}$, the complete coefficients $A$ and $B$ to be used are:
\[
\left\{
\begin{aligned}
\ A_{n;-1}&\ =\ -\dfrac{1}{\sqrt{3}}\,(-)^{n-1}\,N^{(n)}\,N^{(n-1)}\,\CG{n,\,n-1}{n-1,\,1-n}{1,\,0}\,\alpha_{n;-1}\\[2mm]
\ A_{n;+1}&\ =\ -\dfrac{1}{\sqrt{3}}\,\dfrac{(-)^{n}}{2}\,N^{(n)}\,N^{(n+1)}\,\CG{n,\,n-1}{n+1,\,1-n}{1,\,0}\,\alpha_{n;+1}\\[2mm]
\ B_{n;-1}&\ =\ -\dfrac{1}{\sqrt{3}}\,\dfrac{(-)^{n-1}}{2}\,N^{(n)}\,N^{(n-1)}\,\CG{n,\,n-2}{n-1,\,2-n}{1,\,0}\,\beta_{n;-1}\\[2mm]
\ B_{n;+1}&\ =\ -\dfrac{1}{\sqrt{3}}\,{(-)^{n}}\,N^{(n)}\,N^{(n+1)}\,\CG{n,\,n}{n+1,\,-n}{1,\,0}\,\beta_{n;+1}
\end{aligned}
\right.
\]
which become, by substituting the Clebsch-Gordan coefficients by their actual expressions (see~\ref{ann:clebsch}):
\[
\left\{
\begin{aligned}
\ A_{n;-1}&\ =\ (-)^{n}\,N^{(n)}\,N^{(n-1)}\,\dfrac{1}{\sqrt{2n+1}}\\[2mm]
\ A_{n;+1}&\ =\ (-)^{n}\,N^{(n)}\,N^{(n+1)}\,\dfrac{n}{2n+1}\,\dfrac{1}{\sqrt{2n+3}}\\[2mm]
\ B_{n;-1}&\ =\ (-)^{n+1}\,N^{(n)}\,N^{(n-1)}\,\dfrac{n-1}{2n-1}\,\dfrac{1}{\sqrt{2n+1}}\\[2mm]
\ B_{n;+1}&\ =\ (-)^{n+1}\,N^{(n)}\,N^{(n+1)}\,\dfrac{1}{\sqrt{2n+3}}
\end{aligned}
\right.
\]
\subsubsection{The coefficients $\boldsymbol{c_{n;\pm 1}}$}
Finally, we aim to express ${\mathscr T}_{n;\pm 1}$ in the form:
\[
{\mathscr T}_{n;\pm 1}\ =\ a_{n;\pm 1}\,\dfrac{P_n^1(x)}{\sqrt{1-x^2}}\,{\hat k_\rho}\ +\ b_{n;\pm 1}\,\dfrac{P_{n\pm 1}^1(x)}{\sqrt{1-x^2}}\,{\hat k_\lambda}
\qquad\text{where}\qquad
x\ =\ {\hat k_\rho}\cdot{\hat k_\lambda}
\]
in terms of the associated Legendre polynomials. Consequently, the coefficients $a_{n;\pm 1}$ et $b_{n;\pm 1}$ are related to the coefficients $A_{n;\pm 1}$ and $B_{n;\pm 1}$ according to:
\[
A_{n;\pm 1}\ =\ a_{n;\pm 1}\,t_n
\qquad\text{and}\qquad
B_{n;\pm 1}\ =\ b_{n;\pm 1}\,t_{n\pm 1}
\]
Since the $A_{n;\pm 1}$, the $B_{n;\pm 1}$ and the $t_n$ are known, it follows that:
\[
a_{n;-1}\ =\ -\,b_{n;-1}\ =\ \dfrac{(-)^{n+1}}{4\pi\,\sqrt{n}}
\qquad;\qquad
a_{n;+1}\ =\ -\,b_{n;+1}\ =\ \dfrac{(-)^{n+1}}{4\pi\,\sqrt{n+1}}
\]
Finally, if we define:
\[
{c_{n;-1}\ =\ \dfrac{(-)^{n+1}}{4\pi\,\sqrt{n}}}
\qquad\text{and}\qquad
{c_{n;+1}\ =\ \dfrac{(-)^{n+1}}{4\pi\,\sqrt{n+1}}}
\]
we arrive at the following expression:
\[
{
{\mathscr T}_{n;\pm 1}\ =\ c_{n;\pm 1}\,\left[\dfrac{P_n^1(x)}{\sqrt{1-x^2}}\,{\hat k_\rho}\ -\ \dfrac{P_{n\pm 1}^1(x)}{\sqrt{1-x^2}}\,{\hat k_\lambda}\right]
}
\]
\subsection{Jacobi coordinates}\label{ann:jacobi}
The usual method to construct the set of Jacobi coordinates for a system of three points located at $\vec r_i$ and having the masses $m_i$ is:
\begin{orderalph}
\item one chooses two arbitrary points (for example $\vec r_2$ and $\vec r_3$) which defines the first coordinate $\vec \rho = \vec r_2 - \vec r_3$
\item one then introduces the barycenter of the two points used to define $\vec\rho$ and the vector which links this barycenter to the third point defines the second coordinate $\vec\lambda = \dfrac{m_2\vec r_2 + m_3\vec r_3}{m_2+m_3} - \vec r_1$
\item the third coordinate is the position of the barycenter of the complete system $\vec R = \dfrac{m_1\vec r_1 +m_2\vec r_2 + m_3\vec r_3}{m_1+m_2+m_3}$
\end{orderalph}
In our study the following convention is taken. The index 1 refers to the heavy quark (mass $m_1$) while the indices 2 and 3 refer to the light quarks (having the same mass $m_2=m_3=m$).\par
Therefore, this construction defines the following coordinates:
\[
\vec R = \dfrac{m_1\vec r_1 +m\bigl(\vec r_2 + \vec r_3\bigr)}{m_1+2m}
\qquad;\qquad
\vec \rho = \vec r_2 - \vec r_3
\qquad;\qquad
\vec\lambda = \dfrac12\bigl(\vec r_2 +\vec r_3 - 2\,\vec r_1\bigr)
\]
For simplification purposes, it will be more convenient to adopt as Jacobi coordinates:
\[
{
\vec R = \dfrac{m_1\vec r_1 +m\bigl(\vec r_2 + \vec r_3\bigr)}{m_1+2m}
\qquad;\qquad
\vec \rho = \vec r_2 - \vec r_3
\qquad;\qquad
\vec\lambda = \vec r_2 +\vec r_3 - 2\,\vec r_1
}
\]
We then deduce the corresponding Jacobi momenta:
\[
{
\left\{
\begin{aligned}
\vec p_{R} &= \vec p_1 + \vec p_2 + \vec p_3\\[2mm]
\vec p_\rho &= \dfrac{1}{2}\,\bigl(\vec p_2 - \vec p_3\bigr)\\[2mm]
\vec p_\lambda &= \dfrac12\,\dfrac{1}{m_1+2m}\,\bigl(
m_1\vec p_2 + m_1\vec p_3
- 2m\vec p_1
\bigr)
\end{aligned}
\right.
}
\]
Finally, the wave functions $\varphi(\vec k_2,\,\vec k_3)$ which appear in the Bakamjian-Thomas matrix elements are in fact the internal wave functions of the quark system expressed in the rest frame $\vec p_{R} = \vec 0$. With this constraint, the Jacobi coordinates become
\[
{
\left\{
\begin{aligned}
\vec p_\rho &= \dfrac{1}{2}\,\bigl(\vec p_2 - \vec p_3\bigr)\\[2mm]
\vec p_\lambda &= \dfrac12\,\bigl(\vec p_2 + \vec p_3\bigr) = -\dfrac12\,\vec p_1
\end{aligned}
\right.
}
\]
\begin{maliste}
    \item{\underline{Generalization to four-vectors}:}\par
    One generalizes the previous transformation in terms of four-vectors according to 
    \begin{equation}\label{eq:definition}
    {
     p_\rho = \dfrac{1}{2}\,\bigl( p_2 -  p_3\bigr)
    }
    \qquad\text{and}\qquad
    {
     p_\lambda = \dfrac12\,\bigl( p_2 +  p_3\bigr)
    }
    \end{equation}
    It can be inverted to give
    \[
    {
     p_2 =  p_\lambda +  p_\rho
    }
    \qquad\text{and}\qquad
    {
     p_3 =  p_\lambda -  p_\rho
    }
    \]
   An important remark: although $p_2$ and $ p_3$ are on the mass shell ($m^2$), $p_\lambda$ and $p_\rho$ are not on the mass shell
    \begin{gather*}
    { p_\lambda\cdot p_\lambda = (p^o_\lambda)^2 - (\vec p_\lambda)^2 = \dfrac12\bigl(m^2 + p_2\cdot p_3\bigr)}
    \qquad\text{and also}\qquad
    { p_\rho\cdot p_\rho = (p^o_\rho)^2 - (\vec p_\rho)^2 = \dfrac12\bigl(m^2 - p_2\cdot p_3\bigr)}\\[2mm]
    \Longrightarrow\qquad{
        p_\lambda\cdot p_\lambda + p_\rho\cdot p_\rho = m^2
    }
    \end{gather*}
    \item{\underline{Relations between momenta} : }\par
    From what precedes, it follows 
    \[
    {
    \left\{
    \begin{aligned}
    \vec p_\rho = \dfrac{1}{2}\,\bigl(\vec p_2 - \vec p_3\bigr)\quad &; \quad p_\rho^{o} = \dfrac{1}{2}\,\bigl(p_2^o - p_3^o\bigr)\\[2mm]
    \ \vec p_\lambda = \dfrac{1}{2}\,\bigl(\vec p_2 + \vec p_3\bigr)\quad &; \quad p_\lambda^{o} = \dfrac{1}{2}\,\bigl(p_2^o + p_3^o\bigr)
    \end{aligned}
    \right.
    }\qquad
    {
    \left\{
    \begin{aligned}
    \ \vec p_2 = \vec p_\lambda + \vec p_\rho
    \quad &; \quad
    p_2^o = p_\lambda^{o} + p_\rho^{o}\\[2mm]
    \vec p_3 = \vec p_\lambda - \vec p_\rho
    \quad &; \quad
    p_3^o = p_\lambda^{o} - p_\rho^{o}
    \end{aligned}
    \right.
    }
    \]
    \item{\underline{Some useful relations} : }\par
    One has for example
    \begin{multline*}
    {
     p_2\cdot p_3 = p_\lambda\cdot p_\lambda - p_\rho\cdot p_\rho = \Bigl[(p^{o}_\lambda)^2 - (\vec p_\lambda)^2\Bigr]
    - \Bigl[(p^{o}_\rho)^2 - (\vec p_\rho)^2\Bigr]
    }\\[3mm]
    \qquad\text{and}\qquad
    {
     p_\lambda\cdot p_\rho = 0\quad\Longrightarrow\quad\vec p_\lambda\cdot\vec p_\rho = p^o_\lambda\,p^o_\rho
    }
    \end{multline*}
 as well as
    \[
    {
    \vec p_\lambda\cdot\vec p_\rho =  \dfrac14\,\bigl(\vecsq{p}_2 - \vecsq{p}_3\bigr) = \dfrac14\,\Bigl[(p^o_2)^2 - (p^o_3)^2\Bigr]
    }\qquad\text{and}\qquad
    {
    \vec p_2\cdot\vec p_3 = \vec p_\lambda^{\,2} - \vec p_\rho^{\,2}
    }
    \]
    This implies in particular:
    \begin{gather*}
        {
         p_2\cdot p_\rho = -\, p_3\cdot p_\rho =  p_\rho\cdot p_\rho =  \dfrac{1}{2}\,\bigl(m^2 -  p_2\cdot p_3\bigr)
        }\\[2mm]
        {
             p_2\cdot p_\lambda =  p_3\cdot p_\lambda =  p_\lambda\cdot p_\lambda =  \dfrac{1}{2}\,\bigl(m^2 +  p_2\cdot p_3\bigr)
        }
    \end{gather*}
    The Jacobian of this change of variables for the integrals is:
    \[
    {
    \dd\vec p_2\,\dd\vec p_3 = 8\,\dd\vec p_\lambda\,\dd\vec p_\rho
    }
    \]
    and we also have the following expressions:
    \[
    {
    \left\{
    \begin{aligned}
    \vec p_2^{\,2} &= \vec p_\lambda^{\,2} + \vec p_\rho^{\,2} + 2\,\vec p_\lambda\cdot\vec p_\rho\\[2mm]
    \vec p_3^{\,2} &= \vec p_\lambda^{\,2} + \vec p_\rho^{\,2} - 2\,\vec p_\lambda\cdot\vec p_\rho\\[2mm]
    \end{aligned}
    \right.
    }
    \]
    This allows to express the coefficients $p_2^o$, $p_3^o$, $p_\lambda^o$ and $p_\rho^o$ in terms of the vectors $\vec p_\lambda$ and $\vec p_\rho$ because:
    \[
    {
    \left\{
    \begin{aligned}
    p_2^o &= \sqrt{m^2 + \vec p_2^{\,2}}\\[2mm]
    p_3^o &= \sqrt{m^2 + \vec p_3^{\,2}}
    \end{aligned}
    \right.
    \qquad\text{and}\qquad
    \left\{
    \begin{aligned}
    p_\lambda^o &= \dfrac{1}{2}\,\bigl(p_2^o + p_3^o\bigr)\\[2mm]
    p_\rho^o &= \dfrac{1}{2}\,\bigl(p_2^o - p_3^o\bigr)\end{aligned}
    \right.
    }
    \]
    We find explicitely:
    \begin{gather*}
    {
        p_\lambda^o = \dfrac12\,\biggl\{
        \sqrt{m^2 + {\vec p_\lambda}^{\,2} + {\vec p_\rho}^{\,2} + 2\,\vec p_\lambda\cdot\vec p_\rho }  +
        \sqrt{m^2 + {\vec p_\lambda}^{\,2} + {\vec p_\rho}^{\,2} - 2\,\vec p_\lambda\cdot\vec p_\rho }
        \biggr\}
    }\\[2mm]
    {
        p_\rho^o = \dfrac{1}{2}\,\biggl\{
        \sqrt{m^2 + {\vec p_\lambda}^{\,2} + {\vec p_\rho}^{\,2} + 2\,\vec p_\lambda\cdot\vec p_\rho }  -
        \sqrt{m^2 + {\vec p_\lambda}^{\,2} + {\vec p_\rho}^{\,2} - 2\,\vec p_\lambda\cdot\vec p_\rho }
        \biggr\}
    }
    \end{gather*}
\end{maliste}
\subsection{Bakamjian-Thomas normalization of the states}\label{ann:normalisation}
We recall here that, in the expression of the transition amplitudes, the baryon states are normalized according to the non-relativistic normalization:
\[
\ket{\Psi(p)}\qquad\text{such that}\qquad
\braket{\Psi(p^\prime)}{\Psi(p)}\ =\ (2\pi)^3\,\delta^{(3)}({\vec p}^{\,\prime} - \vec p)
\]
while the usual relativistic normalization reads:
\[
\ket{\Psi(p)}_{R}\qquad\text{with}\qquad
{}_R\braket{\Psi(p^\prime)}{\Psi(p)}_R\ =\ (2\pi)^3\,2p^0\,\delta^{(3)}({\vec p}^{\,\prime} - \vec p)
\]
The relationship between both types of normalized states is therefore:
\[
{\ket{\Psi(p)}\ =\ \dfrac{1}{\sqrt{2p^0}}\,\ket{\Psi(p)}_R\ =\ \dfrac{1}{\sqrt{2\,m_\Psi}}\,\dfrac{1}{\sqrt{v^0}}\,\ket{\Psi(p)}_R}
\]
in the inifinite mass limit.
\subsection{Definition of the Isgur-Wise function $\boldsymbol{\sigma_\Lambda(w)}$}\label{ann:isgur}
The definition of the Isgur-Wise function $\sigma_\Lambda(w)$ is given in terms of the baryon states normalized according to the usual relativistic normalization:
\[\dfrac{{}_R\brakket{\Psi^\prime,\,m^\prime}{J}{\Psi,\,m}_R}{\sqrt{m_\Psi\,m_{\Psi^\prime}}}\ =\ \dfrac{\sigma_\Lambda(w)}{\sqrt{3}}\,\left\{
\bar\chi^{(m')}_{ v^{\prime}}\,
\gamma_5\,(\fmslash{ v} + w)
\,J\,
\chi^{(m)}_{ v}
\right\}
\]
But since in the BT formalism the states are normalized non-relativistically, we need to use the relationship given in the appendix~\ref{ann:normalisation}:
\[
\ket{\Psi(p)}\ =\ \dfrac{1}{\sqrt{2p^0}}\,\ket{\Psi(p)}_R\ =\ \dfrac{1}{\sqrt{2\,m_\Psi}}\,\dfrac{1}{\sqrt{v^0}}\,\ket{\Psi(p)}_R
\]
so that the definition of the Isgur-Wise function $\sigma_\Lambda(w)$ in the BT formalism becomes:
\begin{equation}\label{eq:IW}
{
\brakket{\Psi^\prime,\,m^\prime}{J}{\Psi,\,m}\ =\ \dfrac{1}{2\sqrt{3}}\,\dfrac{1}{\sqrt{v^o\,v^{\prime o}}}\,\sigma_\Lambda(w)
\,\left\{
\bar\chi^{(m')}_{ v^{\prime}}\,
\gamma_5\,(\fmslash{ v} + w)
\,J\,
\chi^{(m)}_{ v}
\right\}
}
\end{equation}
\subsection{Determination of the coefficients of the projectors}\label{ann:projecteur}
Starting from a covariant decomposition of the form:
\[
{\mathscr I}^{\mu\nu(\sigma)} = \sum\limits_id_i\,\Lambda_i^{\mu\nu(\sigma)}
\]
the goal is to find a projector ${\mathscr P}^{\mu\nu(\sigma)}_j$ such that:
\[
{\mathscr P}^{\mu\nu(\sigma)}_j\,{\mathscr I}_{\mu\nu(\sigma)} = d_j
\]
To this aim, we can decompose the projector on the same structure of covariant tensors as ${\mathscr I}^{\mu\nu(\sigma)}$ (they form a basis), that is, in other words, find the coefficients $\{c_i^j\}$ where:
\[
{\mathscr P}^{\mu\nu(\sigma)}_j = \sum\limits_ic_i^j\,\Lambda_i^{\mu\nu(\sigma)}
\]
by solving the matrix equation:
\[
\Biggl(
\Lambda_{i,\mu\nu(\sigma)}\,{\mathscr P}^{\mu\nu(\sigma)}_j
\Biggr) \ =\ \Bigl(d_j\Bigr)
\]
where $\Bigl(d_j\Bigr)$ is a column matrix consisting of zeros except on the $j^\text{th}$ row, which contains a 1.
\subsubsection*{Calculation of $\boldsymbol{d_5}$}
We start with:
\[
{\mathscr I}^{\mu\nu} = \sum\limits_id_i\,\Lambda_i^{\mu\nu} = d_1\,v^\mu\,v^\nu + d_2\,v^{\prime\mu}\,v^{\nu} + d_3\,v^{\mu}\,v^{\prime\nu} + d_4\,v^{\prime\mu}\,v^{\prime\nu} + d_5\,g^{\mu\nu}
\]
and we look for the projector:
\[
{\mathscr P}^{\mu\nu}_5 = \sum\limits_ic^5_i\,\Lambda_i^{\mu\nu} = c^5_1\,v^\mu\,v^\nu + c^5_2\,v^{\prime\mu}\,v^{\nu} + c^5_3\,v^{\mu}\,v^{\prime\nu} + c^5_4\,v^{\prime\mu}\,v^{\prime\nu} + c^5_5\,g^{\mu\nu}
\qquad\text{such that}\qquad
{\mathscr P}^{\mu\nu}_5{\mathscr I}_{\mu\nu} = d_5
\]
We will thus solve:
\[
\begin{pmatrix}
\Lambda_{1,\mu\nu}\,{\mathscr P}^{\mu\nu}_5\\
\Lambda_{2,\mu\nu}\,{\mathscr P}^{\mu\nu}_5\\
\Lambda_{3,\mu\nu}\,{\mathscr P}^{\mu\nu}_5\\
\Lambda_{4,\mu\nu}\,{\mathscr P}^{\mu\nu}_5\\
\Lambda_{5,\mu\nu}\,{\mathscr P}^{\mu\nu}_5
\end{pmatrix}
=
\begin{pmatrix}
	c^5_4 w^2+(c^5_2+c^5_3) w+c^5_1+c^5_5\\
	c^5_3 w^2+(c^5_1+c^5_4+c_5) w+c^5_2\\
	c^5_2 w^2+(c^5_1+c^5_4+c^5_5) w+c^5_3\\
	c^5_1 w^2+(c^5_2+c^5_3)w+c^5_4+c^5_5\\
	(c^5_2+c^5_3) w+c^5_1+c^5_4+4 c^5_5
\end{pmatrix}
= \begin{pmatrix}0\\ 0\\ 0\\ 0\\ 1\end{pmatrix}
\qquad\Longrightarrow\qquad
\left\{
\begin{aligned}
	c^5_1 &= \dfrac{1}{2\left(w^2-1\right)}\\[2mm]
	c^5_2 &= -\dfrac{w}{2\left(w^2-1\right)}\\[2mm]
	c^5_3 &= -\dfrac{w}{2 \left(w^2-1\right)}\\[2mm]
	c^5_4 &= \dfrac{1}{2\left(w^2-1\right)}\\[2mm]
	c^5_5 &= \frac{1}{2}
\end{aligned}
\right.
\]
Consequently:
\[
{
{\mathscr P}^{\mu\nu}_5 =
-\,\dfrac12\,\biggl[\dfrac{1}{1-w^2}\bigl(v_\mu\,v_\nu + v^\prime_\mu\,v^\prime_\nu\bigr) - \dfrac{w}{1-w^2}
\bigl(v_\mu\,v^{\prime}_\nu + v^{\prime}_\mu\,v_{\nu}\bigr) - g_{\mu\nu}\biggr]
}
\qquad\text{which satisfies}\qquad
{\mathscr P}^{\mu\nu}_5{\mathscr I}_{\mu\nu} = d_5
\]
\subsubsection*{Calculation of $\boldsymbol{c_9 - c_{11}}$}
In this section, the name of the coefficients has changed from $\{d_j\}$ to $\{c_j\}$. Hence, the starting point is: 
\begin{align*}
    {\mathscr I}^{\mu\nu\sigma} =\begin{multlined}[t][145mm]
	c_1\,v^\mu\,v^\nu\,v^\sigma + c_2\,v'^\mu\,v^\nu\,v^\sigma + c_3\,v^\mu\,v'^\nu\,v^\sigma + c_4\,v'^\mu\,v'^\nu\,v^\sigma +c_5\,v^\mu\,v^\nu\,v'^\sigma\\[2mm]
+  c_6\,v'^\mu\,v^\nu\,v'^\sigma + c_7\,v^\mu\,v'^\nu\,v'^\sigma + c_8\,v'^\mu\,v'^\nu\,v'^\sigma + c_9\,g^{\mu\nu}v^\sigma + c_{10}\,g^{\mu\nu}\,v^{\prime\sigma}\\[2mm]
+ c_{11}\,g^{\mu\sigma}v^\nu + c_{12}\,g^{\mu\sigma}\,v^{\prime\nu} + c_{13}\,g^{\nu\sigma}v^\mu + c_{14}\,g^{\nu\sigma}\,v^{\prime\mu}
\end{multlined}
\end{align*}
We need to determine the two projectors ${\mathscr P}^{\mu\nu\sigma}_9$ et ${\mathscr P}^{\mu\nu\sigma}_{11}$ and compute their difference in order to get the solution to the problem. The method used remains the same:
\[
\left(
\begin{array}{c}
 c^j_8 w^3+(c^j_4+c^j_6+c^j_7) w^2+(c^j_2+c^j_3+c^j_5+c^j_{10}+c^j_{12}+c^j_{14}) w+c^j_1+c^j_9+c^j_{11}+c^j_{13} \\
 c^j_7 w^3+(c^j_3+c^j_5+c^j_8+c^j_{10}+c^j_{12}) w^2+(c^j_1+c^j_4+c^j_6+c^j_9+c^j_{11}+c^j_{13}) w+c^j_2+c^j_{14} \\
 c^j_6 w^3+(c^j_2+c^j_5+c^j_8+c^j_{10}+c^j_{14}) w^2+(c^j_1+c^j_4+c^j_7+c^j_9+c^j_{11}+c^j_{13}) w+c^j_3+c^j_{12} \\
 c^j_5 w^3+(c^j_1+c^j_6+c^j_7+c^j_{11}+c^j_{13}) w^2+(c^j_2+c^j_3+c^j_8+c^j_{10}+c^j_{12}+c^j_{14}) w+c^j_4+c^j_9 \\
 c^j_4 w^3+(c^j_2+c^j_3+c^j_8+c^j_{12}+c^j_{14}) w^2+(c^j_1+c^j_6+c^j_7+c^j_9+c^j_{11}+c^j_{13}) w+c^j_5+c^j_{10} \\
 c^j_3 w^3+(c^j_1+c^j_4+c^j_7+c^j_9+c^j_{13}) w^2+(c^j_2+c^j_5+c^j_8+c^j_{10}+c^j_{12}+c^j_{14}) w+c^j_6+c^j_{11} \\
 c^j_2 w^3+(c^j_1+c^j_4+c^j_6+c^j_9+c^j_{11}) w^2+(c^j_3+c^j_5+c^j_8+c^j_{10}+c^j_{12}+c^j_{14}) w+c^j_7+c^j_{13} \\
 c^j_1 w^3+(c^j_2+c^j_3+c^j_5) w^2+(c^j_4+c^j_6+c^j_7+c^j_9+c^j_{11}+c^j_{13}) w+c^j_8+c^j_{10}+c^j_{12}+c^j_{14} \\
 (c^j_6+c^j_7) w^2+(c^j_2+c^j_3+c^j_5+c^j_8+4 c^j_{10}+c^j_{12}+c^j_{14}) w+c^j_1+c^j_4+4 c^j_9+c^j_{11}+c^j_{13} \\
 (c^j_2+c^j_3) w^2+(c^j_1+c^j_4+c^j_6+c^j_7+4 c^j_9+c^j_{11}+c^j_{13}) w+c^j_5+c^j_8+4 c^j_{10}+c^j_{12}+c^j_{14} \\
 (c^j_4+c^j_7) w^2+(c^j_2+c^j_3+c^j_5+c^j_8+c^j_{10}+4 c^j_{12}+c^j_{14}) w+c^j_1+c^j_6+c^j_9+4 c^j_{11}+c^j_{13} \\
 (c^j_2+c^j_5) w^2+(c^j_1+c^j_4+c^j_6+c^j_7+c^j_9+4 c^j_{11}+c^j_{13}) w+c^j_3+c^j_8+c^j_{10}+4 c^j_{12}+c^j_{14} \\
 (c^j_4+c^j_6) w^2+(c^j_2+c^j_3+c^j_5+c^j_8+c^j_{10}+c^j_{12}+4 c^j_{14}) w+c^j_1+c^j_7+c^j_9+c^j_{11}+4 c^j_{13} \\
 (c^j_3+c^j_5) w^2+(c^j_1+c^j_4+c^j_6+c^j_7+c^j_9+c^j_{11}+4 c^j_{13}) w+c^j_2+c^j_8+c^j_{10}+c^j_{12}+4 c^j_{14} \\
\end{array}
\right)
=
\left(
\begin{array}{c}
 0 \\
 0 \\
 0 \\
 0 \\
 0 \\
 0 \\
 0 \\
 0 \\
 1 \\
 0 \\
 0 \\
 0 \\
 0 \\
 0 \\
\end{array}
\right)
\ \text{or}\ 
\left(
\begin{array}{c}
 0 \\
 0 \\
 0 \\
 0 \\
 0 \\
 0 \\
 0 \\
 0 \\
 0 \\
 0 \\
 1 \\
 0 \\
 0 \\
 0 \\
\end{array}
\right)
\]
After calculations, we find:
\begin{gather*}
{
{\mathscr P}^{\mu\nu\sigma}_{9-11} = 
    -\,\dfrac{1}{2}\,\dfrac{1}{1-w^2}\,
    \biggl[
    v^\nu\,\bigl(g^{\mu\sigma} - v^{\prime\mu}\,v^{\prime\sigma}\bigr) - v^\sigma\,\bigl(g^{\mu\nu} - v^{\prime\mu}\,v^{\prime\nu}\bigr)
    + w\,\bigl(g^{\mu\nu}v^{\prime\sigma} - g^{\mu\sigma}v^{\prime\nu}\bigr)
    \biggr]
}
\intertext{which satisfies:}
{\mathscr P}^{\mu\nu\sigma}_{9-11}{\mathscr I}_{\mu\nu\sigma} = c_9 - c_{11}
\end{gather*}
\subsection{Change of angular variables and integration}\label{ann:integration}
The integrals to be computed depend on two three-vectors $\vec a_1$ and $\vec a_2$ and the integrands are functions of the norms $a_i = \norm{\vec a_i}$ and of the scalar product $\vec a_1\cdot\vec a_2$. We would like to perform only the angular integration on these vectors (in spherical coordinates for example).\par\medskip
The first thing that seems sensible to do for the integrations is to find the directions of the $\vec a_i$ with respect to an axis $\vec e_z$ in spherical coordinates $(\theta_1,\,\phi_1,\,\theta_2,\,\phi_2)$, but the writing of $\vec a_1\cdot\vec a_2$ with these coordinates is involved (elliptic integrals will appear with a further difficulty because we are in a multidimensional situation \ldots).
\par\medskip
A second idea would be to use another way of finding the directions of the $\vec a_i$ : keep $(\theta_1,\,\phi_1)$ to find $\vec a_1$ relatively to $\vec e_z$ but find $\vec a_2$ with respect to $\vec a_1$ in spherical coordinates $(\theta_{12},\,\phi_{12})$. This would be equivalent if one begins by integrating on the $(\theta_{12},\,\phi_{12})$ then on the $(\theta_{1},\,\phi_{1})$. To go from $(\theta_1,\,\phi_1,\,\theta_{12},\,\phi_{12})$ to $(\theta_1,\,\phi_1,\,\theta_2,\,\phi_2)$, one performs a rotation of angle $-\theta_1$ around the axis $\vec e_{\phi_1}$ followed by a rotation of angle $-\phi_1$ around the axis $\vec e_z$. The change of coordinates is therefore given by:
\begin{equation}\label{eq:change}
    {
\left\{
\begin{aligned}
&\cos\theta_2 = \cos\theta_{12}\,\cos\theta_1 - \sin\theta_{12}\,\cos\phi_{12}\,\sin\theta_1\\
&\sin\theta_2\,\cos\phi_2 = (\cos\theta_{12}\,\sin\theta_1 + \sin\theta_{12}\,\cos\phi_{12}\,\cos\theta_1)\,\cos\phi_1 - \sin\theta_{12}\,\sin\phi_{12}\,\sin\phi_1\\
&\sin\theta_2\,\sin\phi_2 = (\cos\theta_{12}\,\sin\theta_1 + \sin\theta_{12}\,\cos\phi_{12}\,\cos\theta_1)\,\sin\phi_1 + \sin\theta_{12}\,\sin\phi_{12}\,\cos\phi_1
\end{aligned}
\right.
}
\end{equation}
Conversely, we have:
\[
\left\{
\begin{aligned}
&\cos\theta_{12} = \sin\theta_1\,\sin\theta_2\,\cos(\phi_1 - \phi_2) + \cos\theta_1\,\cos\theta_2\\
&\sin\theta_{12}\,\cos\phi_{12} = \cos\theta_1\,\sin\theta_2\,\cos(\phi_1 - \phi_2) - \cos\theta_2\,\sin\theta_1\\
&\sin\theta_{12}\,\sin\phi_{12} = \sin\theta_2\,\sin(\phi_2 - \phi_1)
\end{aligned}
\right.
\]
This allows to obtain two useful relations. The first one is obtained by multiplying the first equality by $\sin\theta_1$ and the second by $\cos\theta_1$ and adding the result. The second useful relation is just the third equality here above. In summary:
\begin{subequations}\label{eq:changebis}
\begin{gather}
{\sin\theta_2\,\cos(\phi_1 - \phi_2) = \sin\theta_1\,\cos\theta_{12} + \cos\theta_2\,\sin\theta_{12}\,\cos\phi_{12}}
\intertext{and}
	{
    \sin\theta_2\,\sin(\phi_1 - \phi_2) = -\, \sin\theta_{12}\,\sin\phi_{12}  
}
\end{gather}
\end{subequations}
\medskip
We consider now the angular integration on the angles $(\theta_{2},\,\phi_{2})$
\[
\dd\Omega = \alpha\,\dd\theta_2\,\dd\phi_2 = \alpha\,
\begin{vmatrix}
    \dfrac{\dr\theta_2}{\dr\theta_{12}}&\dfrac{\dr\theta_2}{\dr\phi_{12}}\\[3mm]
    \dfrac{\dr\phi_2}{\dr\theta_{12}}&\dfrac{\dr\phi_2}{\dr\phi_{12}}
\end{vmatrix}\,\dd\theta_{12}\,\dd\phi_{12} = \dfrac{1}{\alpha^2}\,
\underbrace{\begin{vmatrix}
    \alpha\,\dfrac{\dr\theta_2}{\dr\theta_{12}}&\alpha\,\dfrac{\dr\theta_2}{\dr\phi_{12}}\\[3mm]
    \alpha^2\,\dfrac{\dr\phi_2}{\dr\theta_{12}}&\alpha^2\,\dfrac{\dr\phi_2}{\dr\phi_{12}}
\end{vmatrix}}_{J}
\,\dd\theta_{12}\,\dd\phi_{12}
\qquad\text{where}\qquad
\alpha = \sin\theta_2
\]
We need to compute the determinant $J$. The particular form used will simplify the calculations that will follow. 
One begins by taking the derivative of the relations~\eqref{eq:change} with respect to the angles $\theta_{12}$ and $\phi_{12}$ keeping $\theta_1$ constant. The first equality gives
\begin{gather*}
    \sin\theta_2\,\dfrac{\dr\theta_2}{\dr\theta_{12}} = \cos\theta_1\,\sin\theta_{12} + \sin\theta_1\,\cos\phi_{12}\,\cos\theta_{12}\\[2mm]
\sin\theta_2\,\dfrac{\dr\theta_2}{\dr\phi_{12}} = -\, \sin\theta_1\,\sin\phi_{12}\,\sin\theta_{12}
\end{gather*}
while the two following provide two different expressions for each derivative of $\phi_{12}$ with respect to the angles $\theta_{12}$ and $\phi_{12}$. Combining these relations two by two and using the relations~\eqref{eq:changebis}, we obtain: 
\begin{gather*}
    \sin^2\theta_2\,\dfrac{\dr\phi_2}{\dr\theta_{12}} = \sin\theta_1\,\sin\phi_{12}\\[2mm]
    \sin^2\theta_2\,\dfrac{\dr\phi_2}{\dr\phi_{12}} = \sin\theta_1\,\sin\theta_{12}\,\cos\theta_{12}\,\cos\phi_{12} + \cos\theta_1\,\sin^2\theta_{12}
\end{gather*}
One can then compute directly the determinant $J$ which, after some trigonometric manipulations, reads:
\[
J = \sin\theta_{12}\,\biggl[1 - \bigl(\cos\theta_1\,\cos\theta_{12} - \sin\theta_1\,\cos\phi_{12}\,\sin\theta_{12}\bigr)^2\biggr]    
\]
From ~\eqref{eq:change}, we recognize, in the squared term, the $\cos\theta_2$.\par
Finally
\begin{equation*}
    J = \sin\theta_{12}\,\sin^2\theta_2
    \qquad\Longrightarrow\qquad
    {\dd\Omega = \sin\theta_2\,\dd\theta_2\,\dd\phi_2 =
    \sin\theta_{12}\,\dd\theta_{12}\,\dd\phi_{12}} 
\end{equation*}
which is a very simple result.\par\medskip
To summarize, the angular integrations will be done according to the following scheme:
\begin{equation}\label{eq:mesure}
{
    \int\dd\vec a_1\,\dd\vec a_2\ =\ 
    \int a_1^2\,\dd a_1\times a_2^2\,\dd a_2
    \int\sin\theta_1\,\dd\theta_1\,\dd\phi_1
    \int\sin\theta_{12}\,\dd\theta_{12}\,\dd\phi_{12}
}
\end{equation}
where $(\theta_1,\,\phi_1)$ are the angles of $\vec a_1$ with respect to the axis ${\vec e}_z$ and $(\theta_{12},\,\phi_{12})$ are the angles of $\vec a_2$ with respect to the vector ${\vec a}_1$.


\begin{thebibliography}{99}

    \bibitem{IW-1} N. Isgur and M. Wise, {\itshape Weak decays of heavy mesons in the static quark approximation}, Phys. Lett. B {\bfseries 232}, 113 (1989); {\itshape Weak transition form factors between heavy mesons}, Phys. Lett. B {\bfseries 237}, 527 (1990).
    
    \bibitem{BJORKEN} J. D. Bjorken, invited talk at Les Rencontres de la Vall\'ee d'Aoste, La Thuile, SLAC-PUB-5278, 1990.
    
    \bibitem{IW-2} N. Isgur and M. Wise, {\itshape Excited charm mesons in semileptonic $\overline{B}$ decay and their contributions to a Bjorken sum rule}, Phys. Rev. D {\bfseries 43}, 819 (1991).
    
    \bibitem{URALTSEV-1} N. Uraltsev, {\itshape New exact heavy quark sum rules}, Phys. Lett. B {\bfseries 501}, 86 (2001), arXiv:hep-ph/0011124; {\itshape A few aspects of heavy quark expansion}, J. Phys. G {\bfseries 27}, 1081 (2001), arXiv:hep-ph/0012336.
    
    \bibitem{LOR-1} A. Le Yaouanc, L. Oliver and J.-C. Raynal, {\itshape Sum rules in the heavy quark limit of QCD}, Phys. Rev. D {\bfseries 67}, 114009 (2003), arXiv:hep-ph/0210233.
    
    \bibitem{LOR-2} A. Le Yaouanc, L. Oliver and J.-C. Raynal, {\itshape Bounds on the derivatives of the Isgur–Wise function from sum rules in the heavy quark limit of QCD}, Phys. Lett. B {\bfseries 557}, 207 (2003), arXiv:hep-ph/0210231.
    
    \bibitem{LOR-3} A. Le Yaouanc, L. Oliver and J.-C. Raynal, {\itshape Lower bounds on the curvature of the Isgur-Wise function}, Phys. Rev. D {\bfseries 69}, 094022 (2004), arXiv:hep-ph/0307197.
    
    \bibitem{LOR-4} A. Le Yaouanc, L. Oliver and J.-C. Raynal, {\itshape Bound on the curvature of the Isgur-Wise
    function of the baryon semileptonic decay $\Lambda_b \to \Lambda_c \ell \overline{\nu}_\ell$ }, Phys. Rev. D {\bfseries 79}, 014023 (2009), arXiv:0808.2983 [hep-ph].
    
    \bibitem{COVARIANT-QM} A. Le Yaouanc, L. Oliver, O. P\`ene, J.-C. Raynal, {\itshape Covariant quark model of form-factors in the heavy mass limit}, Phys. Lett. B {\bfseries 365} (2), 319-326 (1996), arXiv:hep-ph/9507342.
    
    \bibitem{MORENAS-1} V. Mor\' enas, A. Le Yaouanc, L. Oliver, O. P\`ene and J.-C. Raynal,  {\itshape $B \to D^{**}$ semileptonic decay in covariant quark models \`a la Bakamjian-Thomas}, Phys. Lett. B {\bfseries 386}, 315-327 (1996), arXiv:hep-ph/9605206.
    
    \bibitem{GODFREY-ISGUR} S. Godfrey and N. Isgur, {\itshape Mesons in a Relativized Quark Model with Chromodynamics}, Phys. Rev. D{\bfseries 32}, 189 (1985).
    
    \bibitem{MORENAS-2} V. Mor\' enas, A. Le Yaouanc, L. Oliver, O. P\`ene and J.-C. Raynal, {\itshape Quantitative predictions for $B$ semileptonic decays into $D, D^*$ and the orbitally excited $D^{**}$ in quark models \`a la Bakamjian-Thomas}, Phys. Rev. D {\bfseries 56}, 5668-5680 (1997), arXiv:hep-ph/9706265.
    
    \bibitem{MORENAS-3} V. Mor\' enas, A. Le Yaouanc, L. Oliver, O. P\`ene and J.-C. Raynal, {\itshape Decay constants in the heavy quark limit in models à la Bakamjian and Thomas}, Phys. Rev. D {\bfseries 58}, 114019 (1998), arXiv:hep-ph/9710298.

    \bibitem{LEYAOUANC} D. Bečirević, V. Mor\' enas, A. Le Yaouanc, L. Oliver, {\itshape Simple operator formulation of the Bakamjian-Thomas approach to heavy quark current, with generalisation to HQET, and with applications to transitions of $\Lambda_b$}, arXiv:1907.11613 [hep-ph].

    \bibitem{CAPSTICK-ISGUR} S. Capstick and N. Isgur, {\itshape Baryons in a relativized quark model with chromodynamics}, Phys. Rev. D {\bfseries 34}, 2809 (1986).
    
    \bibitem{SILVESTRE-BRAC} B. Silvestre-Brac, {\itshape The cluster model and the generalized Brody-Moshinsky coefficients}, J. Physique {\bfseries 46}, 1087 (1985).

    \bibitem{LEGENDRE} S.N. Samaddar, {\itshape Some integrals involving associated Legendre functions},  Mathematics of Computation, Vol. 28, Number 125 (1974).

    \bibitem{clebsch} M. Kibler, C. Campigotto and Yu.F. Smirnov, {\itshape Recursion Relations for Clebsch-Gordan Coefficients of $U_q(su_2)$ and $U_q(su_{1,1})$}, arXiv:hep-th/9407065.
    
    \bibitem{DIQ} D. Bečirević, A. Le Yaouanc, V. Morénas and L. Oliver, {\itshape Heavy baryon wave functions, Bakamjian-Thomas approach to form factors, and observables in ${\Lambda_b \to \Lambda_c\left({1 \over 2}^\pm \right) \ell \overline{\nu}}$ transitions}, Phys. Rev. D {\bfseries 102}, 094023 (2020), arXiv:hep-ph/2006.07130.


\end{thebibliography}
\end{document}